\documentclass[fleqn,usenatbib]{mnras}
\usepackage{newtxtext,newtxmath}

\usepackage[T1]{fontenc}
\usepackage{ae,aecompl}
\pdfoutput=1

\usepackage{graphicx}	
\usepackage{amsmath}
\usepackage{amssymb}
\usepackage{enumitem} 
\newcommand{\FIREurl}{\href{http://fire.northwestern.edu}{\url{http://fire.northwestern.edu}}}
\newcommand{\gizmourl}{\href{http://www.tapir.caltech.edu/~phopkins/Site/GIZMO.html}{\url{http://www.tapir.caltech.edu/~phopkins/Site/GIZMO.html}}}
\newcommand\altaffilmark[1]{$^{#1}$}
\newcommand\altaffiltext[1]{$^{#1}$}

\title[Gas on FIRE]{\vspace{-0.5cm} Gas kinematics, morphology, and angular momentum in the FIRE simulations\vspace{-0.5cm}}

\vspace{-0.2cm}
\author[El-Badry et al.]{
\parbox[t]{\textwidth}{ 
Kareem El-Badry\thanks{E-mail: kelbadry@berkeley.edu}\altaffilmark{1},
Eliot Quataert\altaffilmark{1}, 
Andrew Wetzel\altaffilmark{2,3,4}\thanks{Caltech-Carnegie Fellow},
Philip F.~Hopkins\altaffilmark{2},
Daniel R. Weisz\altaffilmark{1}, 
T.~K.\ Chan\altaffilmark{5},
Alex Fitts\altaffilmark{6},
Michael Boylan-Kolchin\altaffilmark{6},
Du\v{s}an Kere\v{s}\altaffilmark{5}, 
Claude-Andr{\'e} Faucher-Gigu{\`e}re\altaffilmark{7}, 
and Shea Garrison-Kimmel\altaffilmark{2}
} 
\vspace*{6pt} \\
\altaffiltext{1}{Department of Astronomy and Theoretical Astrophysics Center, University of California Berkeley, Berkeley, CA 94720} \\
\altaffiltext{2}{TAPIR, Mailcode 350-17, California Institute of Technology, Pasadena, CA 91125} \\
\altaffiltext{3}{The Observatories of the Carnegie Institution for Science, Pasadena, CA 91101} \\
\altaffiltext{4}{Department of Physics, University of California, Davis, CA 95616} \\
\altaffiltext{5}{Department of Physics, Center for Astrophysics and Space Sciences, University of California at San Diego, La Jolla, CA 92093} \\ 
\altaffiltext{6}{Department of Astronomy, The University of Texas at Austin, Austin, TX 78712} \\
\altaffiltext{7}{Department of Physics and Astronomy and CIERA, Northwestern University, Evanston, IL 60208} \\ 
\vspace{-0.7cm}
}

\date{Submitted to MNRAS, May 2017\vspace{-0.5cm}}

\pubyear{2017}

\begin{document}
\label{firstpage}
\pagerange{\pageref{firstpage}--\pageref{lastpage}}
\maketitle

\begin{abstract}
We study the $z=0$ gas kinematics, morphology, and angular momentum content of isolated galaxies in a suite of cosmological zoom-in simulations from the FIRE project spanning $\rm M_{star} = 10^{6-11}M_{\odot}$. 
Gas becomes increasingly rotationally supported with increasing galaxy mass. In the lowest-mass galaxies ($\rm M_{star} < 10^{8}\,M_{\odot}$), gas fails to form a morphological disk and is primarily dispersion and pressure supported. At intermediate masses ($\rm M_{star} = 10^{8 - 10}\,M_{\odot}$), galaxies display a wide range of gas kinematics and morphologies, from thin, rotating disks, to irregular spheroids with negligible net rotation.  All the high-mass ($\rm M_{star} = 10^{10-11}\,M_{\odot}$) galaxies form rotationally supported gas disks.
Many of the halos whose galaxies fail to form disks harbor high angular momentum gas in their circumgalactic medium. 
The ratio of the specific angular momentum of gas in the central galaxy to that of the dark-matter halo increases significantly with galaxy mass, from $\langle j_{\rm gas} \rangle / \langle j_{\rm DM} \rangle \sim 0.1$ at $\rm M_{\rm star}=10^{6-7} \rm M_{\odot}$ to $\langle j_{\rm gas} \rangle / \langle j_{\rm DM} \rangle \sim 2$ at $\rm M_{star}=10^{10-11}\,M_{\odot}$.
The reduced rotational support in the lowest-mass galaxies owes to (a) stellar feedback and the UV background suppressing the accretion of high-angular momentum gas at late times, and (b) stellar feedback driving large non-circular gas motions.
We broadly reproduce the observed scaling relations between galaxy mass, gas rotation velocity, size, and angular momentum, but may somewhat underpredict the incidence of disky, high-angular momentum galaxies at the lowest observed masses ($\rm M_{star} = (10^{6} - 2\times 10^{7})\,M_{\odot}$).
Stars form preferentially from low-angular momentum gas near the galactic centre and are less rotationally supported than gas. The common assumption that stars follow the same rotation curve as gas thus substantially overestimates the simulated galaxies' stellar angular momentum, particularly at low masses. 
\end{abstract}

\begin{keywords}
galaxies: kinematics and dynamics -- galaxies: irregular -- galaxies: dwarf
\vspace{-0.2cm}
\end{keywords}

%%%%%%%%%%%%%%%%%%%%%%%%%%%%%%%%%%%%%%%%%%%%%%%%%%

%%%%%%%%%%%%%%%%% BODY OF PAPER %%%%%%%%%%%%%%%%%%

\section{Introduction}
\label{sec:intro}
Star-forming low-mass $(\rm M_{\rm star}\lesssim 10^{9.5} \rm M_{\odot})$ galaxies in the local Universe exhibit a rich diversity of morphology and kinematic structure. Even in isolated environments, many low-mass galaxies have irregular gas distributions and disordered velocity fields showing significant non-circular motions \citep{Begum_2008, Walter_2008, Ott_2012, Hunter_2012, Roychowdhury_2013}. While late-type galaxies at higher masses fall on a relatively tight scaling relation between rotation velocity and mass, many low-mass galaxies scatter off the relation to lower rotation velocities \citep{Cortese_2014}. Low-mass galaxies are also less likely to form disks than more massive galaxies in comparable environments \citep{Simons_2015}. The observed decrease in rotational support at low masses remains imperfectly understood theoretically, but it may be related to low-mass galaxies' susceptibility to disruption by feedback processes.

Angular momentum has long been recognized as a fundamental quantity in galaxy formation and evolution \citep{Fall_1980, Fall_1983}. Simple semi-analytic models \citep[e.g.][]{Dalcanton_1997, Mo_1998, Hernandez_2006, Romanowsky_2012, Fall_2013} show that many observed galaxy scaling relations arise naturally in the CDM framework if angular momentum is approximately conserved during galaxy formation and galaxies inherit the specific angular momentum of their dark matter halos. Hydrodynamic simulations of galaxy formation in large volumes \citep{Teklu_2015, Genel_2015, Rodriguez_2016, Zavala_2016, Lagos_2017, Penoyre_2017, Zjupa_2017, Grand_2017} have also converged on a paradigm in which the structural parameters of galaxies, such as size, colour, and morphology, follow, to zeroth order, from the mass and angular momentum of their host halos. 

Observational studies \citep[e.g.][]{Emsellem_2007, Romanowsky_2012, Obreschkow_2014, Cortese_2016} have detected the rough scaling relations between galaxy mass, angular momentum, and morphology predicted for hierarchical galaxy formation in the CDM paradigm, though uncertainties remain in the precise form of these relations, particularly at lower masses  \citep{Butler_2016, Chowdhury_2017}. Additional works have clarified how galaxies' angular momentum evolution is affected by environment and merger history \citep[e.g.][]{Naab_2014, Rodriguez_2016, Sokolowska_2017, Penoyre_2017, Lagos_2017b}, and by the geometry of gas accretion from large scale structure \citep{White_1984, Keres_2005, Sales_2012, Stewart_2013, Danovich_2015}. 

However, uncertainty persists in the effects of internal feedback processes on galaxies' angular momentum content and morphology, especially in low-mass galaxies. A host of theoretical works \citep[e.g.][]{SommerLarsen_1999, Binney_2001, Thacker_2001, Governato_2007, Governato_2010, Dutton_2009, Agertz_2011, Guedes_2011, Brook_2011, Brook_2012, Ubler_2014, Genel_2015, Agertz_2016, Sokolowska_2017, DeFelippis_2017} have shown that feedback-driven outflows can preferentially remove low angular momentum gas from galaxies, particularly at high redshift. This mechanism can suppress bulge formation and increase the average specific angular momentum of galaxies' retained baryons, allowing for the formation of centrifugally supported disks with high specific angular momentum. Efficient feedback has been heralded as the key to forming bulgeless disk galaxies similar to those observed in the local Universe \citep[e.g.][]{Kormendy_2010, Kormendy_2016}, as early simulations with weak feedback \citep[e.g.][]{Katz_1991, Navarro_1997, Steinmetz_1999, Kaufmann_2007, Stinson_2010} produced galaxies with too little angular momentum that were too small and bulge-dominated.  

On the other hand, if feedback persists at late times, it can drive turbulence and large-scale outflows, preventing gas from settling into a rotationally supported disk \citep{Stinson_2006, Kaufmann_2007_disk, Gonzalez_2014, Roskar_2014, Muratov_2015, ElBadry_2016, Agertz_2016, Read_2016, Dutton_2016, Hayward_2017, Wheeler_2017, DiCintio_2017, Verbeke_2017}. This is especially true for low-mass galaxies, which are most susceptible to feedback effects. Because low-mass galaxies are typically not well-resolved in large-volume simulations, comparatively few theoretical studies of galaxy angular momentum to date have focused on low-mass galaxies. 

In this paper, we study the angular momentum, gas morphology, and kinematics of a suite of cosmological zoom-in simulations from the Feedback in Realistic Environments (FIRE)  project\footnote{See the FIRE project website: \FIREurl} spanning five decades in stellar mass ($\rm M_{\rm star} = 10^{6 - 11} \rm M_{\odot}$). We show that, due to a combination of low angular momentum and stellar feedback, most low-mass galaxies in our simulations do not form gas or stellar disks but instead remain dispersion-supported at late times. In addition, the conventional wisdom that feedback preferentially removes low-angular momentum material, leading galaxies to have higher specific angular momentum than their host halos, does not hold at low masses: the majority of our low-mass halos have \textit{depleted} angular momentum in their baryonic components (particularly within the central galaxy) compared to the dark matter.

We organize this paper as follows. In Section~\ref{sec:simulations}, we describe the FIRE simulations and summarize the properties of our sample. We then quantify rotational support in our galaxies by studying their gas kinematics (Section~\ref{sec:kinematics}) and shape (Section~\ref{sec:shapes}). In Sections~\ref{sec:ang_mom} and \ref{sec:mass_dists}, we investigate the angular momentum and mass profiles of gas in galaxies' diffuse halos. We compare our simulations' predictions to observed galaxy scaling relations in Section~\ref{sec:obs}. We discuss our results in Section~\ref{sec:discussion} and conclude in Section~\ref{sec:summary}.

\section{Simulations}
\label{sec:simulations}

\begin{table*}
\centering
\caption{Summary of the simulations at $z=0$}
\label{tab:properties}
\begin{tabular}{p{0.8cm} | p{1.3cm}| p{1.4 cm} |  p{1.3 cm} | p{0.8 cm} | p{0.8 cm} | p{0.8 cm} | p{0.8cm}| p{1.8 cm} | p{2 cm} | p{1.4 cm}} 

Name &  $\log (\rm M_{\rm star})$  $(\rm M_{\odot})$ &  $\log (M_{200 \rm m})$ $(\rm M_{\odot})$ &  $\log (M_{\rm HI})$ $(\rm M_{\odot})$ & $R_{\rm HI}$ $(\rm kpc)$  & $f_{\rm HI}$ & $m_{\rm b}$ $(\rm M_{\odot})$ & $m_{\rm DM}$ $(\rm M_{\odot})$ & kinematic gas disk fraction & Morphological gas disk? & Reference \\ 
\hline
\texttt{m10e}   &  6.3  &  10.1  & 7.4    &   1.6   & 0.93  & 500   & 2500   & 0.29   & no       & A \\
\texttt{m10q}   &  6.3  &  9.9   & 6.6    &   0.7   & 0.69  & 260   & 1300   & 0.21   & no       & B \\
\texttt{m10g}   &  6.7  &  9.9   & 6.9    &   0.8   & 0.61  & 500   & 2500   & 0.28   & no       & A \\ 
\texttt{m10h}   &  6.8  &  10.2  & 7.2    &   1.0   & 0.70  & 500   & 2500   & 0.34   & no       & A \\
\texttt{m10j}   &  6.9  &  10.1  & 6.7    &   0.6   & 0.40  & 500   & 2500   & 0.14   & no       & A \\
\texttt{m10k}   &  7.0  &  10.1  & 7.1    &   1.0   & 0.54  & 500   & 2500   & 0.38   & no       & A \\
\texttt{m10y}   &  7.0  &  10.2  & 7.1    &   0.9   & 0.57  & 260   & 1250   & 0.41   & no       & B \\
\texttt{m10f}   &  7.0  &  10.2  & 7.5    &   1.6   & 0.75  & 500   & 2500   & 0.45   & no       & A \\
\texttt{m10l}   &  7.1  &  10.1  & 6.9    &   0.9   & 0.43  & 500   & 2500   & 0.31   & no       & A \\
\texttt{m10m}   &  7.1  &  10.1  & 7.3    &   1.1   & 0.57  & 500   & 2500   & 0.62   & no       & A \\
\texttt{m10z}   &  7.6  &  10.6  & 8.0    &   2.9   & 0.74  & 260   & 1250   & 0.57   & marginal & B \\
\texttt{m11b}   &  8.0  &  10.7  & 9.0    &   10.1  & 0.90  & 2100  & 10000  & 0.95   & yes      & B \\
\texttt{m11a}   &  8.1  &  10.7  & 8.1    &   3.8   & 0.54  & 2100  & 10000  & 0.11   & no       & B \\
\texttt{m11q}   &  8.6  &  11.2  & 8.1    &   3.1   & 0.26  & 880   & 4400   & 0.43   & no       & B \\
\texttt{m11c}   &  9.0  &  11.2  & 8.7    &   4.2   & 0.37  & 2100  & 10375  & 0.46   & no       & B \\
\texttt{m11i}   &  9.0  &  10.9  & 8.9    &   6.8   & 0.48  & 7070  & 35200  & 0.67   & marginal & D \\
\texttt{m11e}   &  9.1  &  11.2  & 9.0    &   9.5   & 0.43  & 7070  & 35200  & 0.55   & marginal & D \\
\texttt{m11h}   &  9.6  &  11.3  & 9.4    &   11.5  & 0.39  & 7070  & 35200  & 0.92   & yes      & D \\
\texttt{m11d}   &  9.6  &  11.5  & 9.4    &   18.3  & 0.35  & 7070  & 35200  & 0.72   & marginal & D \\
\texttt{m11f}   &  10.4 &  11.7  & 9.8    &   15.1  & 0.22  & 12000 & 83000  & 0.99   & yes      & B \\
\texttt{m11g}   &  10.7 &  11.8  & 9.7    &   12.1  & 0.10  & 12000 & 83000  & 0.99   & yes      & E \\
\texttt{m12i}   &  10.8 &  12.1  & 10.1   &   25.7  & 0.17  & 7070  & 35200  & 1.00   & yes      & C \\
\texttt{m12f}   &  10.9 &  12.2  & 10.3   &   31.8  & 0.18  & 7070  & 35200  & 1.00   & yes      & B \\
\texttt{m12m}   &  11.1 &  12.2  & 9.9    &   18.0  & 0.06  & 7070  & 35200  & 0.98   & yes      & B \\

\hline
\end{tabular}
\begin{flushleft}
$\rm M_{\rm star}$ is the stellar mass within $3\times R_{1/2}$, where $R_{1/2}$ is the 3D stellar half-mass radius. $M_{200 \rm m}$ is the total mass within $R_{\rm 200m}$, where $R_{200 \rm m}$ is the radius within which the matter density is $200 \times$ the mean matter density. $M_{\rm HI}$ is the mass of neutral hydrogen within $0.1R_{\rm 200m}$. $R_{\rm HI}$ is the radius of the largest annulus inside which the HI surface density $\Sigma_{\rm HI}$ exceeds $\rm 1\, \rm M_{\odot}\,pc^{-2}$ (Section~\ref{sec:D_HI}). $f_{{\rm HI}}=M_{{\rm HI}}/(M_{{\rm HI}}+M_{{\rm star}})$ is the neutral gas fraction. $m_{\rm b}$ and $m_{\rm DM}$ are the average baryon and dark matter particle masses. The ``kinematic gas disk fraction'' is the fraction of HI gas in rotationally supported orbits, which we define as $\epsilon > 0.5$ (see Section~\ref{sec:circ}). We also provide a (strictly qualitative) summary of whether or not each galaxy has a gas disk based on visual morphology; mock HI moment maps for all our galaxies can be found in Appendix~\ref{sec:idvid_gals}. References: A: \citet{Fitts_2016}; B: \citet{Hopkins_2017}; C: \citet{Wetzel_2016}; D: this work; E: Chan et al., in prep. 
\end{flushleft}
\end{table*}

We study baryonic cosmological zoom-in simulations from the FIRE project \citep{Hopkins_2014}. All the halos studied in this work were simulated with the \texttt{GIZMO}\footnote{A public version of this code is available at \gizmourl} hydrodynamics code \citep{Hopkins_2015} in the Lagrangian meshless finite mass (MFM) mode, using the FIRE-2 model for galaxy formation and feedback \citep{Hopkins_2017}. We briefly summarise the simulations here, directing the reader to \citet{Hopkins_2017} for extensive description of the FIRE physics models, numerical methods, and resolution tests. FIRE-2 implements the same primary star formation and stellar feedback physics models as the original FIRE simulations, but with MFM hydrodynamics and several other numerical improvements.

\texttt{GIZMO} uses an improved version of the TreePM gravity solver from \texttt{GADGET} \citep{Springel_2005}. Force softening for baryons follows the adaptive algorithm from \citet{Price_2007}. Radiative cooling is implemented using rates from \texttt{CLOUDY} \citep{Ferland_2013} across $10-10^{10}$ K; these include ionized, atomic, and molecular cooling as well as metal-line cooling for 11 elements. \texttt{GIZMO} incorporates ionization and heating from the spatially uniform, redshift dependent UV background computed in \citet{FaucherGiguere_2009}.  

The FIRE feedback model incorporates the effects of stellar winds, radiation pressure from massive stars, local photo-ionization and photoelectric heating, and core-collapse and type Ia supernovae, as described in \citet{Hopkins_2014} and \citet{Hopkins_2017}. Energy, momentum, mass, and metal returns are calculated particle-by-particle from stellar evolution models at each timestep, as computed by \texttt{STARBURST99} \citep[v7.0;][]{Leitherer_1999, Leitherer_2010, Leitherer_2014} for a \citet{Kroupa_2001} IMF.  Star formation occurs only in gas which is dense ($n_{\rm H} > 1000\,\rm cm^{-3}$), self-gravitating (following \citealt{Hopkins_2013}), self-shielding and molecular (following \citealt{Krumholz_2011}), and Jeans unstable, and proceeds with an instantaneous efficiency of 100\% per local free-fall time. We use the zoom-in technique \citep{Porter_1985, Onorbe_2014} to re-simulate individual halos taken from a large-volume, low-resolution dark matter-only simulation at high resolution in a cosmological context. Initial conditions are generated at $z\approx 100$ with \texttt{MUSIC} \citep{Hahn_2011}. 

Galaxies simulated with the FIRE model have been shown to reproduce a wide range of observables, including the $\rm M_{\rm star} - M_{\rm halo}$ relation \citep{Hopkins_2014}, the $\rm M_{\rm star}-$metallicity relation \citep{Ma_2016}, realistic outflows and CGM enrichment \citep{Muratov_2015,Muratov_2016,AnglesAlcazar_2016}, the dense HI content of high-redshift galaxy halos \citep{FaucherGiguere_2015, FaucherGiguere_2016}, cored density profiles in low-mass galaxies \citep{Chan_2015, Onorbe_2015}, dispersion-supported stellar kinematics \citep{Wheeler_2017, ElBadry_2017}, the $\rm M_{\rm star}-$size relation \citep{ElBadry_2016, Fitts_2016}, the Kennicutt-Schmidt law \citep{Orr_2017}, thin and thick disks in MW-mass galaxies \citep{Ma_2017}, and realistic populations of satellite dwarf galaxies around MW-mass hosts \citep{Wetzel_2016}.

At the resolution of the simulations used in this work, most of the galaxy-wide properties that we explore here are insensitive to resolution and sub-grid numerical parameters \citep[see][for extensive tests]{Hopkins_2017}. Of particular relevance to this work, \texttt{GIZMO}'s MFM mode conserves angular momentum well, integrating the evolution of gaseous disks for hundreds of orbits with sub-percent level losses \citep{Hopkins_2015}. This is critical for studies of angular momentum and rotational support in galaxies, as previous works have found significant spurious angular momentum loss due to numerical effects, particularly at low resolution \citep{Governato_2004, Kaufmann_2007}. MFM resolves a number of numerical issues that have been shown to cause spurious angular momentum loss in galaxy formation simulations \citep{Zhu_2016, Hopkins_2017}, including the unphysical cooling of cold low-angular momentum ``blobs'' from the hot halo into galaxies and numerical torques red between galaxies and their halos \citep[e.g.][]{Okamoto_2003, Agertz_2007, Torrey_2012, Keres_2012, Few_2016}. 

Table~\ref{tab:properties} summarises our sample of simulated halos. We study 24 galaxies with stellar masses $10^{6} \rm M_{\odot} \lesssim \rm M_{\rm star} \leq 10^{11} \rm M_{\odot}$ and halo masses $10^{10} \rm M_{\odot} \lesssim M_{\rm 200m} \leq 10^{12} \rm M_{\odot}$. All the halos we study in this work were selected to be isolated at $z=0$, meaning that they have no more massive neighbours within at least $3R_{\rm 200m}$. Most of the simulations in our sample have already been presented in previous works. Simulations of four intermediate-mass halos (\texttt{m11i, m11d, m11e}, and \texttt{m11h}) are presented here for the first time. All the galaxies in our sample, including those presented here for the first time, were run with the identical simulation code and physics, as described in \citet{Hopkins_2017}. Besides our stellar mass range and the requirement that halos be isolated at $z=0$, we have not imposed any selection criteria on galaxy or halo properties; our sample spans a representative range of galaxy morphologies and halo spin, concentration, and formation time. 

We identify galaxy centres using an iterative ``shrinking-spheres'' method \citep{Power_2003, Fitts_2016} wherein we recursively compute the centre-of-mass of dark matter particles in a spherical region, reducing the sphere's radius by 50\% and re-centering on the new barycenter at each iteration. We have confirmed that, for all the galaxy properties studied in this work, centering on star particles as opposed to dark matter yields identical results. 

Throughout this work, we focus our kinematic analysis primarily on neutral atomic hydrogen (``HI''). We identify HI gas using the neutral hydrogen abundance tracked by \texttt{GIZMO}. Following \citet{Orr_2017}, we exclude gas with $T<300\,\rm K$ and $n_{\rm H} > 10\,\rm cm^{-3}$, as such gas is likely to be molecular. This is always a small fraction of the galaxies' neutral gas mass, and we have verified that rejecting or including it when calculating HI properties has a negligible effect on our results. 

In the next sections, we investigate the kinematics and morphologies of our full sample of galaxies in detail. We summarise whether galaxies are kinematically and  morphologically disky in columns 8 and 9 of Table~\ref{tab:properties}. Galaxies which are rotationally supported have high kinematic disk fractions (Section~\ref{sec:circ}). Galaxies whose HI zeroth moment (column density) maps show clear rotational flattening are designated as having a morphological disk. We also indicate which systems are marginally flattened (typically very ``puffy'' disks). 

\section{Gas Kinematics}
\label{sec:kinematics}

\begin{figure*}
\includegraphics[width=\textwidth]{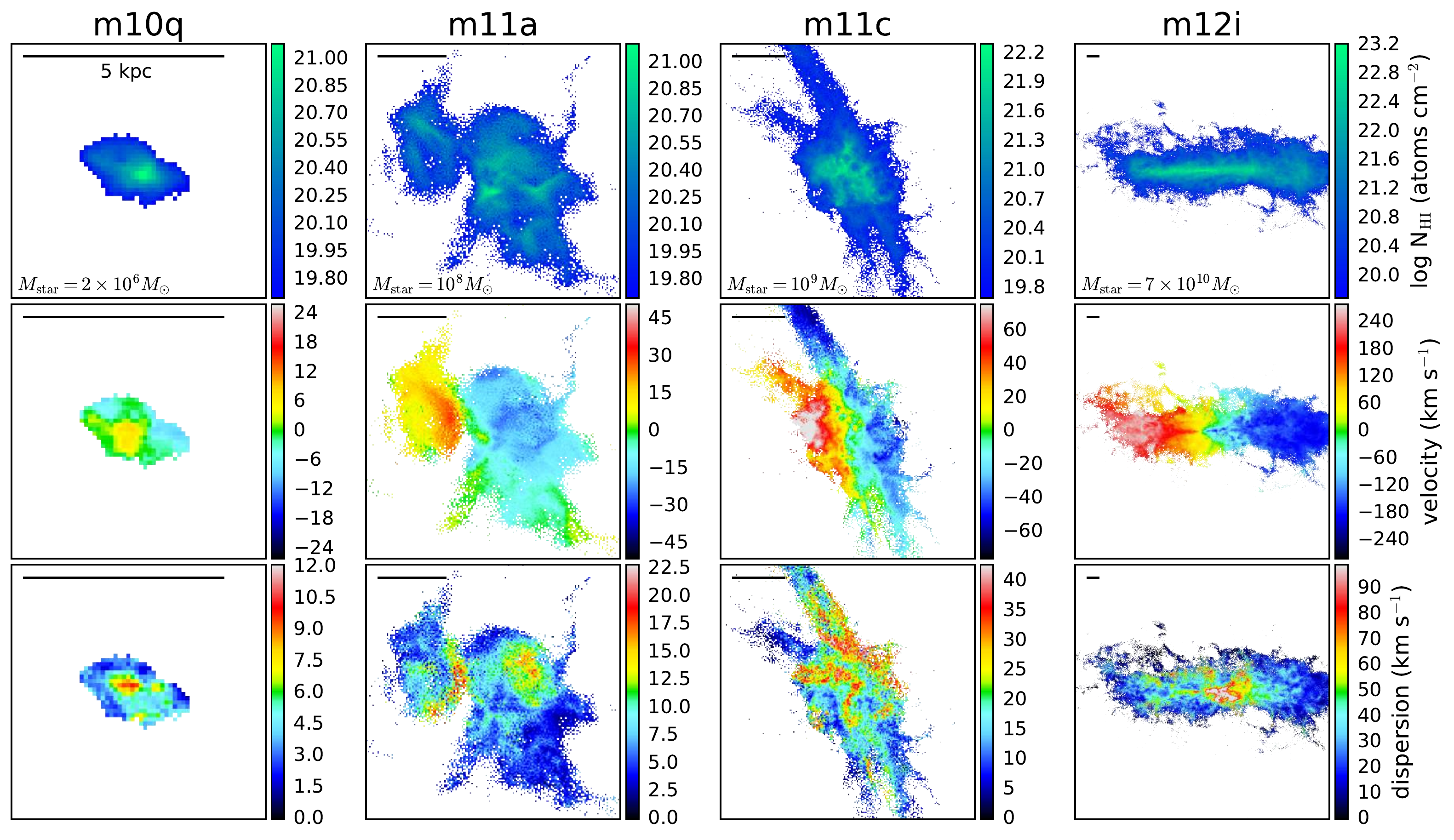}
\caption{HI column density (top), line-of-sight velocity (middle), and velocity dispersion (bottom) for four galaxies spanning the range of masses and morphologies in our sample at $z=0$. Each galaxy is viewed ``edge on''; i.e., along a line-of-sight perpendicular to its net HI angular momentum vector. We only show pixels with column densities $N_{\rm HI}>5\times 10^{19}$ cm$^{-2}$, comparable to the sensitivity of resolved studies of nearby galaxies. Horizontal line represents a length of 5 kpc in all panels. Galaxies become increasingly rotationally supported with increasing mass; of these galaxies, only \texttt{m12i} (a $\sim$MW-mass galaxy) has a clear morphological disk. Many low and intermediate-mass galaxies in our sample (like \texttt{m11c}) fail to form morphological disks despite exhibiting clear signs of rotation in their velocity fields.}
\label{fig:vel_maps}
\end{figure*}

Figure~\ref{fig:vel_maps} shows mock HI moment maps of four galaxies spanning the range of masses and morphologies found in our suite, with $\rm M_{\rm star}$ increasing from left to right. Similar maps for all the galaxies in our sample are presented in Appendix~\ref{sec:idvid_gals}. We generate mock data cubes with a fixed spatial resolution of 100 pc and a limiting sensitivity of $N_{\rm HI} = 5\times 10^{19}\,\rm cm^{-2}$, similar to resolved HI surveys of nearby galaxies conducted with the VLA \citep[e.g.][]{Walter_2008, Hunter_2012, Ott_2012}. The galaxies are viewed ``edge-on'' in order to highlight their maximal velocity gradients; i.e., we orient our coordinate system so that the $\hat{\mathbf{z}}$ axis is parallel to the net angular momentum vector of HI. For the mean velocity (first moment) maps, we chose the colour stretch to range over $\pm v_{\rm halo}$, where $v_{\rm halo}$ is the maximum circular velocity of the host halo in the radial range where gas kinematics are measured.

The two lowest-mass galaxies in Figure~\ref{fig:vel_maps} do not exhibit significant global velocity gradients and show little evidence of net rotation. Indeed, the typical velocity dispersion\footnote{Note that the dispersion shown in Figure~\ref{fig:vel_maps} represents only the dispersion in the relative velocities of different gas particles. The velocity dispersion inferred from observed line widths would include an additional component due to thermal broadening within each gas particle; i.e., $\sigma_{\rm obs}^2= \sigma^2 + c_s^2$, where $c_s \sim (5-10)\,\rm km\,s^{-1}$ for typical ISM temperatures in our galaxies.} in each pixel is comparable to the galaxies' global velocity gradients, indicating that the gas is supported primarily by dispersion or pressure, not rotation. The intermediate-mass galaxy, \texttt{m11c}, shows clear signs of rotation in its velocity field. Despite this, it is not morphologically disky: feedback-driven turbulence and outflows  disrupt the gas, preventing it from settling into a rotationally flattened disk. Finally, the gas in \texttt{m12i} is both rotationally supported and morphologically disky. This galaxy's properties are similar to those of the Milky Way \citep{Wetzel_2016}. 

\subsection{Rotation curves}
\label{sec:rot_curves}

\begin{figure*}
\includegraphics[width=\textwidth]{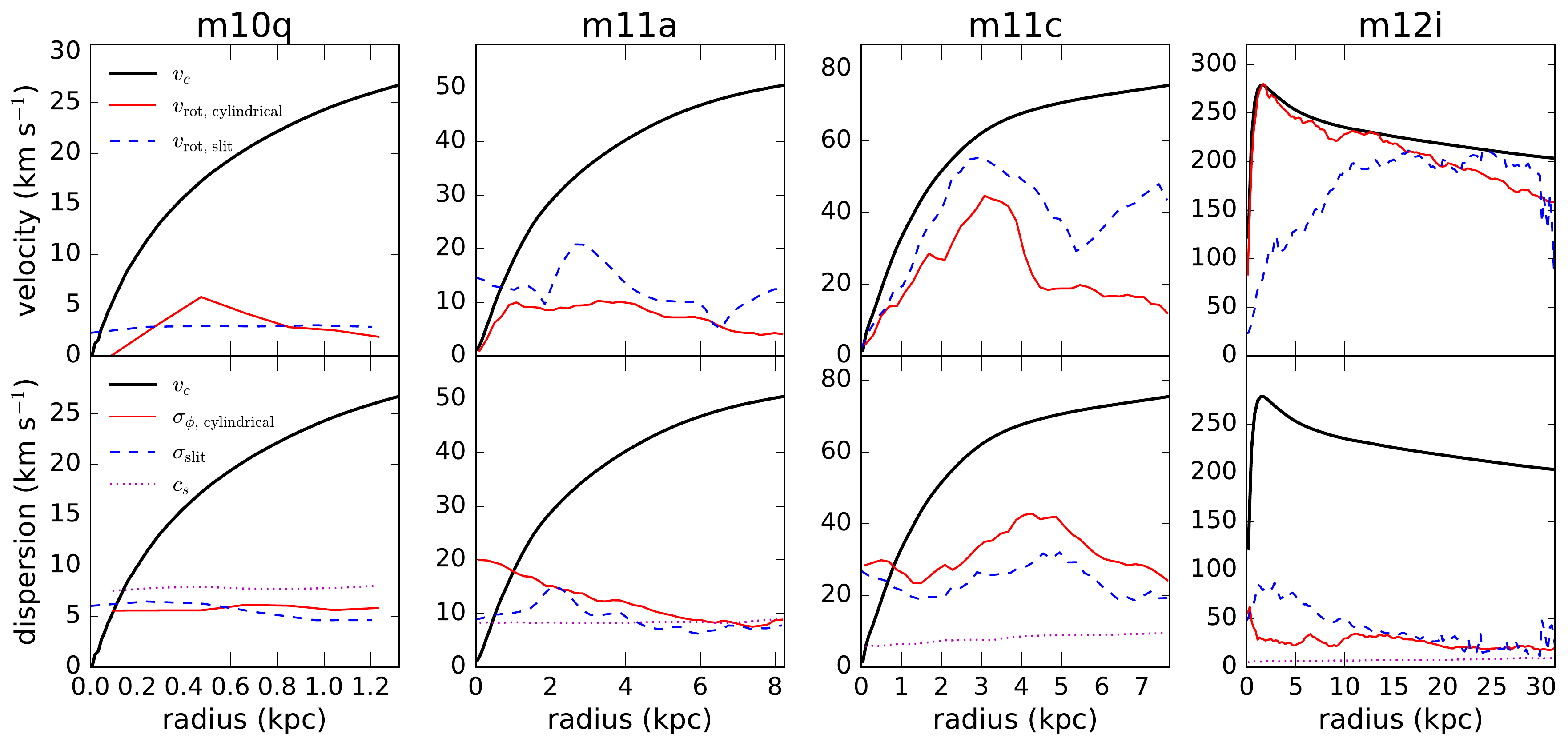}
\caption{Rotation and dispersion curves for gas in four galaxies spanning the range of masses and morphologies represented in our sample. For each galaxy, we compare the theoretical circular velocity ($v_c=\sqrt{GM(<r)/r}$; solid black line) to the mean HI gas rotation velocity (top) and HI velocity dispersion (bottom). We compare quantities measured in cylindrical bins (red) to quantities measured along a slit (averaging over opposite sides of the slit; blue); see Section~\ref{sec:rot_curves} for details. In most of our low-mass galaxies, the gas rotation velocity is well below the circular velocity and $v_{\rm rot} \sim \sigma$, indicating that the gas is primarily supported by dispersion, not rotation. In the lowest-mass galaxies, the mean gas sound speed $c_s$ is comparable to the circular velocity, indicating that thermal pressure support is important. Galaxies become increasingly rotationally supported with increasing mass.}
\label{fig:rot_curves}
\end{figure*}

Galaxy rotation curves provide a straightforward kinematic diagnostic of whether or not a  galaxy is rotationally supported. For a centrifugally supported disk in a spherically symmetric potential, gas at radius $r$ should, by definition, rotate with speed $v_{\rm rot}(r)=v_c(r)=\sqrt{GM(<r)/r}$, where $v_c(r)$ denotes the circular velocity and $M(<r)$ the total mass enclosed within radius $r$. Gas cannot rotate faster than $v_c(r)$ without moving to larger radii; gas rotating slower than $v_c(r)$ is indicative of additional non-rotational support. 

To compute $v_{\rm rot}(r),$ we first orient our coordinate system so that the $\hat{\textbf{z}}$ axis is parallel to the net angular momentum vector of HI in the galaxy. We then compute the azimuthal velocity $v_{\phi}=(xv_{y}-yv_{x})/\sqrt{x^{2}+y^{2}}$ for each gas particle and define $v_{\rm rot}$ in a given cylindrical bin to be the mean $v_{\phi}$ of particles in that bin, weighting by the particles' HI mass. We also define $\sigma_{\phi}=(\langle v_{\phi}^{2}\rangle -\langle v_{\phi}\rangle ^{2})^{1/2}$ as the one-dimensional azimuthal velocity dispersion in a given bin. 

Although these definitions are physically sensible for determining whether a galaxy is rotationally supported, they are not directly comparable to observable quantities. We therefore also compute a ``mock slit'' rotation and velocity dispersion curve for each galaxy, in order to mimic the most readily available observational data. To this end, we align a 5 kpc wide slit with the major axis of the edge-on galaxy and compute the line-of-sight velocities of each gas particle in the slit. We then calculate the mean line-of-sight velocity and velocity dispersion in horizontal bins along the slit. Finally, we mirror line-of-sight velocity and line-of-sight velocity dispersion values across the middle of the slit, defining $v_{\rm rot,\,slit}(r)$ and $\sigma_{\rm slit}(r)$ as the arithmetic mean of the (absolute) mean velocity and dispersion measured on opposite sides of the slit at a projected distance $r$ from the centre. We emphasise that $v_{\rm rot,\,slit}(r)$ simply quantifies the mean line-of-sight velocity at a projected radius $r$; it should not be compared to studies that fit a model to slit-based observations.

In Figure~\ref{fig:rot_curves}, we plot the halo circular velocity $v_c(r)$, actual HI gas rotation velocity $v_{\rm rot}(r)$, HI gas dispersion profile $\sigma(r)$, and the HI-mass weighted mean of the isothermal sound speed, $c_s = \sqrt{kT/\mu m_p}$, for the same four galaxies shown in Figure~\ref{fig:vel_maps}.  Since HI-mass weighted quantities are only well-defined where there is  HI, we plot all quantities out to $r_{\rm max\, HI}$, which we define as the radius of the largest annulus in which the HI surface density exceeds $\rm 0.1\,\rm M_{\odot}\,pc^{-2}$. This is comparable to the radial extent of HI observations in resolved studies of nearby galaxies \citep{Walter_2008}. Note that $r_{\rm max,\, HI}$ is larger than $R_{\rm HI}$ as defined in Table~\ref{tab:properties}. We present similar curves for all the galaxies in our sample in Appendix~\ref{sec:idvid_gals}.

Consistent with the 2D kinematic maps shown in the previous section, the two lower mass galaxies in Figure~\ref{fig:rot_curves}, \texttt{m10q} and \texttt{m11a}, exhibit very little net rotation: they have $v_{\rm rot}(r) \ll v_c(r)$ and $v_{\rm rot}(r) \sim \sigma(r)$ at all radii. In \texttt{m10q}, the sound speed $c_s$ is comparable to $v_c$, especially at small radii, indicating the importance of thermal pressure support. $c_s$ remains at $\lesssim 10\,\rm km\,s^{-1}$ in the higher mass galaxies, but pressure support becomes less important dynamically in higher mass galaxies' deeper gravitational potentials. \texttt{m11c} is somewhat more rotationally supported, with $v_{\rm rot}$ comparable to $v_c$ in the central few kpc; its rotation velocity and dispersion are similar to values measured in the SMC \citep{Stanimirovic_2004}. The galaxy becomes increasingly dispersion-supported at large radii. Inspecting its edge-on velocity field in Figure~\ref{fig:vel_maps}, it is clear that the galaxy's gas is rotating. However, it rotates at well below the local circular velocity in the outer regions, indicating that rotation is not the only source of dynamical support. Such a rotation curve is expected if an initially small, rotating disk is significantly expanded by feedback. In this case, material will maintain its initial angular momentum, so the velocity field will still exhibit coherent rotation, but, by conservation of angular momentum, the rotation velocity will be much lower than before the gas was blown outward. Finally, \texttt{m12i} has $v_{\rm rot} \approx v_c$ and $v_{\rm rot} \gg \sigma$ at all radii, indicating that the galaxy is rotationally supported and gas kinematics trace the gravitational potential. 

For all galaxies in Figure~\ref{fig:rot_curves}, our broad conclusions regarding rotation or dispersion support are unchanged irrespective of whether we consider the three-dimensionally calculated, cylindrically averaged velocities (solid red lines) or line-of-sight, slit-averaged velocities (dashed blue lines). There are nevertheless some qualitative differences between the cylindrically-averaged and mock-slit rotation and dispersion curves. In \texttt{m12i}, $v_{\rm rot,\, slit} < v_{\rm rot,\,cylindrical}$ and $\sigma_{\rm  rot,\, slit} > \sigma_{\rm rot,\,cylindrical}$ near the galactic centre; at large radii, the cylindrical and slit-averaged quantities begin to converge.\footnote{This can be understood geometrically: each annulus in the disk intersects a sight line through the disk at a different angle and thus has a different projected line-of-sight velocity, even if all annuli have approximately the same $v_{\phi}$. Sight lines passing near the galactic centre are contaminated by annuli at large radii, which are nearly perpendicular to the line of sight. These sight lines thus have a higher line-of-sight velocity dispersion and lower mean line-of-sight velocity than sight lines at larger projected radii.} On the other hand, in \texttt{m11a} and \texttt{m11c}, the slit-averaged mean line-of-sight velocity $v_{\rm  rot,\, slit}$ is somewhat \textit{higher} than $v_{\rm rot,\,cylindrical}$. Comparing all our galaxies' cylindrical and mock-slit rotation curves (see Appendix~\ref{sec:idvid_gals}), we find that in disky, rotationally supported galaxies, $v_{\rm rot,\, slit}$ is always less than $v_{\rm rot,\,cylindrical}$ in the central regions. In the dispersion-supported galaxies, the two quantities are comparable in most galaxies, but there are several highly disordered systems in which $v_{\rm rot,\, slit} > v_{\rm rot, \,cylindrical}$.   

We note that attempts to measure the dynamical mass profiles of galaxies like \texttt{m10q} and \texttt{m11a} (and possibly even \texttt{m11c}) from their rotation curves, as is commonly done with the HI rotation curves of observed galaxies \citep[e.g.][]{Oh_2015}, would likely lead to large systematic errors, because $v_{\rm rot}$ is much lower than $v_c$. Of course, this ignores the correction to $v_{\rm rot}(r)$ due to pressure support or ``asymmetric drift'' that is commonly applied to the rotation curves of observed galaxies in studies attempting to measure dynamical mass profiles \citep[e.g.][]{Tully_1978, Meurer_1996, Adams_2014}. However, this correction is predicated on the assumption that galaxies are to first order rotating disks, and that non-circular motions represent a small perturbation \citep{Dalcanton_2010,KuzioDeNaray_2011,Oh_2011}. It cannot be expected to yield reliable results in highly dispersion-supported galaxies, where the perturbation dominates over any net rotation. 

Particularly in the lowest-mass systems, gas is significantly supported through both noncircular bulk gas motions and thermal pressure. We find that in many of the galaxies at $\rm M_{\rm star} \lesssim 10^{8} \rm M_{\odot}$, the mean speed of HI gas particles at a given radius, $\left\langle \sqrt{\mathbf{v}_{{\rm gas}}^{2}}\right\rangle$, is less than the local circular velocity $v_c(r)$ by as much as a factor of two. This suggests that thermal pressure and pressure gradients play an important role in the support of these systems against gravity, and thus that gas kinematics do not directly trace the gravitational potential. Dynamical mass estimates derived from gas kinematics in these galaxies will consequently be unreliable if pressure support and asymmetric bulk motions are not incorporated in the dynamical model.

Several works using other simulations \citep[e.g.][]{Valenzuela_2007, Pineda_2017, Read_2016, Verbeke_2017} have pointed out that $v_{\rm rot} < v_{c}$ is common in dwarf galaxies due to pressure support, particularly near the galactic centre, and have suggested that this may explain the apparent tension between the approximately linearly rising rotation curves observed in most dwarf galaxies and the steeper circular velocity curves predicted by many $\Lambda$CDM simulations \citep[e.g.][]{Moore_1994, Flores_1994, Oman_2015}. The low gas rotation velocities found in our low-mass galaxies support this possibility. That said, the morphologies and velocity fields of many of our low-mass galaxies are likely sufficiently irregular that they would not pass the selection cuts for observational mass-modeling studies; these tend to explicitly select disky galaxies, for which it is easier to construct dynamical models. 

\begin{figure}
\includegraphics[width=\columnwidth]{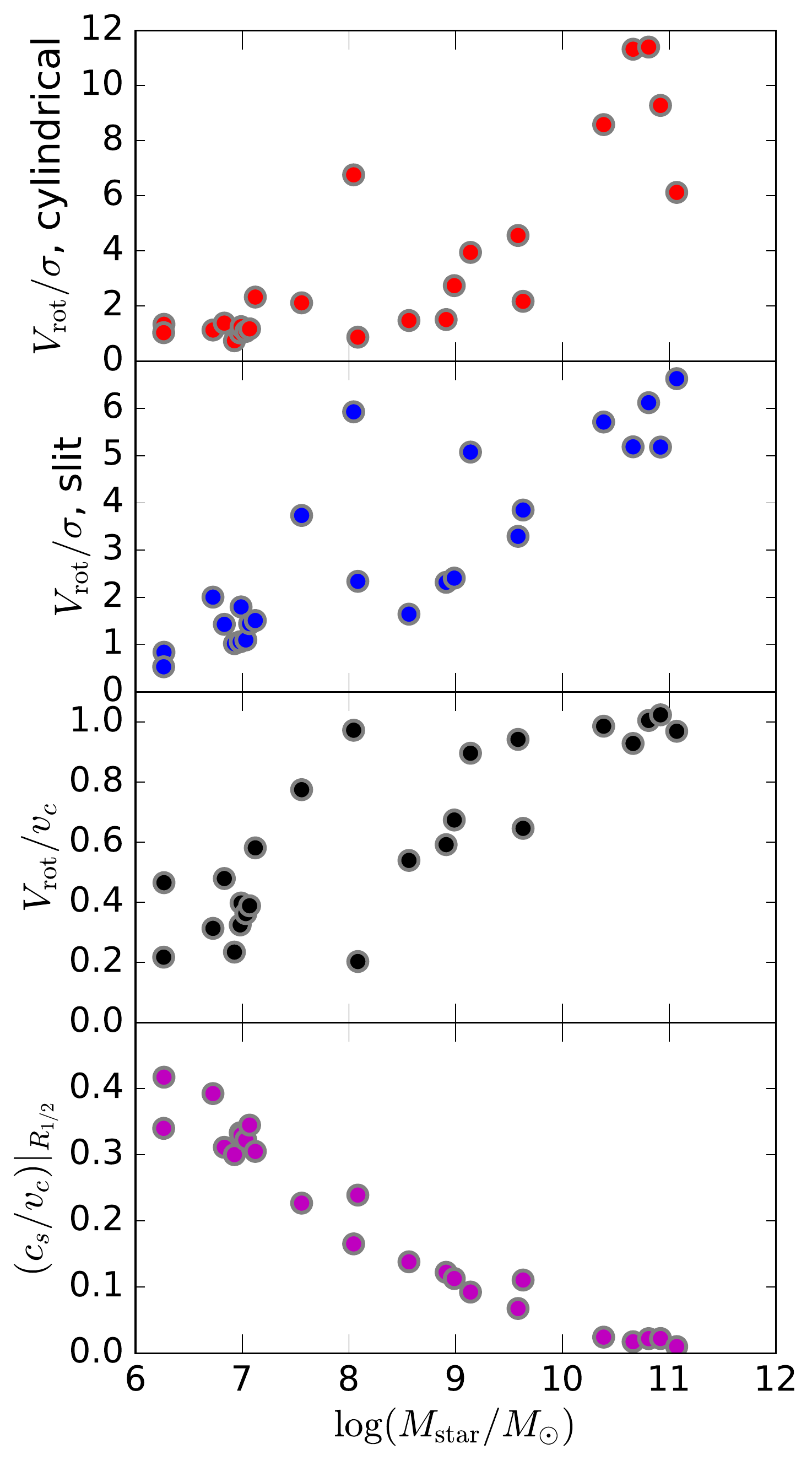}
\caption{\textbf{Top}: HI gas $V_{\rm rot}/\sigma$ vs. $\rm M_{\rm star}$, for all the galaxies in our sample. We define $V_{\rm rot}$ as the maximum of the gas rotation curve in the region with HI (red lines in the top panels of Figure~\ref{fig:rot_curves}) and $\sigma$ as the median of the velocity dispersion profile (red lines in the bottom panels of Figure~\ref{fig:rot_curves}). \textbf{Panel 2}: Same as top, but for line-of-sight velocities measured in a mock slit rather than $v_{\phi}$ measured in cylindrical bins. \textbf{Panel 3}: Ratio of the maximum (cylindrical) gas rotation velocity to the maximum halo circular velocity $v_c = \sqrt{GM(<r)/r}$ over the radial range probed by HI. \textbf{Bottom}: Ratio of the mean isothermal sound speed of HI gas, $c_{s}=\sqrt{kT/\mu m_{p}}$, to the circular velocity, with both quantities measured at the stellar half-mass radius. By all three metrics shown ($V_{\rm rot}/\sigma$, $V_{\rm rot}/v_c$, and $c_s/v_c$), galaxies become more rotationally supported at higher $\rm M_{\rm star}$. }
\label{fig:all_v_sig}
\end{figure}

From the four rotation curves shown in Figure~\ref{fig:rot_curves}, it appears that galaxies become more rotationally supported with increasing $\rm M_{\rm star}$. We investigate this scaling explicitly in Figure~\ref{fig:all_v_sig}, where we quantify the rotation versus dispersion support of all the galaxies in our sample. In the first and second panels, we plot HI gas $V_{\rm rot}/\sigma$ for the cylindrically averaged and mock-slit rotation curves, respectively. We define $V_{\rm rot}$ here as the \textit{maximum} of the rotation curve $v_{\rm rot}(r)$ and $\sigma$ as the \textit{median} of the dispersion profile $\sigma(r)$. In the third panel, we plot the ratio of the maximum cylindrically averaged rotation velocity to the maximum circular velocity in the radial range where we measure the rotation curve;\footnote{We measure the gas rotation curve out to the radius where the HI surface density drops below $0.1\,\rm M_{\odot}\,\rm pc^{-2}$; this is typically $\lesssim 0.1 R_{\rm 200m}$. In low-mass halos, the maximum circular velocity over this radial range is usually somewhat less than the global maximum of $v_c(r)$.} this will be $\sim 1$ for a rotationally supported system and $\ll 1$ for a system supported entirely by dispersion. In the bottom panel, we plot the ratio of the isothermal sound speed to the circular velocity, $c_s/v_c$, measured at the stellar half-mass radius. This quantifies the importance of thermal pressure support. 

Both $V_{\rm rot}/\sigma$ and $V_{\rm rot}/v_c$ increase with stellar mass. Most of the galaxies in our sample with $\rm M_{\rm star} \lesssim 10^8 \rm M_{\odot}$ show only mild rotational support; they have $V_{\rm rot} \sim \sigma$ and rotate at well below the circular velocity of their host halo. At higher masses, an increasing fraction of our galaxies rotate at near the halo circular velocity, with lower relative velocity dispersions. In particular, all five galaxies with $\rm M_{\rm star} > 10^{10} \rm M_{\odot}$ have $V_{\rm rot}/v_c \sim 1$ and $V_{\rm rot}/\sigma >5$. Galaxies with intermediate masses show a range of rotation curves: \texttt{m11b}, \texttt{m11h}, and \texttt{m11e} show significant rotational support, \texttt{m11a} and \texttt{m11q} are dispersion-supported at all radii, and \texttt{m11d} and \texttt{m11i} are intermediate cases (see the rotation curves of individual galaxies in Appendix~\ref{sec:idvid_gals}). This mass-scaling is in good agreement with observational studies \citep{Simons_2015}, which find that rotation-supported disks are ubiquitous in star forming galaxies at $\rm M_{\rm star} \gtrsim 10^{9.5} \rm M_{\odot}$, while kinematically and morphologically irregular systems become increasingly prevalent at lower masses.

Finally, the bottom panel of Figure~\ref{fig:all_v_sig} shows that the ratio of the sound speed to the circular velocity decreases rapidly with increasing $\rm M_{\rm star}$: thermal pressure support is important for the lowest-mass galaxies in our sample ($\rm M_{\rm star} \lesssim 10^{8} \rm M_{\odot}$) but becomes subdominant at higher masses.

\subsection{Kinematic disk/spheroid decomposition}
\label{sec:circ}
\begin{figure}
\includegraphics[width=\columnwidth]{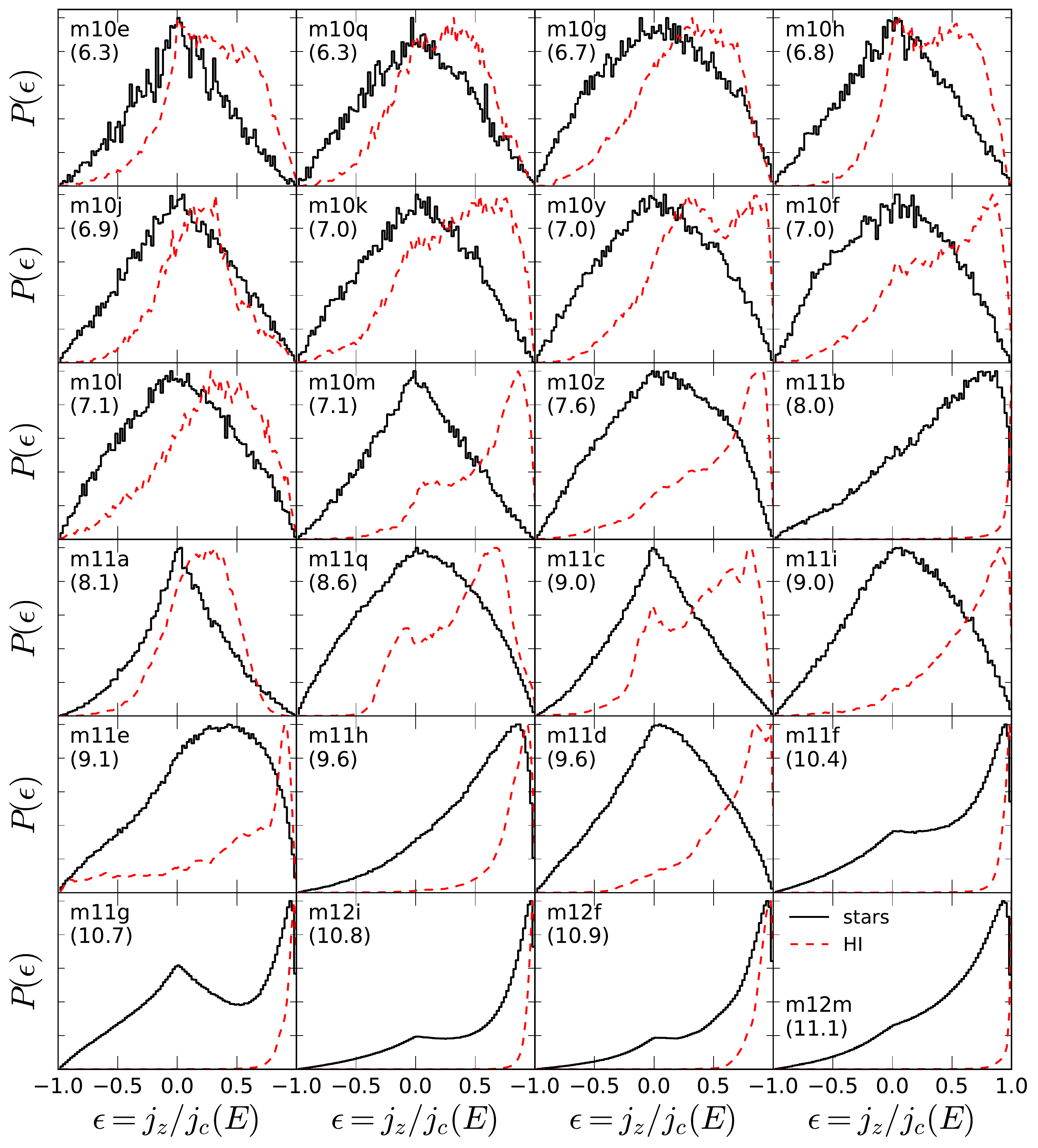}
\caption{Mass-weighted probability distributions (linear $y-$scale, arbitrary normalization) of the orbital circularity parameter $\epsilon \equiv j_z/j_c(E)$ of stars (solid black line) and HI gas (red dashed line) for all the galaxies in our sample. The number in parenthesis below each simulation name indicates $\log(\rm M_{\rm star}/\rm M_{\odot})$. 
$\epsilon=1$ corresponds to circular orbits in the plane perpendicular to the galaxy's net angular momentum vector, $\epsilon=0$ to radial or isotropic orbits, and $\epsilon=-1$ to counter-rotating circular orbits (see Section~\ref{sec:circ}). In low-mass galaxies, both stars and gas have broad circularity distributions with a significant disordered component ($\epsilon \sim 0$). At higher masses, galaxies form gas disks with skewed $P(\epsilon)$ distributions peaking at $\epsilon \to 1$. Stars have more disordered kinematics than gas at all masses; even galaxies with thin gas disks exhibit a significant disordered stellar (bulge) component.}
\label{fig:circularity_distributions}
\end{figure}

To quantify the fraction of stars and gas in our galaxies on rotationally supported orbits, we next analyse the distribution of orbital circularity, defined below, for both stars and gas. We first calculate each galaxy's net specific angular momentum vector:
\begin{equation}
\label{eqn:ang_mom}
\mathbf{j}_{\rm net}=\frac{\mathbf{J}_{\rm net}}{M}=\frac{\sum_{i}m_{i}\mathbf{v}_{i}\times\mathbf{r}_{i}}{\sum_{i}m_{i}},
\end{equation}
where the sum is over all star or HI gas particles, depending on the component of interest. For each star or gas particle (indexed $i$), we compute $j_{z,i}$, the component of that particle's specific angular momentum vector parallel to the galaxy's net angular momentum:
\begin{equation}
\label{eqn:j_z}
j_{z,i}=\frac{\mathbf{j}_{i}\cdot\mathbf{j}_{{\rm net}}}{\left|\mathbf{j}_{{\rm net}}\right|}.
\end{equation}
Note that $\mathbf{j}_{\rm net}$ represents the net angular specific momentum of all stars or HI gas in the galaxy, while $\mathbf{j}_{i}=\mathbf{v}_{i}\times\mathbf{r}_{i}$ is the specific angular momentum vector of a single star or gas particle.

Following \citet{Abadi_2003}, we then define the orbital ``circularity'' of each particle as 
\begin{equation}
\label{eqn:circularity}
\epsilon_{i} = \frac{j_{z,i}}{j_c(E_{i})}, 
\end{equation}
where $j_c(E_{i})$ is the specific angular momentum of a circular orbit with the same specific energy $E_{i}$ as the true orbit. Because a circular orbit has maximal angular momentum for a given energy, the circularity parameter $\epsilon$ ranges between -1 and 1, with $\epsilon=1$ corresponding to prograde circular orbits in the galaxy's plane of rotation, $\epsilon = 0$ to radial or randomly distributed isotropic orbits, and $\epsilon = -1$ to circular retrograde orbits. $\epsilon$ is commonly used in kinematic bulge/disk decomposition of simulated galaxies\footnote{Note that some authors define $\epsilon$ in terms of a circular orbit with the same \textit{radius} (rather than energy) as a particular orbit. This yields qualitatively similar results to our definition, but with the differences that this $\epsilon$ is not a constant of motion and is in principle unbounded (rather than being restricted to $-1 \leq \epsilon \leq  1$).} \citep[e.g.][]{Martig_2012, Aumer_2013, Kannan_2015, Zavala_2016, Obreja_2016, Sokolowska_2017}, where a disordered bulge-like component can be recognized by a symmetric distribution of $\epsilon$ values centered on 0, and a distribution of $\epsilon$ skewed toward $\epsilon=1$ signifies a disk.
Thin and thick stellar disks typically have $\epsilon \sim 0.8$ and $\epsilon \sim 0.5$, respectively \citep{Abadi_2003, Knebe_2013, Okamoto_2010}; most stars in the Milky Way thin disk have $\epsilon > 0.7$ \citep{Nordstrom_2004, Governato_2007}.

To compute $j_c(E_i)$, we first calculate the true specific orbital energy $E_i$ of each particle: 
\begin{equation}
\label{eqn:energy}
E_{i}=\phi\left(r_{i}\right)+\frac{1}{2}v_{i}^{2};\qquad \phi\left(r\right)=-G\int_{r}^{\infty}\frac{M\left(<r'\right)}{r'^{2}}\,{\rm d}r',
\end{equation}
where $\phi(r)$ is the spherically-averaged gravitational potential. The specific energy of a particle on a circular orbit at radius $r_c$ is 

\begin{equation}
\label{eqn:circ_energy}
E(r_c)=\frac{GM\left(<r_c\right)}{2r_c}+\phi\left(r_c\right),
\end{equation}
with corresponding specific angular momentum 
\begin{equation}
\label{eqn:j_circ}
j_c = r_c v_c= \sqrt{GM(<r_c)r_c}.
\end{equation}
Thus, to calculate $j_c(E_i)$, we set Equation~\ref{eqn:circ_energy} equal to $E_i$ for each particle, numerically solve for $r_{c,i}$, and then substitute the result into Equation ~\ref{eqn:j_circ}. 

In Figure~\ref{fig:circularity_distributions}, we plot orbital circularity distributions, $P(\epsilon)$, of HI gas and stars for all the galaxies in our sample. Galaxies are ordered by increasing $\rm M_{\rm star}$. For both gas and stars, we include all particles within $r_{\rm max,\,HI}$, the radius where $\Sigma_{\rm HI}$ falls below $0.1\,\rm M_{\odot}\,\rm pc^{-2}$. Each gas particle is weighted by its neutral HI mass; the adopted normalization is arbitrary.

Our sample exhibits significant diversity in stellar and gas circularity distributions. For both gas and stars, average circularity increases with galaxy mass. Most of the galaxies with $\rm M_{\rm star} < 10^{8} \rm M_{\odot}$ (all galaxies up to \texttt{m11b} in Figure~\ref{fig:circularity_distributions}) have broad $P_{\rm HI}(\epsilon)$ distributions, with a significant fraction of gas on isotropic and counter-rotating orbits (i.e., $\epsilon \lesssim 0$). In contrast, all five of our most massive galaxies (\texttt{m11f}, \texttt{m11g}, \texttt{m12i}, \texttt{m12f}, and \texttt{m12m}) exhibit unambiguous gas disks, with $\left\langle \epsilon\right\rangle \sim 0.9$ and essentially no gas on orbits with $\epsilon \lesssim 0.6$. Intermediate-mass galaxies display a variety of gas circularity distributions, ranging from the almost completely isotropic distribution in \texttt{m11a} to the dynamically cold, rotationally supported disk in \texttt{m11b}.  

\subsubsection{Reduced rotational support in stars}
Across all galaxies, stars have more symmetric circularity distributions than gas and lower average circularity, indicating that their kinematics are more disordered and less rotationally supported. Indeed, none of the galaxies with $\rm M_{\rm star} < 10^{8} \rm M_{\odot}$ show \textit{any} evidence of rotation in their stellar circularity distributions; they all have $\left\langle \epsilon\right\rangle \sim 0$, consistent with completely isotropic orbits. Intriguingly, some of the low-mass galaxies (e.g. \texttt{m10m} and \texttt{m11i}) clearly have a significant rotating gas component, as evidenced by their skewed $P_{\rm HI}(\epsilon)$ distributions, but show no hint of ordered rotation whatsoever in their stellar kinematics. 

\begin{figure}
\includegraphics[width=\columnwidth]{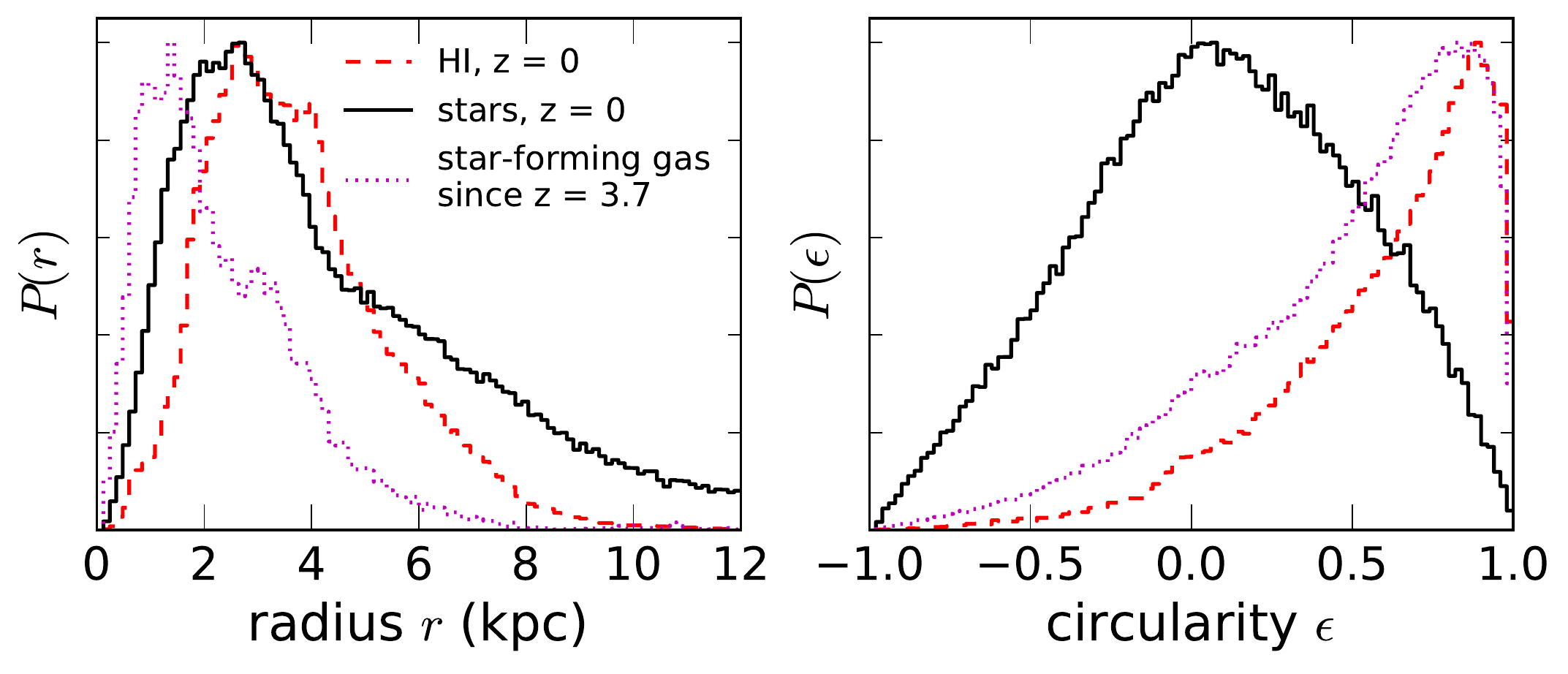}
\caption{Distribution of galactocentric radii, $r$, and circularity, $\epsilon$, (Equation~\ref{eqn:circularity}) for gas and stars in \texttt{m11i}, an example galaxy with rotationally-supported gas but dispersion-supported stars. Along with HI gas and stars at $z=0$, we show ``star-forming gas'' for all stacked simulation snapshots since $z=3.7$. Star-forming gas is identified as gas particles which will turn into star particles by the next simulation snapshot. Stars form from gas which has similar circularity to all gas at $z=0$ but is concentrated near the galactic centre. Stars migrate to larger radii and are dynamically heated over time, so that by $z=0$, they are much less rotationally supported than when they formed.}
\label{fig:sf_gas_circ}
\end{figure}

There are two main reasons why stars have lower mean circularity than gas. First, the orientation of a galaxy's net angular momentum vector can change over time as new gas is accreted. Because gas is collisional, gas particles can re-align their orbits with the galaxy's net angular momentum vector when this occurs, maintaining high circularity. In contrast, stars that formed prior to changes in a galaxy's net angular momentum orientation will end up with lower circularity on average. 

Second, stars form preferentially from low-angular momentum gas. We investigate this in Figure~\ref{fig:sf_gas_circ}, where we compare for \texttt{m11i} the spatial distribution and circularity of star-forming gas to the corresponding distributions for stars and all HI at $z=0$. We identify star-forming gas particles in each simulation snapshot between $z=3.7$ and $z=0$ as those which have turned into star particles by the following snapshot;\footnote{Simulation snapshots have a maximum spacing of 27 Myr; our approach allows us to tag gas particles $\sim 10 \rm \, Myr$ before they turn into stars on average. This means that for typical radial velocities of $10\,\rm km\,s^{-1}$, we measure gas particles within $100\,\rm pc$ of the radius where they form stars.} the distributions for star forming gas are thus not measured at $z=0$, but are the aggregate of all gas particles that turn into stars between $z=3.7$ and $z=0$, which comprise more than 90\% of the stars in the galaxy at $z=0$. We chose the simulation \texttt{m11i} because it is one in which gas is significantly rotationally supported but stars are not.  

The left panel of Figure~\ref{fig:sf_gas_circ} compares the spatial distribution of HI gas, stars, and star-forming gas. Star-forming gas is significantly more centrally concentrated than either HI or stars; i.e., stars form preferentially near the galactic centre, where it is easier for gas to accumulate at high densities. Since gas near the galactic centre has low angular momentum ($j\sim v\times r $), stars born from this material have, on average, lower angular momentum than most of the gas. After they form, stars migrate to larger radii as they gain energy from feedback-driven potential fluctuations \citep{ElBadry_2016}, but this process is approximately angular-momentum conserving \citep{Pontzen_2012}. Stars at $z=0$ thus end up at large radii with low angular momentum, supported primarily by dispersion, not rotation.  

In the right panel of Figure~\ref{fig:sf_gas_circ}, we compare the orbital circularity distribution of star forming gas to the distributions for all HI and stars at $z=0$. We compute $\epsilon$ for star-forming gas particles in each snapshot relative to the galaxy's net HI angular momentum vector in that snapshot. Star-forming gas has a similar circularity distribution to all HI, indicating that when stars form, they have similar angular momentum at fixed radius to HI. However, the combination of post-formation migration and gradual dynamical heating makes stars lose their rotational support, so that at $z=0$, they have broad, symmetric circularity distributions. 

We note that while Figure~\ref{fig:sf_gas_circ} only shows the $z=0$ circularity distribution for all HI, we have verified that star-forming gas in any given snapshot is \textit{always} more centrally-concentrated than all HI in the same snapshot; i.e., star-forming gas \textit{never} fairly samples the HI radial and angular momentum distribution. 

\subsubsection{Stellar rotational support as a function of age}
Even the galaxies with unambiguous, thin gas disks in Figure~\ref{fig:circularity_distributions} have a significant fraction of stars on isotropic bulge-like orbits. Indeed, in the five most massive galaxies, the $P_{\rm star}(\epsilon)$ distribution can be visually decomposed into two components: a symmetric, disordered component that peaks at $\epsilon=0$, consisting mostly of stars in the bulge, and a skewed, rotating component, consisting of dynamically colder stars in the disk. 

\begin{figure}
\includegraphics[width=\columnwidth]{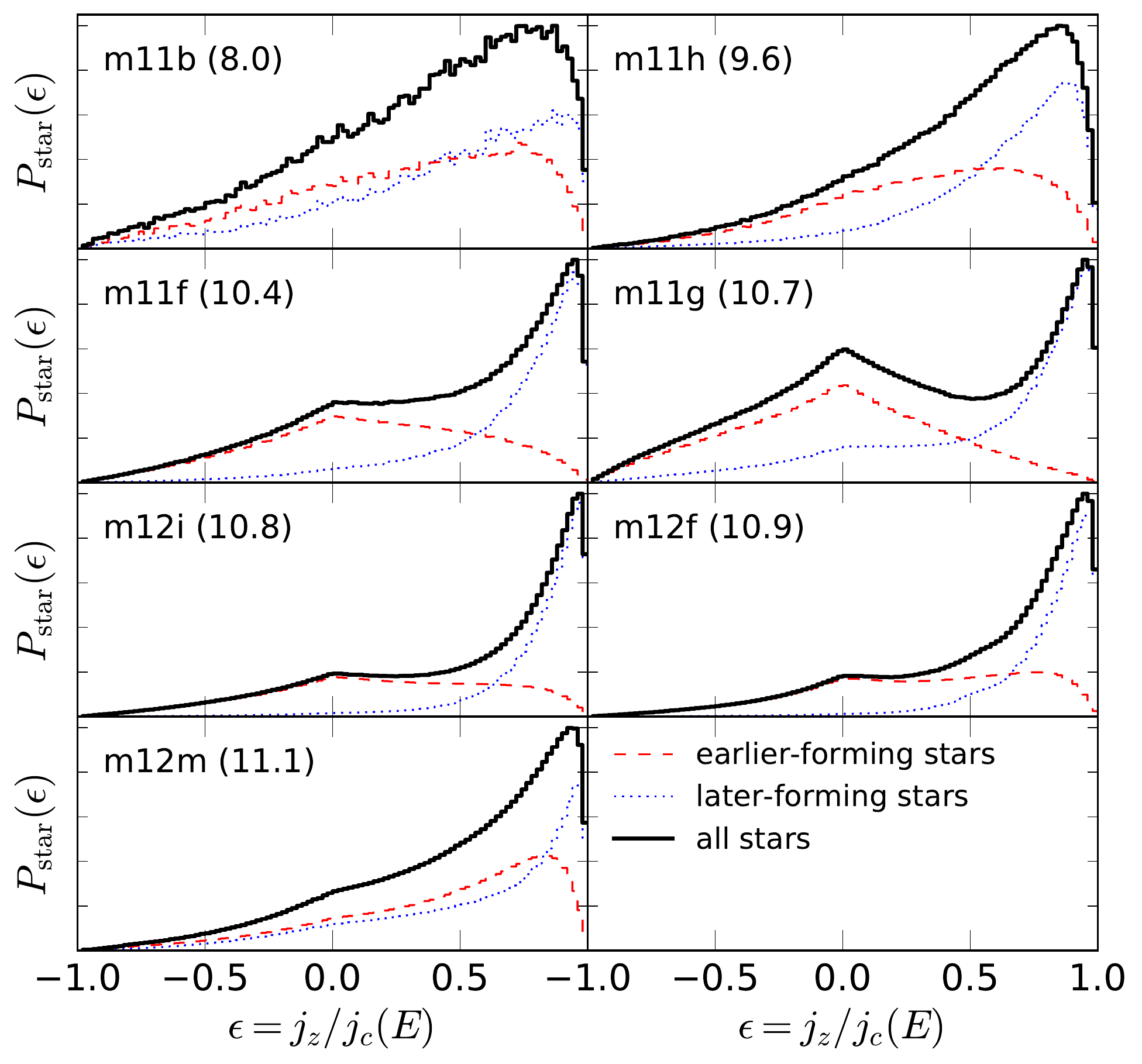}
\caption{Stellar orbital circularity probability distributions (see Section~\ref{sec:circ}) for the seven galaxies in our sample with stellar disks at $z=0$. The number in parenthesis beside each simulation name indicates $\log(\rm M_{\rm star}/\rm M_{\odot})$.  We divide stars into ``earlier-forming'' and ``later-forming'' subsamples based on whether they formed before or after half the stellar mass in the galaxy at $z=0$ was formed. In all galaxies, the old component has lower circularity and less coherent rotation than the young component, implying that the galaxies became more disky at late times.}
\label{fig:young_old_stars}
\end{figure}

We investigate this explicitly in Figure~\ref{fig:young_old_stars}, where we decompose the stellar circularity distributions for the galaxies which have stellar disks into ``earlier-forming'' and ``later-forming'' samples based on whether they formed before or after 50\% of the stellar mass in the galaxy at $z=0$ had formed. This definition is such that the stellar mass in the two components is always equal. In these galaxies, average circularity uniformly decreases with increasing stellar age: the young component is significantly more disk-like than the old component; in several galaxies, the latter is nearly symmetric in $\epsilon$ and shows negligible evidence of rotational support.  

The increased rotational support for later-forming stars can be understood as the result of two processes. First, stars dynamically heat up over time, even if they form in cold, rotationally supported orbits. This occurs both due to global fluctuations in the potential caused by mergers or galaxy-scale outflows \citep{Teyssier_2013, ElBadry_2016, Gozalez_2016}, and due to more gradual phase-mixing resulting from instabilities in the disk and scattering off non-axisymmetric structures such as GMCs and spiral arms \citep{MartinezMedina_2015}. Unlike gas, which can efficiently dissipate its turbulent energy through collisions, stars cannot cool once they are dynamically heated. 

Second, our simulated galaxies build up their disks primary at late times \citep{Ma_2017}; before $z = 1-2$, they are puffy, gas-rich, dispersion-supported systems, dynamically similar to the intermediate-mass galaxies in our sample at $z=0$. These systems' gravitational potentials grow deeper at later times, until stellar feedback can no longer drive global outflows and galaxies become dynamically more stable \citep{Muratov_2015}.\footnote{Note that this picture applies for $L\lesssim L_{\star}$ galaxies; more massive systems can likely form stable disks at earlier times \citep{Simons_2016, Feldmann_2016, Genzel_2017}.} Both observational \citep{Kassin_2012} and theoretical \citep{Aumer_2013, Kassin_2014, Peschken_2017, Hayward_2017, Ma_2017, FaucherGiguere_2017, Ceverino_2017} works have found that the disks of most $\lesssim$MW-mass galaxies are built up between $z=1$ and $z=0$.

\section{Galaxy Shapes}
\label{sec:shapes}
We have thus far used strictly kinematic metrics to classify the diskiness and rotational support of our galaxies. We now turn to classifying their shapes. We expect morphological and kinematic indicators of diskiness to be closely related, since rotation flattens both collisional and collisionless components of galaxies \citep{Fall_1980, Binney_2008}. However, shape and kinematics are not one-to-one, both because triaxial systems can be flattened without rotation, and because rotation will not necessarily flatten galaxies (see, for example, the HI moment maps of \texttt{m11c} in the third column of Figure~\ref{fig:vel_maps}).

We quantify galaxy shape using the HI moment of inertia tensor, defined as 
\begin{equation}
I_{ij}=\int_{\mathcal{V}}\rho_{\rm HI}(\mathbf{x})\left(\delta_{ij}r^{2}-x_{i}x_{j}\right)\,{\rm d}^{3}\mathbf{x}
\end{equation}
Here $i,j\in(x,y,z)$ denote the standard Cartesian coordinates, $r^2 = x^2 + y^2 + z^2$, $\rho_{\rm HI}$ is the HI mass density, and the integral is over all material in a volume $\mathcal{V}$, which we take to be a sphere of radius $R_{\rm HI}$ (see Table~\ref{tab:properties}) centered on the HI centre-of-mass. Diagonalizing $I_{ij}$ yields three coordinate-independent eigenvalues $E_1\leq E_2\leq E_3$. We translate these into the axis ratios of an ellipsoid following the prescription of \citet{GonzalezGarcia_2005}: for a homogeneous ellipsoid with axis lengths $2a$, $2b$, and $2c$, the ratio of the shortest axis to the longest is 
\begin{equation}
\label{eqn:axis_ratio}
\frac{c}{a}=\sqrt{\frac{E_{1}+E_{2}-E_{3}}{E_{3}+E_{2}-E_{1}}}.
\end{equation}
Finally, we define the ``flattening'' $\varepsilon$:
\begin{equation}
\label{eqn:flattening}
\varepsilon = 1 - \frac{c}{a}.
\end{equation}
An infinitely thin disk will have $c/a = 0$ and $\varepsilon=1$, while a uniform sphere will have $c/a = 1$ and $\varepsilon=0$. The galaxies' spatial HI distributions are of course not uniform ellipsoids, so $\varepsilon$ can be interpreted as the flattening of a uniform ellipsoid with the same principle moments of inertia as the galaxy's true HI distribution \citep[see also][]{Obreja_2016}.

\begin{figure}
\includegraphics[width=\columnwidth]{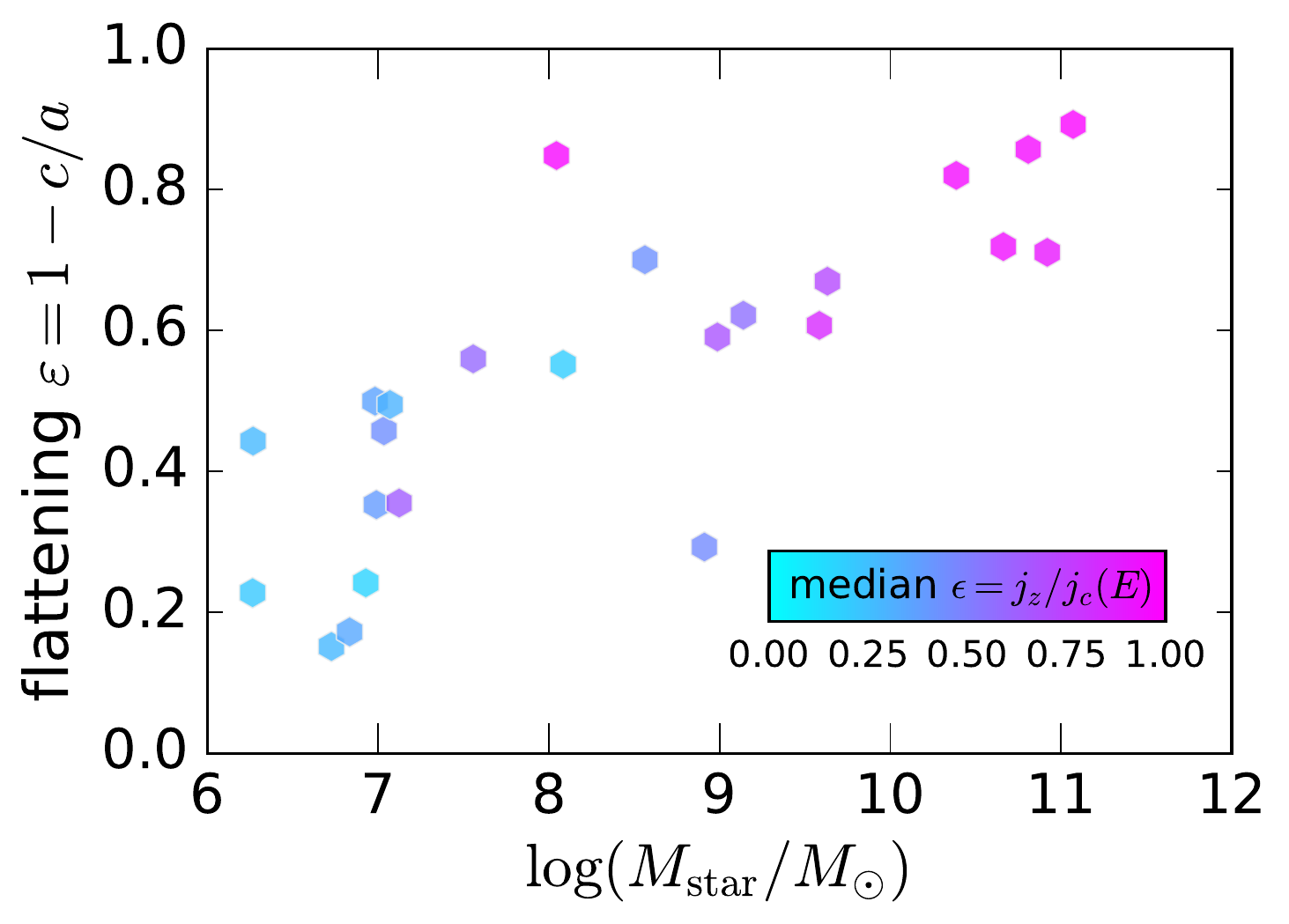}
\caption{Flattening, $\varepsilon$, of galaxies' HI gas (Equation~\ref{eqn:flattening}), vs. $\rm M_{\rm star}$. Points for each galaxy are colored by the median of their HI circularity parameter (Equation~\ref{eqn:circularity}), which quantifies \textit{kinematic} rotational support. Flattening is defined purely in terms of the shape of the gas distribution (Section~\ref{sec:shapes}), with no dependence on kinematics. At higher masses, galaxies become both flatter and more rotationally supported. Most galaxies at $\rm M_{\rm star} \lesssim 10^{8} \rm M_{\odot}$ are not significantly flattened; this is due in part to the increased contribution of thermal pressure support in these systems.}
\label{fig:shape}
\end{figure}

In Figure~\ref{fig:shape}, we show the HI flattening of all the galaxies in our sample as a function of $\rm M_{\rm star}$. As with the kinematic metrics presented in the previous sections, galaxies become flatter on average with increasing $\rm M_{\rm star}$; in particular, all of the massive galaxies ($\rm M_{\rm star}> 10^{10} \rm M_{\odot}$) are strongly flattened. However, there is significant scatter in the relation, likely because feedback-driven outflows can change galaxy shapes significantly on short timescales, especially at low masses.\footnote{For the low-mass dispersion-supported galaxies in our sample, we find that $\epsilon$ can fluctuate by $\pm 0.2$ between snapshots as galaxies go through starburst/outflow cycles. However, the trend in Figure~\ref{fig:shape} is robust.} The trend of decreasing flatness at lower masses is qualitatively consistent with observational studies of low-mass galaxies, which find that both the HI \citep{Roychowdhury_2010, Johnson_2017} and stellar \citep{Yoachim_2006, SanchezJanssen_2010, Roychowdhury_2013, Wheeler_2017} components of galaxies become increasingly puffy at low masses.

We also indicate in Figure~\ref{fig:shape} the mass-weighted median HI orbital circularity; i.e., the 50th percentile of the red histograms in Figure~\ref{fig:circularity_distributions}. This quantity is close to 1 for a rotationally-supported system and close to 0 for a dispersion-supported system. Circularity is somewhat correlated with flattening -- i.e., circularity increases along the $y-$axis of Figure~\ref{fig:shape} on average -- but with significant scatter: some rotating systems are not flattened, and some flattened systems are not rotating. This is not surprising: if an initially flat, rotating disk is puffed up vertically by stellar feedback, it should retain its rotation as long as angular momentum is conserved, but it will no longer be flat. Conversely, if a flattened system is expanded radially by feedback but remains somewhat flattened, it will rotate more slowly by angular momentum conservation.  

\section{Angular momentum profiles}
\label{sec:ang_mom}

\begin{figure}
\includegraphics[width=\columnwidth]{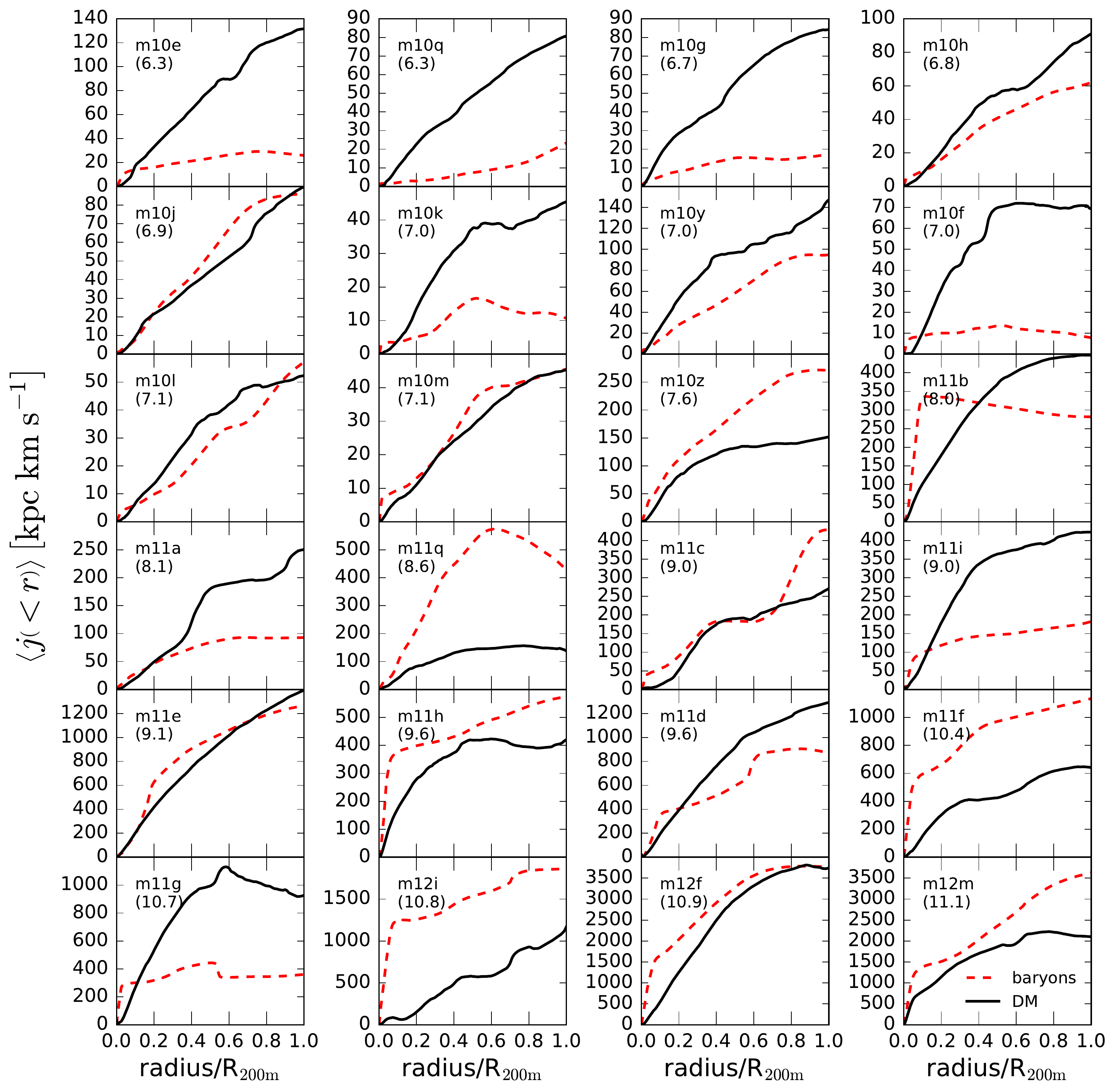}
\caption{Cumulative specific angular momentum profiles for dark matter (black) and baryons (stars + gas; red dashed) for all halos in our sample. The number in parenthesis below each simulation name indicates $\log(\rm M_{\rm star}/\rm M_{\odot})$. We plot all quantities out to $R_{\rm 200m}$; the main galaxy is within $\lesssim 0.1 R_{\rm 200m}$. Galaxies with disks (e.g. \texttt{m11b}) have comparable $\left\langle j_{{\rm bar}}\right\rangle$ within the central galaxy to the outer halo; most galaxies which lack disks (e.g. \texttt{m11q}) have higher $\left\langle j_{{\rm bar}}\right\rangle$ in their halo gas than in the central galaxy. Within $R_{\rm 200m}$, most low-mass galaxies have $\left\langle j_{{\rm bar}}\right\rangle < \left\langle j_{{\rm DM}}\right\rangle$; the opposite is true for most high-mass galaxies.}
\label{fig:cumulative_angmom}
\end{figure}

We now examine the radial distribution of angular momentum in our simulated halos.  Angular momentum is in some sense a more fundamental quantity than the kinematic and morphological parameters studied in the previous sections, since it is to first order conserved as galaxies go through starburst/outflow cycles that change their kinematics and morphology. 

In Figure~\ref{fig:cumulative_angmom}, we plot the \textit{cumulative} specific angular momentum profiles of dark matter and baryons (stars + gas, including ionized gas) for all the halos in our sample. We plot these profiles out to $R_{\rm 200m}$ (approximately the ``edge'' of the halo); the central galaxy is at the left edge of each panel, within $r \lesssim 0.1R_{200m}$. This allows us to survey the full angular momentum content of each halo, including gas that was ejected from the central galaxy and gas that could be (re)-accreted at later times. The cumulative profiles $\left\langle j\left(<r\right)\right\rangle  = \left|\mathbf{j}_{\rm net}(<r)\right|$ are computed using Equation~\ref{eqn:ang_mom} separately for each species, summing over all particles enclosed in a sphere of radius $r$. 

First, we note that in nearly all the galaxies in our sample, $\left\langle j_{{\rm bar}}\left(<r\right)\right\rangle $ increases between the edge of the central galaxy ($\sim0.1 R_{\rm 200m}$) and $R_{\rm 200m}$. That is, the average specific angular momentum of ionized gas in the halo (the ``circumgalactic medium''; CGM) is higher -- sometimes by as much as an order or magnitude -- than that of the cold gas and stars in the central galaxy. This increase is, however, mass dependent, with low-mass halos having a larger fraction of their baryon angular momentum in the CGM. This is likely part of the reason our low-mass galaxies do not form disks: in many cases (e.g. \texttt{m10j}), the gas within $R_{\rm 200m}$ has enough angular momentum that the galaxy would be rotating if the high specific angular momentum gas in the CGM could cool into the central galaxy. But the high-angular momentum gas remains in the halo, and so the gas that \textit{does} end up in the central galaxy has insufficient angular momentum to be rotationally supported.

Other studies have similarly found high specific angular momentum gas in the CGM. In the Illustris simulation, \citet{DeFelippis_2017} recently found that gas loses a significant fraction of its angular momentum between first being accreted into a halo and cooling into the central galaxy. \citet{Brook_2012} and \citet{Ubler_2014} found that, because gas in the CGM has higher specific angular momentum than gas in the galaxy, gas that falls back onto galaxies following a galactic fountain often has higher specific angular momentum than when it was ejected. 

In all the galaxies that \textit{do} form disks, (\texttt{m11b, m11h, m11f, m11g, m12i, m12f}, and \texttt{m12m}), the baryons in the central galaxy have about the same $\left\langle j\right\rangle$ as all the baryons within $R_{\rm 200m}$ (to within a factor of two). This suggests that an important requirement for successful disk formation is that high angular momentum gas in the halo is able to reach the central galaxy. Even in galaxies with large angular momentum reservoirs in their halos, (e.g., \texttt{m11q}), halo gas is primarily supported by pressure, not rotation. It is likely a combination of heating by feedback from the central galaxy and the UV background that keeps this gas in the halo. 

Another striking conclusion which can be drawn from Figure~\ref{fig:cumulative_angmom} is that there is large variation within our sample in the ratio $\langle j_{\rm bar} \rangle / \langle j_{\rm DM} \rangle$ within $R_{\rm 200m}$. Classical galaxy formation models \citep[e.g.][]{Fall_1980, Fall_1983, Mo_1998, Romanowsky_2012} generally work under the assumption that the specific angular momentum of baryons in a galaxy is equal to that of the host dark matter halo, so that the distribution of galaxy specific angular momenta at $z=0$ can be predicted to first order from the spin distribution of dark matter halos. While this is approximately true for our sample on average, $\langle j_{\rm DM} \rangle$ and $\langle j_{\rm bar} \rangle$ are clearly not one-to-one: this can be seen, for example, by comparing the cumulative $\langle j_{\rm bar} \rangle$ and $\langle j_{\rm DM} \rangle$ profiles of \texttt{m10f} and \texttt{m11q}. 

\begin{figure}
\includegraphics[width=\columnwidth]{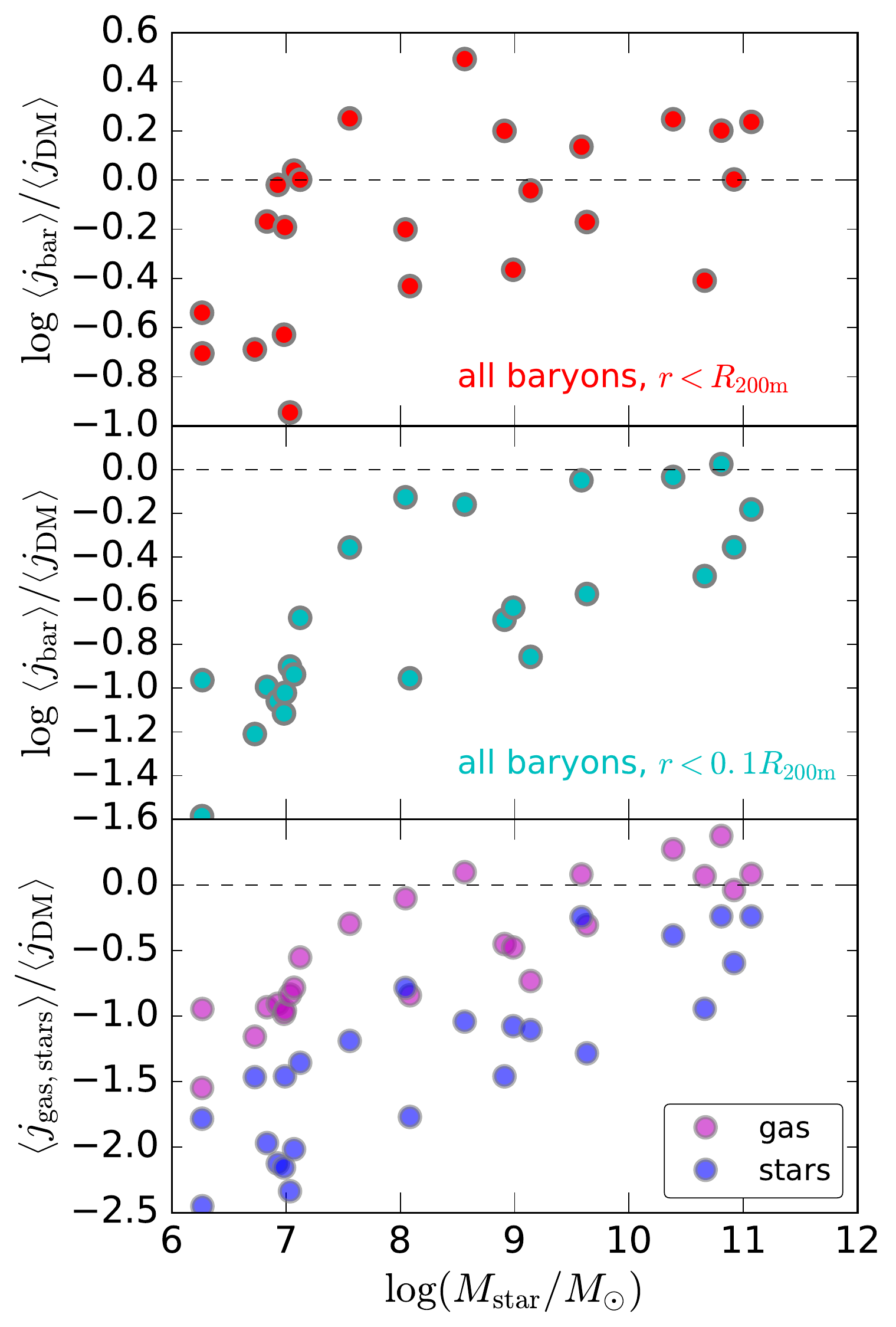}
\caption{Ratio of the $z=0$ mean specific angular momentum of baryons, gas, and stars to $\langle j_{\rm DM} \rangle$, the mean specific angular momentum of dark matter. In all panels, $\langle j_{\rm DM} \rangle$ is measured within $R_{\rm 200m}$. In the top panel, $\langle j_{\rm bar} \rangle$ is measured within $R_{\rm 200m}$; in the middle and lower panels, $\langle j_{\rm bar} \rangle$, $\langle j_{\rm gas} \rangle$,  and $\langle j_{\rm star} \rangle$ are measured within $0.1R_{\rm 200m}$. In the lowest-mass galaxies, the baryons have low specific angular momentum relative to the DM halo. The opposite is true for the most massive galaxies, in which baryons often have \textit{higher} specific angular momentum than the DM halo. The trend of decreasing specific angular momentum in baryons relative to the DM halo is robust for both gas and stars and is particularly strong within $0.1R_{\rm 200m}$.}
\label{fig:jgas_over_jdark}
\end{figure}

In particular, the ratio $\langle j_{\rm bar} \rangle/\langle j_{\rm DM} \rangle$ within $R_{\rm 200m}$ is mass-dependent: most of the low-mass galaxies have $\langle j_{\rm bar} \rangle < \langle j_{\rm DM} \rangle$, while most of the galaxies with $\rm M_{\rm star}>10^{10}\rm M_{\odot}$ have $\langle j_{\rm bar} \rangle > \langle j_{\rm DM} \rangle$. We show this more explicitly in Figure~\ref{fig:jgas_over_jdark}, where we plot the ratio of the average specific angular momentum of the halos' baryonic components to $\langle j_{\rm DM} \rangle$. In all panels, $\langle j_{\rm DM} \rangle$ is the mean specific angular momentum of all dark matter within $R_{\rm 200m}$.

In the top panel, we plot the ratio for all baryons within $R_{\rm 200m}$. There is significant scatter across halos at fixed $\rm M_{\rm star}$, but the average $\langle j_{\rm bar} \rangle/\langle j_{\rm DM} \rangle$ increases by almost an order of magnitude across the mass range probed by our simulations. This indicates that baryons in the low-mass halos have systematically lost (or failed to gain) angular momentum compared to their dark halos. 

In the bottom two panels of Figure~\ref{fig:jgas_over_jdark}, we plot the ratio of the specific angular momentum of all baryons, gas, and stars within $0.1R_{\rm 200m}$ (approximately corresponding to the extent of the central galaxy) to $\langle j_{\rm DM} \rangle$ (still within $R_{\rm 200m}$). Because most of the galaxies have increasing $j_{\rm bar}(<r)$ profiles over $r=(0.1-1)R_{\rm 200m}$ (Figure~\ref{fig:cumulative_angmom}), these ratios are lower for most galaxies than the corresponding ratios for all material within $R_{\rm 200m}$. However, they exhibit a similar positive trend with $\rm M_{\rm star}$, increasing by more than an order of magnitude on average over the mass range probed by our sample. These ratios are more strongly mass dependent than the ratio for all baryons in the halo. This is because low-mass halos generally have less of their angular momentum in the central galaxy relative to the CGM (see Figure~\ref{fig:cumulative_angmom}). 

Across all masses, gas has higher specific angular momentum than stars (and thus, than all baryons). This occurs both because stars form primarily from high-density, low-angular momentum gas near the galactic centre (Figure~\ref{fig:sf_gas_circ}), and because a significant fraction of stars form from gas accreted at early times, which has lower specific angular momentum than gas accreted at late times. In the MW-mass galaxies, stars do form from high-angular momentum gas in the outer disk at late times, but all the galaxies in our sample have a substantial stellar bulge component with low angular momentum (Figure~\ref{fig:circularity_distributions}). 

A number of other works have investigated the relationship between $\langle j_{\rm bar} \rangle$ (or in some cases, $\langle j_{\rm gas} \rangle$) and $\langle j_{\rm DM} \rangle$, primarily for $\sim$MW-mass galaxies, with mixed results. Early hydrodynamic simulations without cooling \citep{vandenBosch_2002, Sharma_2005} found very similar specific angular momentum for gas and DM. Using semi-analytic models based on observed galaxy scaling relations, \citet{Dutton_2012} found that for disk galaxies which obey the Tully-Fisher relation, $\langle j_{\rm bar} \rangle/\langle j_{\rm DM} \rangle$ should be  $\sim 0.6$ for MW-mass galaxies, decreasing slightly at lower masses. In recent years, both large-volume and zoom-in simulations with radiative cooling and feedback \citep{Teklu_2015, Danovich_2015, Genel_2015, Pedrosa_2015, Zavala_2016, Sokolowska_2017, Zjupa_2017} have found mean $\langle j_{\rm bar} \rangle/\langle j_{\rm DM} \rangle$ values of 1-2 for late-type galaxies because (a) gas is accreted along filaments and has a higher quadrupole moment than dark matter, and (b) feedback preferentially removes low-angular momentum gas. These works have also found significant dispersion in $\langle j_{\rm bar} \rangle/\langle j_{\rm DM} \rangle$ values at fixed mass, but not the mass evolution seen in Figure~\ref{fig:jgas_over_jdark}. However, previous works have all focused primarily on $\sim$MW-mass galaxies, and large volume simulations typically do not have the resolution to model low-mass galaxies with high fidelity. Since the downturn in $\langle j_{\rm bar} \rangle/\langle j_{\rm DM} \rangle$ seen in Figure~\ref{fig:jgas_over_jdark} is most significant at $\rm M_{\rm star}\lesssim 10^{8} \rm M_{\odot}$, it may be that it has gone undetected because this mass range has not been studied in detail.

The decrease in $\langle j_{\rm bar} \rangle/\langle j_{\rm DM} \rangle$ for low-mass galaxies is a natural consequence of their baryonic accretion histories. It is well known that on average, low-mass galaxies assemble at earlier times than high-mass galaxies: while MW-mass halos continue to efficiently accrete baryons at late times \citep[e.g.][]{Wetzel_2015}, gas accretion becomes increasingly inefficient at late times in halos with $M_{\rm 200m}\lesssim 10^{11}\rm M_{\odot}$ \citep{Noh_2014}. At the same time, the average specific angular momentum of accreted gas is much higher at late times than at early times \citep{Brooks_2011,Ubler_2014}. MW-mass halos thus obtain much of their high-angular momentum gas at late times, when low-mass galaxies have stopped accreting gas.

\begin{figure}
\includegraphics[width=\columnwidth]{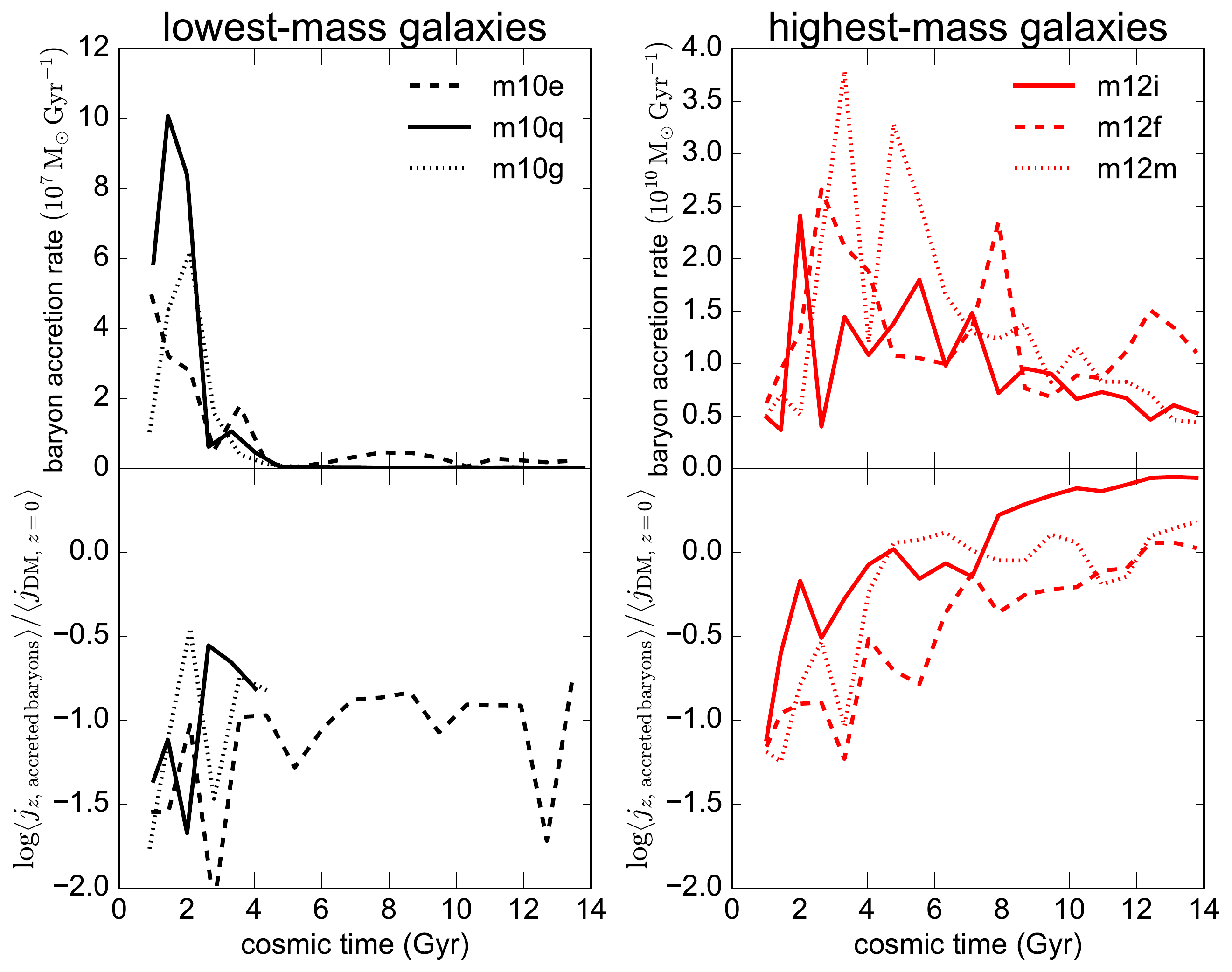}
\caption{Baryonic mass (top) and angular momentum (bottom) accretion histories, measured at fixed physical radius $r=0.1R_{\rm 200m,\,z=0}$, for the three lowest- and highest-mass galaxies in our simulation suite. Low-mass galaxies stop efficiently accreting baryons by $z\sim 2$. MW-mass galaxies continue accreting fresh baryons until $z=0$ and accrete their highest-angular momentum baryons after low-mass galaxies have already shut off accretion.}
\label{fig:j_evolution}
\end{figure}

We show this explicitly in Figure~\ref{fig:j_evolution}, where we plot the mass and angular momentum accretion histories of baryons in the three lowest- and highest-mass galaxies in our sample. Similar to \citet{Brook_2011}, we identify recently accreted gas and star particles in a given snapshot as those which have reached $r<0.1R_{200{\rm m},\,z=0}$ for the first time in that snapshot. The top panels of Figure~\ref{fig:j_evolution} show the average accretion rate of baryons as a function of time, while the bottom panels show $\left\langle j_{z}\right\rangle$, the mean specific angular momentum of recently accreted baryons projected onto the galaxy's net gas angular momentum vector. $\left\langle j_{z}\right\rangle$ is plotted in units of the specific angular momentum of the dark matter halo at $z=0$.

Figure~\ref{fig:j_evolution} shows that, while our simulated low-mass galaxies accrete very little baryonic mass at late times, the MW-mass galaxies continue accreting efficiently until $z=0$. Consistent with previous studies, Figure~\ref{fig:j_evolution} also shows that the mean specific angular momentum of material accreted at late times in MW-mass galaxies is significantly higher than that of material accreted at early times, such that most of their $z=0$ specific angular momentum is accreted at late times. This implies that low-mass galaxies end up with low baryon angular momentum at $z=0$ because gas accretion and cooling are inefficient in low-mass halos at late times. This prevents high-angular momentum gas from cooling from the IGM into the CGM and from the CGM into the central galaxy at low redshifts.

Inefficient cooling in low-mass halos can also explain why the trend in $\langle j_{\rm bar} \rangle/\langle j_{\rm DM} \rangle$ with mass is weaker in the top panel of Figure~\ref{fig:jgas_over_jdark} than in the middle panels. As Figure~\ref{fig:cumulative_angmom} shows, many halos have higher average $j_{\rm bar}$ in the outer halo than in the central galaxy. In low-mass halos, this material cannot cool into the central galaxy at late times; it therefore increases $\langle j_{\rm bar} \rangle(< R_{\rm 200m})$, but not $\langle j_{\rm bar} \rangle(< 0.1 R_{\rm 200m})$.

The suppression of late-time accretion in low-mass halos is likely a consequence of both stellar feedback-driven outflows and photoheating from the UV background \citep[e.g.][]{Pawlik_2009}. We do not attempt to disentangle the separate contributions of different processes in this work; we simply note that their combined effect is to make late-time accretion less efficient at low masses on average.

\section{Radial mass distributions}
\label{sec:mass_dists}

\begin{figure}
\includegraphics[width=\columnwidth]{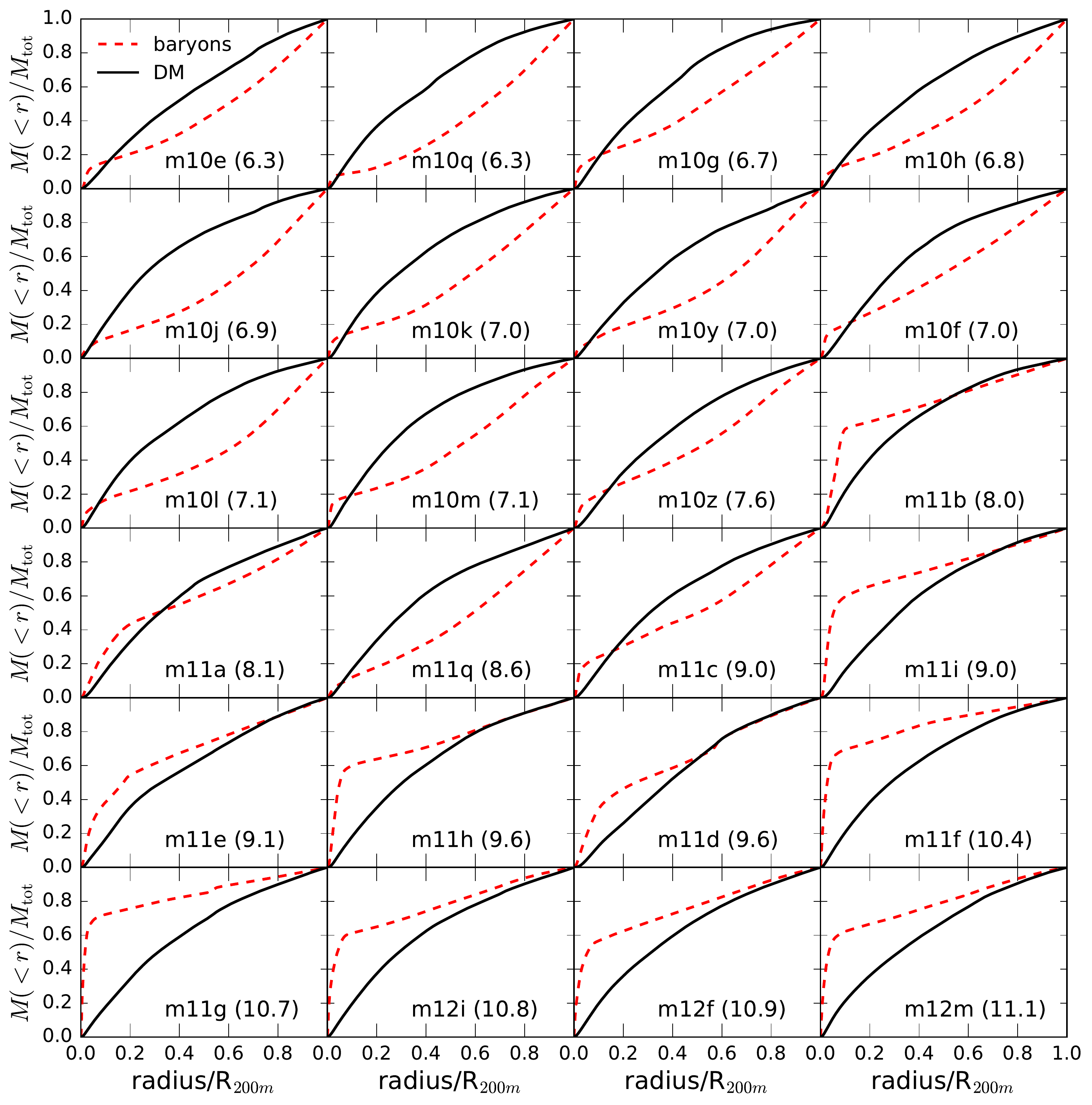}
\caption{Cumulative mass profiles of dark matter (black) and baryons (red dashed) for all the halos in our sample, out to $R_{\rm 200m}$. The number in parenthesis beside each simulation name indicates $\log(\rm M_{\rm star}/\rm M_{\odot})$. In all the simulations, but especially in the low-mass halos, a large fraction of the halo's baryons reside in the ionized gas in the CGM; in many low-mass systems, only $\sim 10\%$ of all the baryons in the halo are in the central galaxy. Baryons become more centrally concentrated with increasing mass: In most of the low-mass galaxies, baryons are \textit{less} centrally concentrated in the inner halo than dark matter; the opposite is true at higher masses. At lower masses, galaxies which have disks (e.g. \texttt{m11b}) also have more centrally concentrated baryons (i.e., more steeply rising $M_{\rm bar}(<r)$ curves).}
\label{fig:cum_mass_profiles}
\end{figure}

Given that many of the galaxies in our sample have gas halos with significantly higher specific angular momentum than their central galaxies, we now investigate the spatial distribution of this gas in more detail. In Figure~\ref{fig:cum_mass_profiles}, we plot each of the halos' cumulative normalized enclosed mass profiles, $M(<r)$, for dark matter and baryons. The edge of the central galaxy can be recognized as the radius where $M_{\rm bar}(<r)$ begins to rise less steeply; typically $r\sim 0.1R_{\rm 200m}$. 

Consistent with recent observational results \citep{Bordoloi_2014, Peeples_2014, Stern_2016, Werk_2016}, all the galaxies in our sample have significant baryonic (gas) mass in the CGM. In nearly all of the halos hosting low-mass galaxies, the CGM mass exceeds the mass of cold gas and stars in the galaxy by a factor of several. In the low-mass galaxies, baryons are significantly less centrally concentrated than dark matter: while the halos' DM half-mass radius is typically $\sim 0.3R_{\rm 200m}$, the baryon half-mass radius is $\sim 0.6R_{\rm 200m}$. This occurs for two reasons. First, feedback is much more efficient at lower masses and continues to eject gas from the central galaxy into the halo at late times \citep{Muratov_2015, Muratov_2016, AnglesAlcazar_2016}. Second, cooling in low-mass halos is inefficiently at late times \citep{Noh_2014}: even relatively cool gas in the outer halo is pressure supported in the presence of a photoionizing background and thus does not cool into the central galaxy.

Consistent with this interpretation, we find that even at fixed mass, the galaxies which \textit{do} form disks have (on average) significantly more centrally-concentrated baryon distributions than those which do not. For example, \texttt{m11b}, the only low-mass galaxy in our sample with a clear disk, has a significantly more steeply rising cumulative baryon profile than any of the other systems with similar mass (\texttt{m11a}, \texttt{m11q}, and \texttt{m11c} do not have disks). 
In the galaxies with gas disks, baryons really \textit{are} more centrally concentrated than dark matter (as one would naively always expect, since gas can cool and dark matter cannot).

\begin{figure}
\includegraphics[width=\columnwidth]{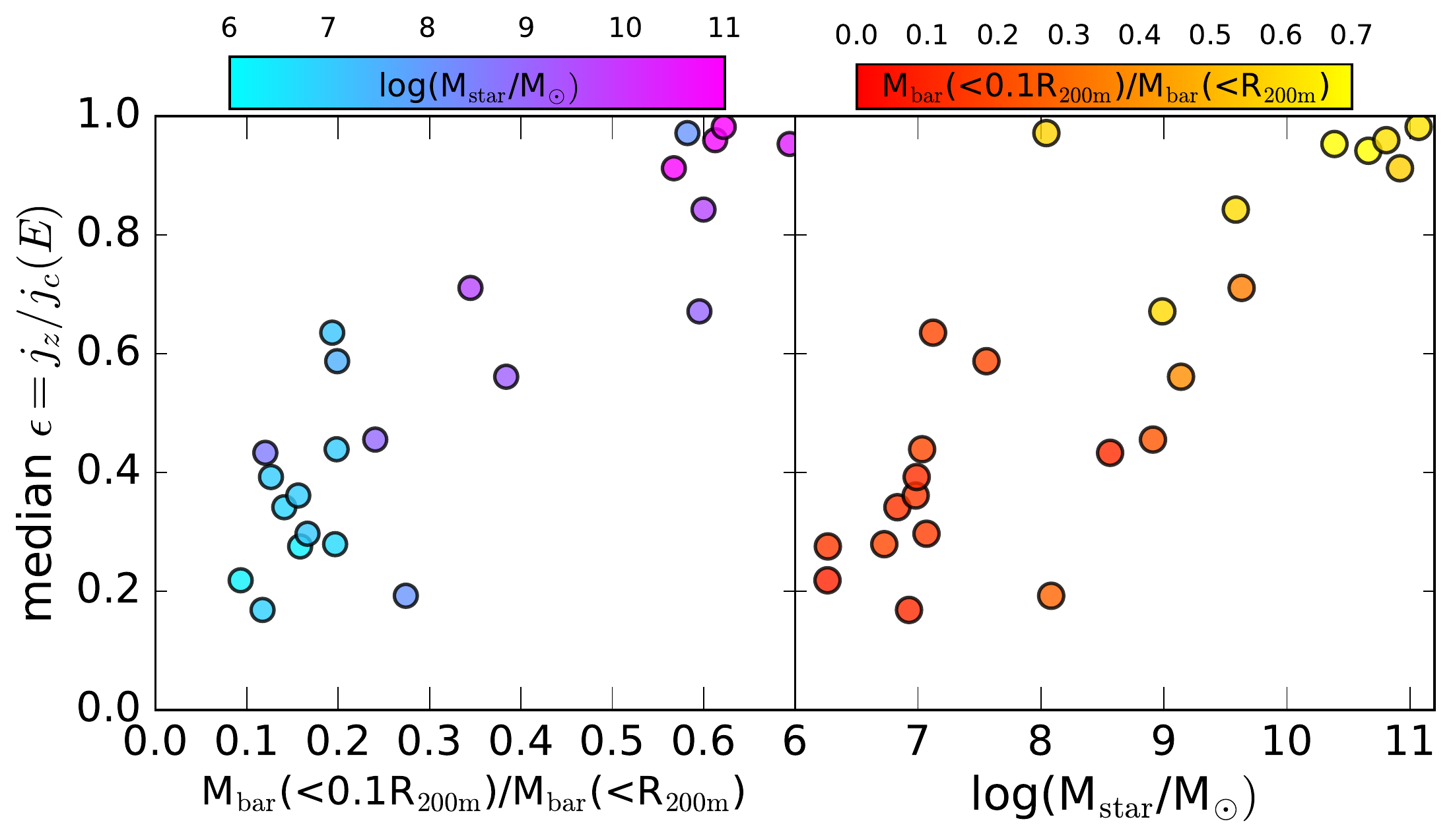}
\caption{\textbf{Left}: Median HI orbital circularity $\epsilon$ (Equation~\ref{eqn:circularity}; a measure of rotational support) vs. the fraction of the total baryons in the halo (within $R_{\rm 200m}$) that are in the central galaxy (within $0.1R_{\rm 200m}$). \textbf{Right}: Circularity vs. stellar mass. Points in each panel are coloured by the $x-$axis of the other panel. Circularity is correlated with both stellar mass and the fraction of the halo's baryons that are in the central galaxy, so the precise causal relationship between these three parameters is difficult to determine. There is, however, one very rotationally-supported low-mass galaxy (\texttt{m11b}) that also has unusually high $M_{\rm bar}(<0.1R_{\rm 200m})/M_{\rm bar}(R_{\rm 200m})$, suggesting that rotational support is related to baryon concentration even at fixed mass.}
\label{fig:mass_frac_vs_circ}
\end{figure}

Because both disk formation and baryon concentration are strongly correlated with mass, it is nontrivial to disentangle the causal relation between these quantities. 
We demonstrate this explicitly in Figure~\ref{fig:mass_frac_vs_circ}. The left panel shows the median HI circularity (a measure of rotational support) as a function of $M_{{\rm bar}}(<0.1R_{{\rm 200m}})/M_{{\rm bar}}(<R_{{\rm 200m}})$, which quantifies how centrally concentrated the halo's baryons are. Median circularity is clearly higher on average for halos with more centrally concentrated baryons. We color points in the left panel by the galaxy's stellar mass; higher mass galaxies have both higher circularity and more centrally-concentrated baryons on average. 

We further investigate the relationship between stellar mass, rotational support, and central baryon concentration in the right panel of Figure~\ref{fig:mass_frac_vs_circ}, where we plot the median HI circularity versus stellar mass, colouring points by how centrally concentrated the halo's baryons are. This panel shows that the correlation between $\epsilon$ and $\rm M_{\rm star}$ is nearly as strong as that between $\epsilon$ and $M_{{\rm bar}}(<0.1R_{{\rm 200m}})/M_{{\rm bar}}(<R_{{\rm 200m}})$, making it difficult to determine conclusively whether rotational support is causally related to the concentration of baryons or whether the correlation between these quantities is simply a consequence of their joint correlation with stellar mass. There is, however, one simulation in our sample which can help break this degeneracy: the halo in \texttt{m11b} hosts a relatively low-mass galaxy ($\rm M_{\rm star} \approx 10^{8} \rm M_{\odot}$) and has an unusually high baryon concentration ($M_{{\rm bar}}(<0.1R_{{\rm 200m}})/M_{{\rm bar}}(<R_{{\rm 200m}})\approx 0.6$); this is the lowest-mass galaxy in our sample to form an unambiguous gas disk. This system falls cleanly on the correlation between circularity and baryon concentration (left panel) but is an outlier on the correlation with $\rm M_{\rm star}$ (right panel). Better statistics at fixed $\rm M_{\rm star}$ are required to determine whether the apparent correlation between rotational support and baryon concentration is robust or \texttt{m11b} is an outlier. 

We also caution that if there is a causal relationship between how centrally concentrated a halo's gas is and rotational support, the precise interplay between disk formation and the distribution of baryons in the halo is still somewhat unclear. On the one hand, having more gas in the central galaxy may make it easier to form a disk, since more energy is required to disrupt a large gas mass.\footnote{Of course, this is only true up to a point; eventually, if the gas mass grows too high, the disk will become globally Toomre unstable. However, for our low-mass galaxies, the typical Toomre parameter is $Q\sim 5$, and high-mass galaxies which \textit{do} host stable disks can have regions where $Q\ll 1$ \citep{Orr_2017}, so this is unlikely to be the primary issue preventing disk formation.} On the other hand, star formation is less bursty in a disk, because high-angular momentum gas cannot accumulate in the galactic nucleus and reach high densities \citep{Torrey_2017}. Less bursty star formation may in turn lead to less efficient outflows, meaning that fewer baryons will be driven from the central galaxy into the CGM. 

\section{Comparison with Observations}
\label{sec:obs}

\begin{figure*}
\includegraphics[width=\textwidth]{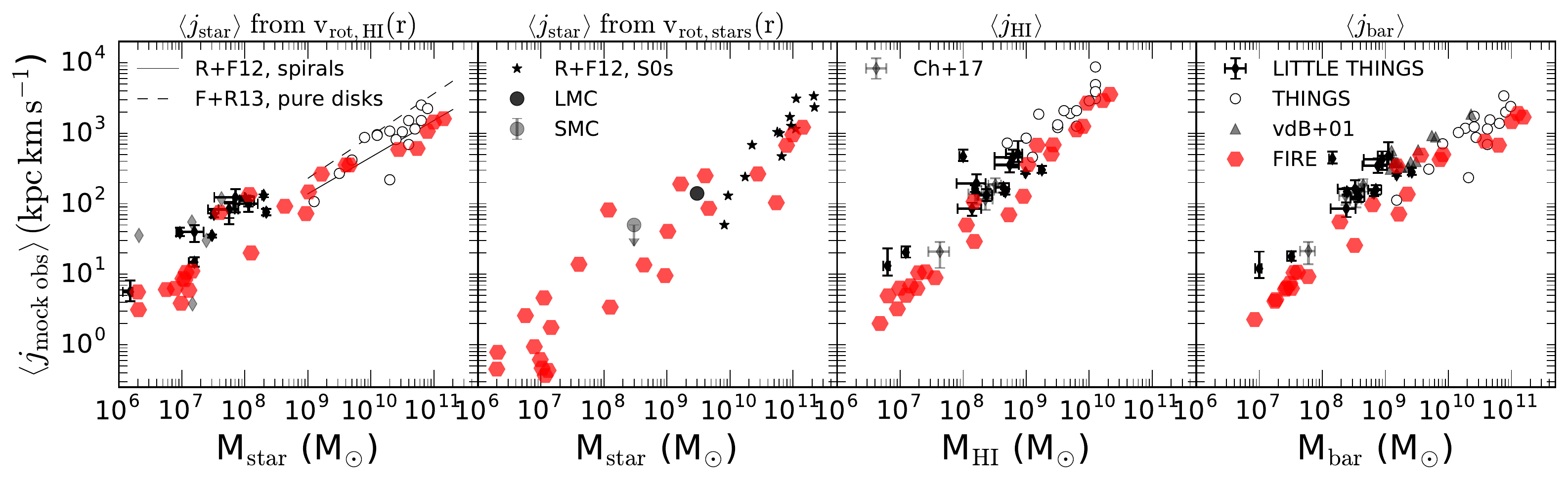}
\caption{Specific angular momentum, $\left\langle j\right\rangle = J/M$ of stars (panels 1 and 2), HI gas (panel 3), and all baryons (stars + gas; panel 4). In each panel, we compare galaxies from the FIRE simulations to available observational data. In panel 1, we calculate $\left\langle j_{{\rm star}}\right\rangle$ under the assumption that stars follow the same rotation curve as HI, and we compare against observational studies that make the same assumption. In panel 2, we calculate $\left\langle j_{{\rm star}}\right\rangle$ from the \textit{stellar} rotation curve and compare to data measured from stellar kinematics. In panels 3 and 4, we assume that all baryons follow the gas rotation curve (as do the plotted observational studies). 
Observational data are taken from the THINGS \citep{Obreschkow_2014} and LITTLE THINGS \citep{Butler_2016} HI surveys, as well as from \citet{vandenBosch_2001}, \citet{Chowdhury_2017}, \citet{Romanowsky_2012}, and \citet{Fall_2013}. The specific angular momentum of the LMC and SMC are calculated from their stellar rotation curves and surface brightness profiles (see \citealt{vanderMarel_2014}). Note that the observational studies shown here explicitly selected rotating, disky galaxies; Figure~\ref{fig:SAMI} shows a comparison to a morphologically blind sample.}
\label{fig:j_vs_mass_baryons}
\end{figure*}

We now compare our simulated galaxies to observations. We begin with the scaling relation between specific angular momentum and mass. In an effort to make a realistic comparison to observations, we do not only calculate the galaxies' ``true'' stellar and gas angular momenta by simply vector-summing the 3D angular momentum vectors of all particles. Instead, we first measure the galaxy rotation curve $v_{\rm rot}(r)$, and then proceed under the assumption that the component of interest (stars or HI) follows this rotation curve. The observationally-inferred specific angular momentum is then 
\begin{equation}
\label{eqn:obs_j}
\left\langle j\right\rangle_{{\rm mock\, obs}}=\frac{\int_{0}^{\infty}v_{{\rm rot}}(r)\Sigma(r)r^{2}\,{\rm d}r}{\int_{0}^{\infty}\Sigma(r)r\,{\rm d}r},
\end{equation}
where $\Sigma(r)$ is the surface density of the component of interest (stars or HI). Essentially all angular momentum studies of late-type galaxies \citep{vandenBosch_2001, Romanowsky_2012,Obreschkow_2014,Butler_2016,Chowdhury_2017} measure $v_{\rm rot}(r)$ from \textit{gas} kinematics and use $v_{\rm rot,\,gas}(r)$ (measured from HI) to calculate \textit{stellar} angular momentum. That is, they assume that stars and gas follow the same rotation curve. This assumption is obviously problematic, but it is ubiquitous in the literature for late-type galaxies largely due to the observational difficulty of measuring the stellar rotation curve. For our mock observations, we begin by also assuming that stars follow the HI rotation curve. We then investigate how the specific angular momentum values inferred in this way compare to the ``true'' stellar angular momentum.  

For $v_{\rm rot,\,gas}(r)$, we use the mean $v_{\phi}$ of HI gas averaged in cylindrical radius bins, not the mock-slit rotation curve (see Section~\ref{sec:rot_curves}). Most of the observational studies against which we compare obtain $v_{\rm rot}(r)$ by fitting a model to the observed 2D velocity field; other theoretical works have found that this approach can usually recover the true gas rotation curve with high fidelity, except in systems viewed at very high or low inclinations \citep{DiTeodoro_2015, Read_2016, Pineda_2017}. One of the observational studies we compare to \citep{Butler_2016} fit analytic profiles to $v_{\rm rot}(r)$ and $\Sigma(r)$ rather than integrating numerically. We do not do this, for consistency with the other studies, but we note that it could change our results in systems with irregular gas distributions and rotation curves.

In Figure~\ref{fig:j_vs_mass_baryons}, we plot the specific angular momentum of stars (panels 1 and 2), HI (panel 3), and all baryons (panel 4) for all the galaxies in our sample. We compare our simulations to a variety of observational data from the literature. Only $\left\langle j_{{\rm star}}\right\rangle$ and $\left\langle j_{{\rm HI}}\right\rangle$ are computed directly from the simulations; following the observational literature, we calculate the total baryon mass as $M_{\rm bar} = \rm M_{\rm star} + 1.33 M_{\rm HI}$ and the baryon specific angular momentum as $\left\langle j_{{\rm bar}}\right\rangle  = (\rm M_{\rm star}\left\langle j_{{\rm star}}\right\rangle + 1.33M_{\rm HI}\left\langle j_{{\rm HI}}\right\rangle )/M_{\rm bar}$, where the factor of 1.33 accounts for helium. We calculate $\left\langle j_{{\rm star}}\right\rangle$ both under the assumption that stars follow the HI rotation curve (panel 1), and using the true stellar rotation curve (panel 2). The primary observational sample to which we compare the stellar kinematics-derived $j_{\rm star}$ values is the sample of gas-poor S0s from \citet{Romanowsky_2012}. We supplement these with $j_{\rm star}$ calculations for the LMC and SMC based on resolved proper motion studies. There are no galaxies with high-quality observationally measured stellar rotation curves with $\rm M_{\rm star}\lesssim 10^{8} \rm M_{\odot}$. 

The simulated galaxies approximately reproduce the observed scalings for stars both when $\left\langle j_{{\rm star}}\right\rangle$ is measured from the HI rotation curve and when it is measured directly from stellar kinematics. However, $\left\langle j_{{\rm star}}\right\rangle$ for the simulated galaxies is systematically higher when it is calculated from the HI rotation curve, particularly in the low-mass galaxies. This is unsurprising, since we showed in Figure~\ref{fig:circularity_distributions} that stars rotate less than gas in all galaxies. 

Both the observed and simulated galaxies exhibit an approximate power law scaling between $j$ and $M$. Classical galaxy formation models predict $\left\langle j\right\rangle \propto M^{\alpha}$, with $\alpha \approx 2/3$ if galaxies approximately inherit the angular momentum of their dark matter halos and halos obtain their angular momentum from tidal torques \citep{Peebles_1969, Heavens_1988, Romanowsky_2012}. Comparing the simulated galaxies to the observational data by eye, we approximately reproduce the observed power law scaling, but the simulated galaxies are offset to lower specific angular momentum by a factor of 2-3, particular in $\left\langle j_{{\rm HI}}\right\rangle $. 

However, in interpreting Figure~\ref{fig:j_vs_mass_baryons}, it is important to note that the observational data plotted is \textit{not} an unbiased, morphologically blind sample of the galaxy population: with the exception of the data from \citet{Romanowsky_2012} in the second panel, all the studies represented in Figure~\ref{fig:j_vs_mass_baryons} specifically selected systems with well-behaved velocity fields and disky morphologies which could be well-fit by a simple tilted-ring model.\footnote{In particular, \citet{Obreschkow_2014} studied 16 spiral galaxies from the 23 THINGS galaxies for which data were available from \citet{Leroy_2008}, rejecting 7 that were too irregular. \citet{Butler_2016} rejected approximately half of the LITTLE THINGS dwarf galaxies for which data were available because they were too compact or irregular, or appeared to have low inclinations (but note that the velocity field of a puffy, dispersion-supported galaxy can be misinterpreted as a nearly face-on disk). \citet{vandenBosch_2001} selected 14 galaxies from the sample of 20 galaxies modeled in \citet{vdBosch2001b}, rejecting six galaxies whose rotation curves were not well-fit by rotating disk models (and these 20 galaxies were themselves selected from a larger sample of 73 galaxies observed by \citet{Swaters_1999} because they had high quality rotation curves without strong asymmetries). All the galaxies studied by \citet{Chowdhury_2017} have velocity fields well-fit by rotating disks.} There are no published observational angular momentum measurements of late-type galaxies with significantly disordered morphologies or kinematics (though such galaxies are quite common; \citealt{Roychowdhury_2010, Simons_2015}), because there is no straightforward way to measure the angular momentum observationally of systems with disordered kinematics. It is therefore possible that the apparent offset toward lower angular momentum for simulated galaxies in Figure~\ref{fig:j_vs_mass_baryons} is due in large part to selection effects in the observed samples. 
We also reiterate that the choice of how to measure galaxies' angular momentum from observational data (e.g., fitting a model to the velocity field, integrating the data numerically, or fitting a function to the rotation curve, etc.) can introduce factor of $\sim 2$ differences in the inferred $\left\langle j\right\rangle $. Given the observational and theoretical uncertainties in measuring $\left\langle j_{{\rm HI}}\right\rangle$, we regard the agreement with observations in Figure~\ref{fig:j_vs_mass_baryons} as quite satisfactory.

\begin{figure}
\includegraphics[width=\columnwidth]{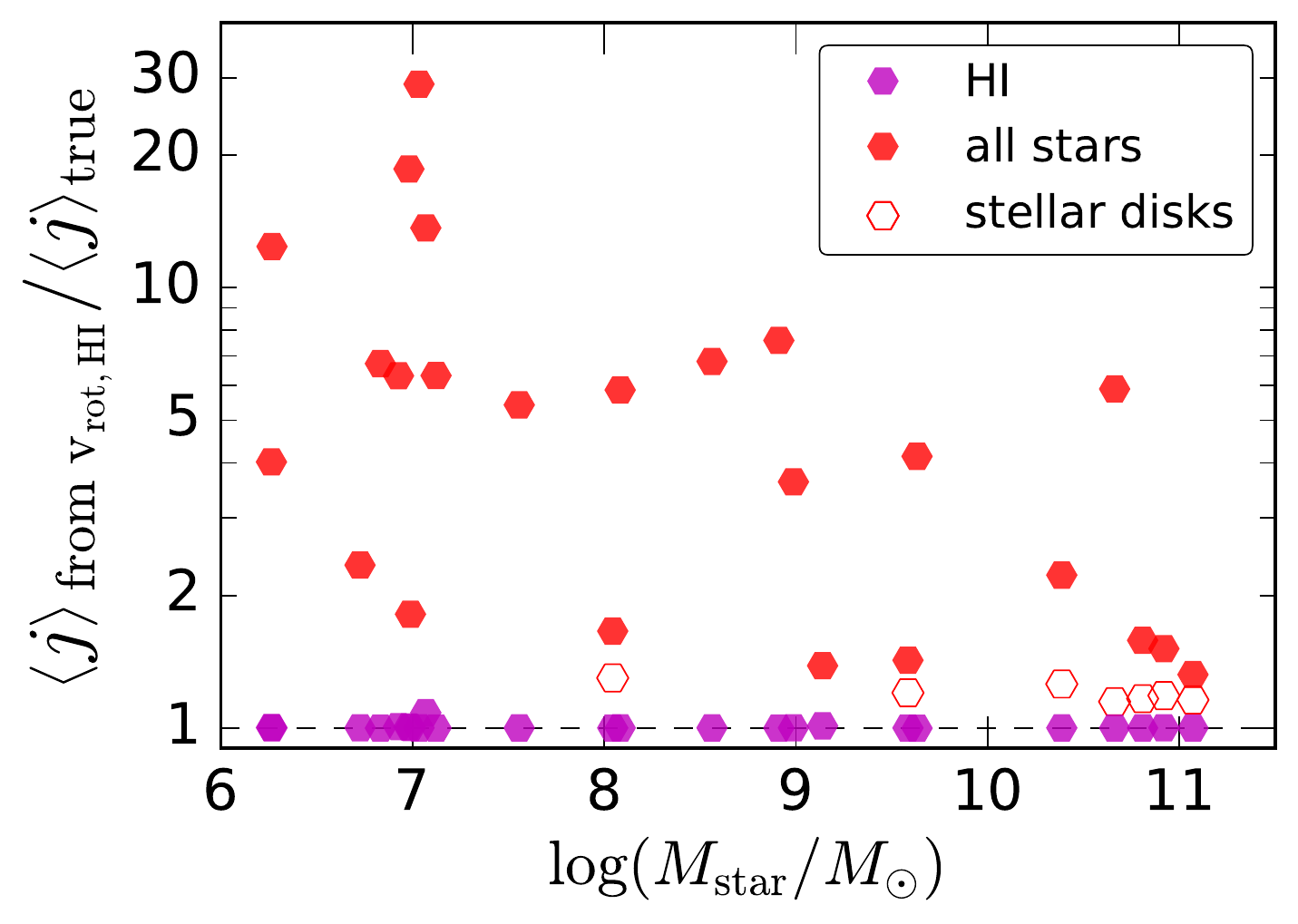}
\caption{Ratio of galaxies' stellar and HI specific angular momentum inferred under the assumption that stars are in a thin disk and follow the same rotation curve as HI gas (Equation~\ref{eqn:obs_j}, with $v_{\rm rot}(r) = v_{\rm rot,\,HI}(r)$), to $\left\langle j\right\rangle_{\rm true}$, the true angular momentum (Equation~\ref{eqn:ang_mom}). For HI, Equation~\ref{eqn:ang_mom} recovers the true angular momentum with high fidelity. In \textit{all} galaxies, but particularly at low masses, the stellar component is less rotationally supported than the gas. Assuming $v_{\rm rot,\,stars}=v_{\rm rot,\,\rm HI}$ can therefore dramatically overestimate the stellar angular momentum in our simulations. For galaxies with stellar disks, Equation~\ref{eqn:obs_j} accurately recovers the specific angular momentum of the disk within $\sim 20\%$ when the bulge is excluded.}
\label{fig:j_star_err}
\end{figure}

We now investigate explicitly the error introduced by approximating galaxies' angular momentum from the HI rotation curve. In Figure~\ref{fig:j_star_err}, we plot for stars and HI the ratio of the angular momentum calculated from Equation~\ref{eqn:obs_j} (under the assumption that all material follows the HI rotation curve) to the true angular momentum, which we calculate with Equation~\ref{eqn:ang_mom}. \textit{For every single galaxy in our sample}, using Equation~\ref{eqn:obs_j} and assuming $v_{\rm rot,\,stars} = v_{\rm rot\,,HI}$ overestimates the true stellar angular momentum. In the MW-mass galaxies, the error is relatively modest ($\sim 0.2$ dex), but in several of the lowest-mass galaxies, the inferred stellar specific angular momentum is more than a factor of 10 too high. This is not surprising: stars rotate less than gas in every galaxy (see Figure~\ref{fig:circularity_distributions}). On the other hand, Equation~\ref{eqn:obs_j} recovers the true specific angular momentum of HI quite accurately, implying that the primary source of error in $\left\langle j_{{\rm star}}\right\rangle$ is the mismatch between the stellar and HI rotation curves (as opposed to coherently counter-rotating stars).

Some observational studies of angular momentum \citep[e.g.][]{Romanowsky_2012, Fall_2013, Obreschkow_2014} attempt to measure the specific angular momentum of galaxies' stellar disks alone, either by explicitly removing the bulge or by studying bulgeless disk galaxies. In order to compare to such studies, we also show in Figure~\ref{fig:j_star_err} the error in the inferred stellar angular momentum when only star particles in the disk are included in the $\left\langle j_{{\rm star}}\right\rangle$ calculation. For the seven galaxies in our sample with stellar disks (the same galaxies shown in Figure~\ref{fig:young_old_stars}), we separate stars in the disk from those in the bulge and halo with the kinematic circularity-based decomposition used in \citet{Abadi_2003} and \citet{Grand_2017}.\footnote{Note that observational studies typically use photometric bulge-disk decompositions which do not necessarily yield the same results as kinematic decompositions \citep{Aumer_2013}. We do not expect the choice of decomposition to significantly affect our results, as the observations against which we compare select galaxies with small bulge-to-disk ratios.} When only disk stars are included in the specific angular momentum calculation, the error introduced by assuming that stars follow the HI rotation curve is modest: the specific angular momentum of the stellar disk is overpredicted by only $\sim 20\%$. This demonstrates that the primary source of error in using Equation~\ref{eqn:obs_j} and measuring $v_{\rm rot}$ from the HI rotation curve is the presence of dispersion-supported bulge/halo components, not asymmetric drift of stars in the disk. If galaxies are dominated by rotationally supported disks, then assuming $v_{\rm rot\,star} = v_{\rm rot\,,HI}$ should lead to only modest errors. However, this approach is likely to seriously overestimate galaxies' stellar angular momentum in galaxies at lower masses with stars that are primarily supported by dispersion. 

%Some previous works \citep[e.g.][]{vandenBosch_2001} have suggested that the observationally-inferred angular momentum distributions of low-mass galaxies pose a challenge for hierarchical galaxy formation models because they a) imply that dwarf galaxies lose the majority of their baryons but retain most of their initial angular momentum, and b) lack the low and high-specific angular momentum tails seen in the specific angular momentum distributions of dark matter halos. This tension may be partially alleviated if observational studies systematically overestimate stellar angular momentum in dwarf galaxies, as our results imply. In this case, the true baryon angular momentum of galaxies would be lower than previously inferred, and the ``missing'' low-angular momentum tail of the baryons would simply have been preferentially converted to stars. 

\subsection{Tully-Fisher Relation}
\label{sec:tfr}

\begin{figure}
\includegraphics[width=\columnwidth]{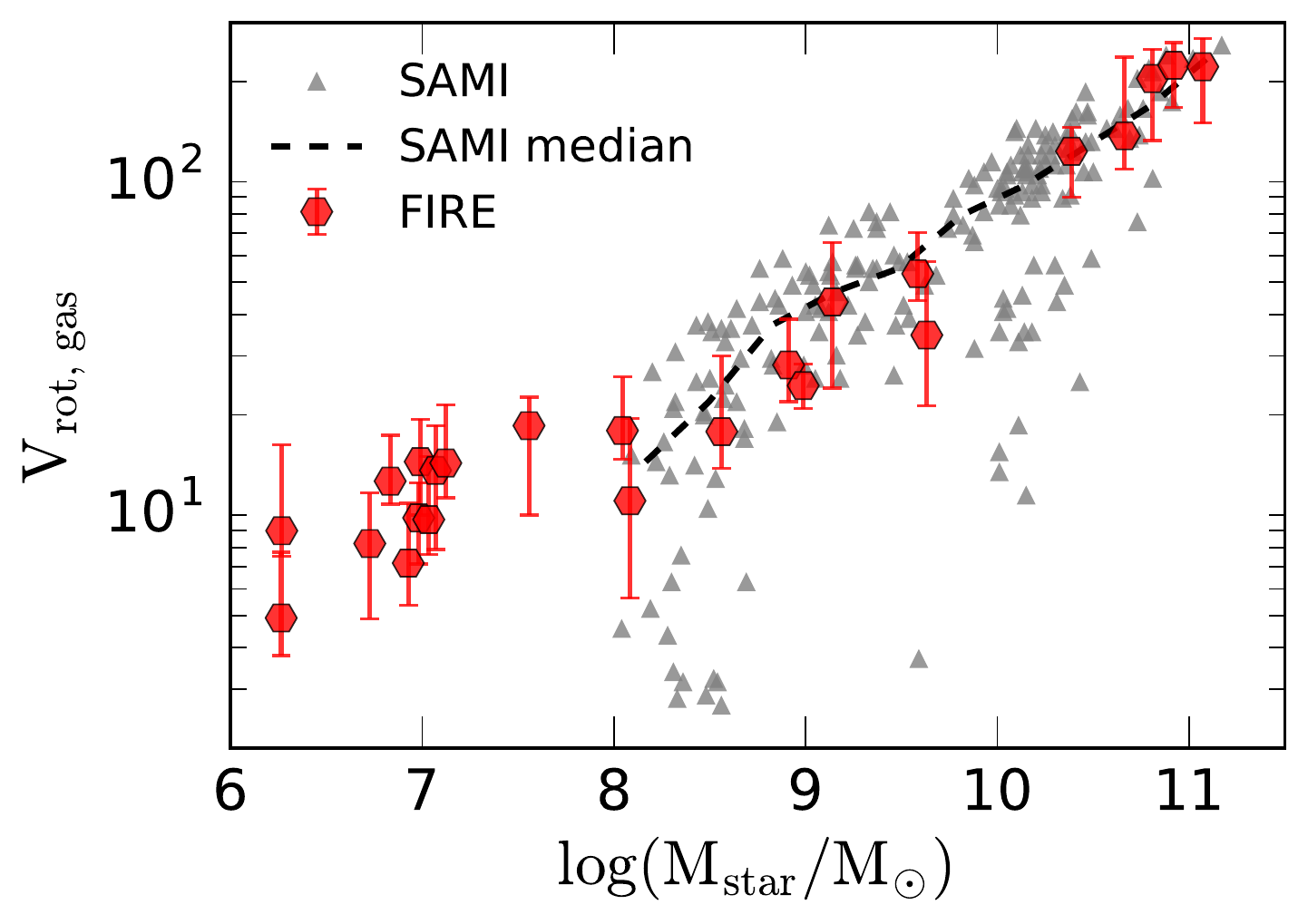}
\caption{Gas rotation velocity versus stellar mass. $V_{\rm rot}$ is measured from gas kinematics within one effective radius using mock-IFU data cubes (see text in Section~\ref{sec:tfr}). Red points and error bars show the median and middle 68\% of $V_{\rm rot,\,gas}$ values measured over 100 randomly distributed viewing angles. We compare to observations from the SAMI galaxy survey \citep{Cortese_2014}. Dashed line shows the binned median of the SAMI galaxies. The SAMI sample, being morphologically blind, includes many galaxies with low rotation velocities that are excluded from angular momentum studies because they lack gas disks. 
The simulations broadly lie on the observed relation, though they appear to slightly underpredict the number of high-$V_{\rm rot}$ systems at $\rm M_{\rm star} = 10^{8-10} \rm M_{\odot}$.}
\label{fig:SAMI}
\end{figure}

We next compare the rotation velocities of our galaxies to the observed scaling relation with galaxy mass. To test whether our galaxies have realistic rotation velocities and fall on a comparable relation to observed galaxies, we compare them to observational results from the SAMI galaxy survey presented in \citet{Cortese_2014}. This work measured gas kinematics within one effective radius for 193 nearby galaxies using integral-field spectroscopy. Unlike the observational angular momentum studies discussed in the previous section, \citet{Cortese_2014} did not explicitly select disk galaxies; their morphologically blind sample contains both dispersion and rotationally supported systems with a range of morphologies. 

We mock-observe our simulations to obtain measurements of $V_{\rm rot}$ that can be compared directly to those measured by \citet{Cortese_2014}. We begin by constructing gas velocity maps of the simulated galaxies similar to those in Figure~\ref{fig:vel_maps}. To mimic observations of galaxies with random orientations, we mock-observe each galaxy along 100 random viewing angles distributed uniformly on the unit sphere. For each viewing angle, we consider only pixels with projected radii $r \leq r_{\rm eff}$, where $r_{\rm eff}$ is the SDSS-$r$ band effective radius. $r_{\rm eff}$ is calculated after assigning luminosities to star particles by interpolating on a grid of stellar models from the Padova group \citep{Bressan_2012}. $r_{\rm eff}$ is often slightly smaller than the 3D half-mass radius, both because young stars are centrally-concentrated in most low-mass galaxies and because the 2D projected half-mass radius is necessarily smaller than the equivalent 3D quantity.

Following \citet{Cortese_2014}, we define the velocity width $W$ as the difference between the 90th and 10th percentile points of the histogram of pixel velocities within $r_{\rm eff}$; i.e., $W = V_{90} - V_{10}$. We then compute $V_{\rm rot}$ as 
\begin{equation}
\label{eqn:vrot_sini}
V_{{\rm rot}}=\frac{W}{2\sin\left(i\right)},
\end{equation}
where $i$ is the inclination for a particular viewing angle as defined by the net angular momentum vector of HI gas. Note that while \citet{Cortese_2014} measure gas kinematics from ionized gas (see \citealt{Ho_2014}), we use HI. Observational works have generally found gas rotation velocities inferred from ionized gas to be in good agreement with HI measurements \citep{Moiseev_2014}, though dispersions measured from ionized gas are often higher than those measured from HI \citep{Andersen_2006}, especially in starburst galaxies. We note that this definition of $V_{\rm rot}$ is sensitive not only to rotation, but to any large-scale gas motions.  

In Figure~\ref{fig:SAMI}, we compare our resulting calculations of $V_{\rm rot}$ to those measured in the SAMI survey. For each galaxy, we plot the median and middle 68\% scatter in $V_{\rm rot}$ values across 100 random viewing angles. We emphasise that this $V_{\rm rot} - \rm M_{\rm star}$ relation is not equivalent to the standard Tully-Fisher Relation (TFR), which is generally presented as the relationship between the \textit{asymptotic} circular velocity and $\rm M_{\rm star}$ for disk galaxies. 

In Figure~\ref{fig:SAMI} (both for the simulations and the observational data), there is significant scatter, particularly at lower masses. Many observational works \citep[e.g.][]{McGaugh_2000} report a TFR with significantly less scatter than the data shown here; the data shown here fall on a less tight correlation both because unlike many TFR studies, \citet{Cortese_2014} did not explicitly select disk galaxies (see \citealt{Bradford_2016}) and because $V_{\rm rot}$ is only measured within one effective radius, where galaxies are often less rotationally supported than at larger radii.

Over the mass range probed by observations, our galaxies' $V_{\rm rot}$ values are in good agreement with the observed sample. All of our galaxies with $\rm M_{\rm star} > 10^{10} \rm M_{\odot}$, which are rotationally supported and morphologically disky, have $V_{\rm rot} > 100\,\rm km\,s^{-1}$, and fall on a tighter relationship than galaxies at lower masses. The increased scatter in the simulated galaxies' $V_{\rm rot}$ values at lower $\rm M_{\rm star}$ is consistent with the observed sample. At $\rm M_{\rm star} = 10^{8-10} \rm M_{\odot}$, the simulated galaxies fall within the parameter space spanned by the observed sample. However, their mean rotation values are slightly lower than those of the observed galaxies, and we note that none of the simulated galaxies in this mass range have median $V_{\rm rot}$ values in the upper $\sim 50\%$ of the observed $V_{\rm rot}$ distribution at fixed mass. This may indicate that the simulations produce too few highly rotationally-supported systems at low mass. It could also stem from the fact that even the rotationally supported galaxies in our suite tend to be dispersion-supported in their nuclear regions; see, for example, the rotation curve of \texttt{m11b} in Figure~\ref{fig:m11b}.

\citet{Cortese_2014} also presented velocity dispersion data for galaxies in the SAMI survey. Their measured dispersions are systematically higher than what we measure in our simulated galaxies, particularly at $\rm M_{\rm star}\gtrsim 10^{10}\rm M_{\odot}$. We do not present a comparison of the simulated galaxies' dispersions because a) these are likely to be significantly higher for gas kinematics traced by ionized gas than for HI, and b) many of the high-mass SAMI galaxies host AGN, which are likely to drive up the measured dispersions and are not modeled in our simulations.  

The comparable $V_{\rm rot}$ values for our simulated galaxies and the observed sample suggests that our simulated galaxies have a realistic amount of rotational support, except perhaps for a shortage of fast-rotating systems at low masses. It is important to note, however, that the measures of $V_{\rm rot}$ in Figure~\ref{fig:SAMI} are \textit{not} equivalent to the asymptotic $V_{\rm flat}$ or $V_{\rm max}$ used in some TFR studies. Because most of our galaxies have rising rotation curves, $V_{\rm rot}$ measured inside one effective radius is almost always lower than the maximum rotation velocity. We opted to compare to the observations from the SAMI survey rather than other studies reaching larger radii because such studies without exception impose morphological selection criteria; i.e., they preferentially select disk galaxies. Because the primary goal of many of these studies is to probe the galaxies' mass distributions and gravitational potentials, they also often attempt to explicitly correct measured rotation curves for pressure support and non-circular motions, making it less straightforward to directly compare them to simulations. The completion of more homogeneous, morphologically blind observational datasets that reach to larger radii \citep[e.g.][]{Glazebrook_2013, Bundy_2015} will make it easier to make statistically fair comparisons between observed gas kinematics and those in simulated galaxies.

\begin{figure}
\includegraphics[width=\columnwidth]{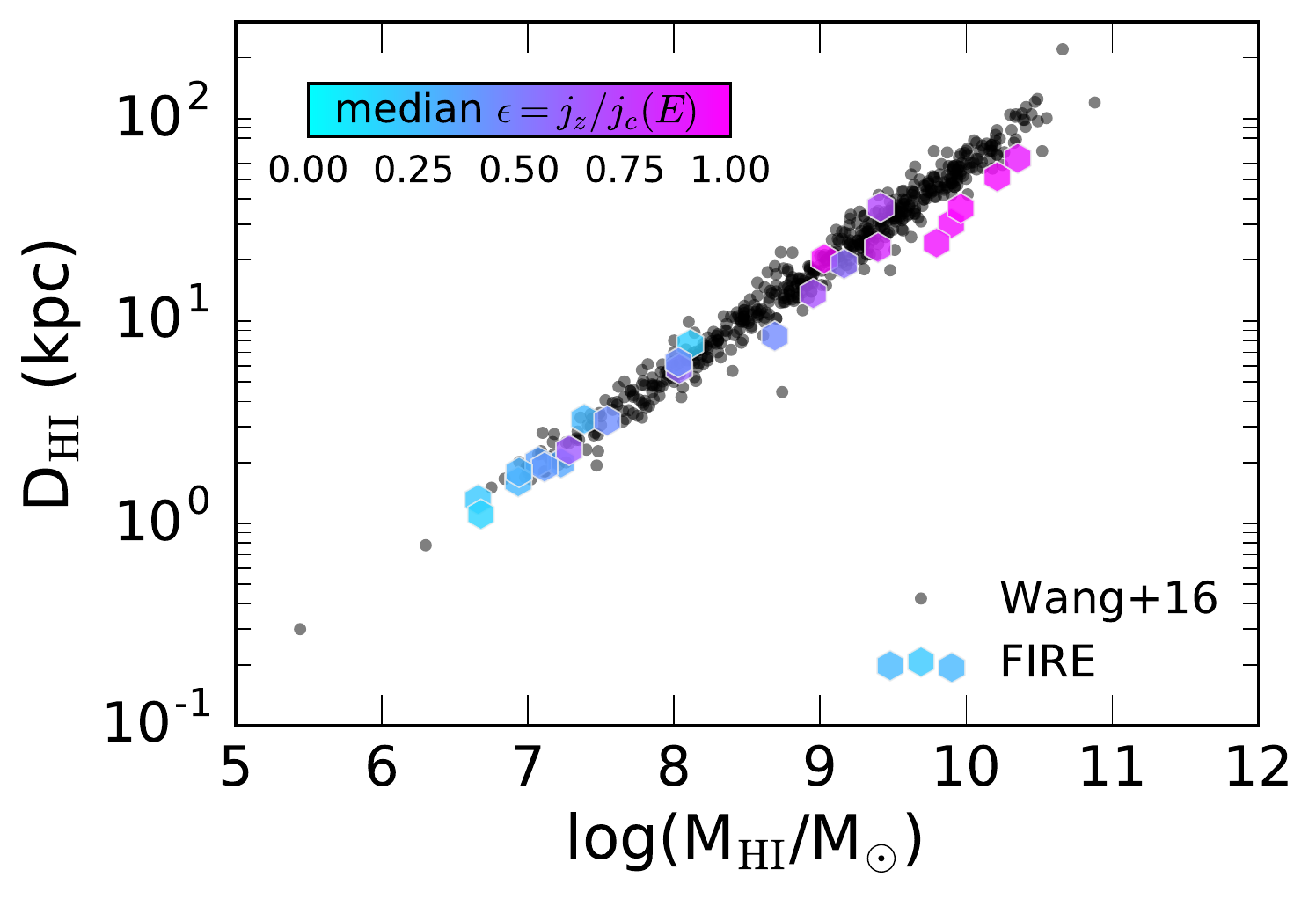}
\caption{HI size -- mass relation for all the galaxies in our sample at $z=0$. $D_{\rm HI}$ is the diameter of the largest annulus with HI surface density exceeding $\rm 1\, \rm M_{\odot}\,\rm pc^{-2}.$ Points are colour-coded by the median orbital circularity (Equation~\ref{eqn:circularity}), a measure of rotational support. We compare to the observational sample presented in \citet{Wang_2016}. The simulated low-mass galaxies fall on a tight power law consistent with observations, irrespective of whether or not they rotate. The most massive galaxies ($M_{\rm HI}\gtrsim 10^{9.5}\rm M_{\odot}$) are slightly smaller (by a factor of $1.5$ on average) than the observed relation.}
\label{fig:D_HI}
\end{figure}

\subsection{HI Size-Mass relation}
\label{sec:D_HI}
Rotation versus dispersion support in galaxies is closely related to size. If galaxies are rotationally supported, those with higher specific angular momentum will be larger \citep{Fall_1980, Mo_1998, Zasov_2017}. At fixed angular momentum, galaxies which rotate more slowly will be larger; for example, if a galaxy's gas is puffed up by feedback processes, its rotation will slow down. We therefore investigate the sizes of our galaxies' gas distributions. Following the observational literature, we quantify galaxy sizes using $D_{\rm HI}$, the diameter of the HI disk. $D_{\rm HI}$ is defined as the diameter of the largest annulus within which the HI surface density exceeds $1\rm\,\rm M_{\odot}\,pc^{-2}$. 

In Figure~\ref{fig:D_HI}, we compare measured $D_{\rm HI}$ values for our simulated galaxies to the observational data from \citet{Wang_2016}, who compiled HI data for a diverse sample of more than 500 spiral and irregular galaxies spanning more than 5 decades in HI mass. Both the observed and simulated galaxies fall on a remarkably tight power law at lower masses. Although our low-mass galaxies exhibit a large diversity of morphologies and most do not form disks, they fall neatly on the observed relation. The five most massive galaxies (which all have thin gas disks) fall slightly below the observed relation. The disagreement is not enormous -- the simulated galaxies are smaller than the observed galaxies by a factor of $\sim 1.5$ on average -- but it does appear to be systematic. Intriguingly, we note that while the isolated MW-mass galaxies in Figure~\ref{fig:D_HI} appear systematically smaller than the observed trend, galaxies at the same halo mass in simulations of Local-Group analogs are systematically larger and more gas-rich (Garrison-Kimmel et al., in prep).  We defer a careful investigation of these galaxies' sizes to other work. We further emphasise that $D_{\rm HI}$, which traces only cool atomic gas, is not one-to-one with size metrics based on all gas.   

We colour points in Figure~\ref{fig:D_HI} by their HI orbital circularity (Equation~\ref{eqn:circularity}), in order to determine whether there are systematic offsets between the sizes of galaxies supported primarily by dispersion and those exhibiting significant rotation. Size does not appear to be significantly correlated with rotational support at fixed mass. Galaxies with low specific angular momentum are supported primarily by turbulent pressure from feedback, while those with high specific angular momentum are supported primarily by rotation, but both types of galaxies reach approximately the same size. This is perhaps somewhat surprising, in the sense that the low-mass galaxies have a range of specific angular momenta, and one might expect those with higher specific angular momentum to be larger. This suggests that, at least in the mass range where most of the galaxies do not form disks, angular momentum is not a primary determinant of HI disk size. 

\section{Discussion}
\label{sec:discussion}

\subsection{Low baryon angular momentum in dwarfs}
\label{sec:lowbardisc}
One of the main results of this work is that the ratio of baryon to dark matter specific angular momentum, $\left\langle j_{{\rm gas}}\right\rangle/\left\langle j_{{\rm DM}}\right\rangle$, decreases systematically with decreasing galaxy mass. This is true both when $\left\langle j_{{\rm bar}}\right\rangle$ refers only to the specific angular momentum of gas or stars in the galaxy, and when it includes all material within the virial radius, though the scatter is lower when $\left\langle j_{{\rm bar}}\right\rangle$ is measured only within the central galaxy.(Figure~\ref{fig:jgas_over_jdark}). The physical origin of this decrease can be better understood by contrasting low-mass galaxies' formation histories with those of higher mass galaxies.

A number of studies \citep{Governato_2010, Teklu_2015, Sokolowska_2016, Stewart_2016, Zjupa_2017} have found that the specific angular momentum of gas -- both inside the central galaxy and in the halo -- is usually \textit{higher} than that of the dark matter halo at $z=0$. Exactly why this occurs remains somewhat contested, but there is broad agreement in the literature that it is related to feedback-driven galactic winds, which increase the average angular momentum of the baryons retained in galaxies \citep{Brook_2011, Brook_2012, Ubler_2014, DeFelippis_2017}. This occurs both because feedback preferentially removes low angular momentum gas and because feedback-driven outflows gain angular momentum from the high angular momentum CGM before being re-accreted. 

Especially in higher mass galaxies, feedback is most efficient at driving outflows at early times, when the galactic potential well is shallower. This is also when the specific angular momentum of accreted material is lowest: gas that is accreted at late times  has been subject to more tidal torques from large scale structure \citep{Ryden_1988, Quinn_1992} and has higher specific angular momentum (by as much as an order of magnitude) than gas accreted earlier \citep{Brook_2011, Pichon_2011, Stewart_2013, Ubler_2014, Stewart_2016}. Disks thus typically form after $z\sim 1$, when feedback becomes inefficient at driving galactic winds and galaxies begin to accrete large quantities of high angular momentum gas. 

This picture breaks down for several reasons in the lowest mass galaxies in our sample ($\rm M_{\rm star}\lesssim 10^{8} \rm M_{\odot}$). First, the gravitational potentials of these galaxies are sufficiently shallow that feedback continues to efficiently drive galactic winds into the halo at late times \citep{Muratov_2015,Muratov_2016,AnglesAlcazar_2016}. As a result, both low and high angular momentum gas (accreted at early and late times, respectively), are efficiently evacuated from the central galaxy into the halo; feedback less preferentially selects low-angular momentum gas for removal. 

Second, and likely more importantly, gas accretion is stymied at late times in low-mass halos \citep{Bullock_2000, Hoeft_2006, FG_2011, Noh_2014}. In the lowest-mass halos ($M_{\rm 200m} \sim 10^{10} \rm M_{\odot}$), the virial temperature is a few $\times 10^{4}\,\rm K$, comparable to the temperature of the photoionized IGM \citep{FaucherGiguere_2009}. For these halos, gas accretion becomes very inefficient after $z\sim 5$, because the gravitational potential can barely retain gas that is heated by the UV background and is not self-shielding. This is true even in the absence of stellar feedback; feedback only makes matters worse, because a) gas that is driven into the halo by galactic winds and would be recycled at higher halo masses cannot cool back into the galaxy, and b) feedback-driven outflows entrain halo gas (which has relatively high specific angular momentum) and push it out of the halo \citep{Muratov_2015}. Most of the gas that ends up in low-mass halos at $z=0$ was thus accreted at early times and has low specific angular momentum. 

Even in intermediate-mass halos ($M_{\rm 200m}=10^{10-11} \rm M_{\odot}$), which have virial temperatures well above the temperature of the IGM, gas accretion is significantly less efficient at late times than in MW-mass galaxies. For example, \citet{Noh_2014} found that on average, halos with $M_{\rm 200m} \sim 10^{11}\rm M_{\odot}$ are unable to accrete unshocked intergalactic gas by $z=0$, with the transition occurring at earlier times in lower mass halos. Since high angular momentum gas is preferentially available for accretion at late times, it is unsurprising that low-mass galaxies, which accrete less efficiently at late times, will end up with lower $\left\langle j_{{\rm gas}}\right\rangle$. 

Inefficient cooling and accretion can thus likely also explain the large reservoirs of high-angular momentum gas in the CGM of many of our galaxies (Figure~\ref{fig:cumulative_angmom}). Many of the low and intermediate-mass galaxies in our sample (e.g. \texttt{m11q}) have significantly higher average $\left\langle j_{{\rm gas}}\right\rangle$ in their CGM and outer halo than in the central galaxy; if this gas could cool efficiently, these galaxies would likely be significantly more rotationally supported. The median cooling time of CGM gas in these halos is only a few Gyr, so much of the gas would likely cool into the galaxy in the absence of external energy sources \citep[e.g.][]{Fielding_2017}. However, heating from the UV background and star formation, likely in combination with feedback-driven outflows, prevent CGM gas from reaching the central galaxy. This also explains why the gas in our low-mass galaxies is significantly less centrally concentrated on average than in the MW-mass galaxies (Figure~\ref{fig:cum_mass_profiles}). 

\citet{Zjupa_2017} studied the relative specific angular momentum of gas and dark matter in a large number of galaxies in the Illustris simulation. They did not detect a decrease in $\left\langle j_{{\rm gas}}\right\rangle /\left\langle j_{{\rm DM}}\right\rangle $ at low masses comparable to what we find in Figure~\ref{fig:jgas_over_jdark}, or any significant mass scaling in $\left\langle j_{{\rm gas}}\right\rangle /\left\langle j_{{\rm DM}}\right\rangle $. While this seems somewhat inconsistent with our results, we emphasise that the decrease in $\left\langle j_{{\rm gas}}\right\rangle /\left\langle j_{{\rm DM}}\right\rangle $ seen in Figure~\ref{fig:jgas_over_jdark} does not become significant until $\rm M_{\rm star} \lesssim 10^{8} \rm M_{\odot}$. In large-volume simulations like Illustris, this mass scale is only marginally resolved, corresponding to fewer than 100 star particles per galaxy. In addition, because Illustris does not resolve a high-density cold ISM, its star formation is less concentrated and episodic than that seen in simulations employing a high density threshold for star formation (see \citealt{Pontzen_2014} for further discussion). Finally, the decrease in  $\left\langle j_{{\rm gas}}\right\rangle /\left\langle j_{{\rm DM}}\right\rangle$ at low masses is stronger when $\left\langle j_{{\rm gas}}\right\rangle$ is measured within the central galaxy ($r<0.1R_{\rm 200m}$); \citet{Zjupa_2017} included all material with the halo. This may explain why these authors did not find the decrease in $\left\langle j_{{\rm gas}}\right\rangle /\left\langle j_{{\rm DM}}\right\rangle$ at lower masses found here. 

\subsection{Absence of disk formation in very low mass galaxies}
As can be seen from our galaxies' HI moment maps, none of the galaxies with $\rm M_{\rm star} \lesssim 10^{8} \rm M_{\odot}$ form thin disks, and most of them (with the exception of \texttt{m10z} at $\rm M_{\rm star} = 10^{7.6} \rm M_{\odot}$) do not exhibit any significant rotational flattening (see Figure~\ref{fig:vel_maps} and Appendix~\ref{sec:idvid_gals}).

A number of studies \citep{Read_2006, Kaufmann_2007_disk, Dalcanton_2010} have provided theoretical arguments for why disk formation should fail in very low-mass galaxies. Even absent coherent galactic winds, thermal and turbulent pressure support (from e.g., supernovae, cosmic rays, or the photoionizing background) in the warm ($T\sim 10^4 \,\rm K$) ISM keeps gas velocities at a significant fraction of the local circular velocity in very low-mass halos (see Figure~\ref{fig:all_v_sig}), so that gas pressure support is always dynamically non-negligible compared to angular momentum support. For galaxies in $M_{\rm 200m}\sim 10^{10}\rm M_{\odot}$ halos, the equilibrium size of a gas distribution supported by thermal pressure is comparable to the expected size of the angular momentum supported disk if the galaxy inherits the specific angular momentum of its host halo \citep{Kaufmann_2007_disk}. 

Additionally, most of our lowest-mass galaxies have sufficiently high gas fractions that the Toomre mass ($M_{\rm Toomre}\sim f_{\rm gas}^2 M_{\rm gas}$), which sets the scale of the largest gas clumps in a disk, is comparable to the entire baryonic mass of the galaxy \citep[see][]{Hopkins_2012}. Very low-mass galaxies are thus susceptible to becoming dominated by a small number of massive clumps, even if they are initially disky \citep{FaucherGiguere_2017}. The inclusion of energetic feedback from star formation further inhibits disk formation and survival. In our lowest-mass galaxies, the energy input from just 10 supernovae is sufficient to coherently accelerate all the gas in the galaxy to $v\sim 10\,\rm km\,s^{-1}$. Due to the very bursty mode of star formation in this mass regime, supernovae are generally highly spatially and temporally clustered, enhancing the effect of their feedback. 

There are thus several reasons to expect that the lowest-mass galaxies in our suite should not form gas disks, irrespective of their angular momentum content. And indeed, we find in idealized experiments that galaxies simulated with the FIRE feedback model \textit{cannot} maintain disks in halos with $v_{\rm max}\lesssim (40-50)\,\rm km\,s^{-1}$; even galaxies initialized with thin rotation-supported disks are rapidly puffed up and destroyed within a few 100 Myr. In slightly more massive halos, disks can survive if they are initialized with sufficient angular momentum; the paucity of disks in cosmologically simulated halos with $M_{\rm 200m}=10^{10-11}\rm M_{\odot}$ is thus likely exacerbated by the diminished gas angular momentum content of these halos. 

A number of other simulations of low-mass galaxies have found a comparable lack of rotational support. For example, \citet{Verbeke_2017} recently studied the gas kinematics of a sample of 10 low-mass galaxies from the ``Moria'' simulations. Their main conclusion was that dynamical mass profiles inferred from modeling gas kinematics tend to systematically underestimate the total mass, because gas is significantly pressure-supported and thus does not cleanly trace the gravitational potential. This is consistent with our results (see Figure ~\ref{fig:rot_curves}). Equally interesting, \citet{Verbeke_2017} were only able to construct kinematic models for 10 of the $\sim 30$ galaxies in their suite -- the remaining $\sim 20$ galaxies had such irregular morphologies and kinematics that it was impossible to construct a kinematic model. Similarly, \citet{Obreja_2016} studied disky galaxies selected from the suite of $100$ ``NIHAO'' zoom-in simulations \citep{Wang_2015} spanning $M_{\rm 200m} = 10^{9.7-12.3}\rm M_{\odot}$; only 18 of the 100 galaxies had stellar disks, and \textit{all} the disky galaxies had $M_{\rm 200m} > 10^{11} \rm M_{\odot}$. 

These theoretical results are consistent with observational studies that have found galaxies' gas disks to become thicker and more dispersion-supported at lower masses \citep{Roychowdhury_2013, Simons_2015, Johnson_2017}. A quantitative comparison of the degree of rotational support found in our simulated galaxies with observed systems is difficult to carry out due to our limited sample size and selection effects in observed samples. 

That being said, there \textit{are} observed galaxies at the lowest mass scales probed by our simulations that appear to be more rotationally supported than any of the galaxies in our sample at the same $\rm M_{\rm star}$. The most comprehensive survey of resolved HI kinematics in low-mass galaxies is LITTLE THINGS \citep{Hunter_2012}. LITTLE THINGS obtained gas kinematics of nine galaxies\footnote{UGC 8508, DDO 53, DDO 154, F564-V3, DDO 126, DDO 216, WLM, and Haro 29.} at the mass scale where none of the galaxies in our sample form disks ($\rm M_{\rm star} = (10^{6} - 2\times 10^{7}) \rm M_{\odot}$); six of these were found to have maximum rotation velocities $V_{\rm rot} \geq 30\,\rm km\,s^{-1}$ \citep{Oh_2015, Iorio_2017}. In our simulations, all ten galaxies at this mass scale have $V_{\rm rot} = (10-20)\rm\,km\,s^{-1}$. This may indicate that the simulations do not capture all the relevant physics for forming highly rotationally supported galaxies at these mass scales. 

However, the rotation velocities for LITTLE THINGS galaxies are derived by fitting a model to galaxies' velocity maps, which will not necessarily yield identical results to directly measuring $v_{\rm rot}(r)$ from the simulations. A detailed comparison with such observations requires that the same modeling procedure be applied to the simulated galaxies. In addition, the LITTLE THINGS galaxies are a highly-rotating subset of the observed galaxy population; they are selected for dynamical modeling studies precisely because they are rotationally supported. We therefore caution that comparisons of simulations to observational studies of stellar and gas kinematics should compare to morphologically blind observational samples whenever possible. 

\section{Summary}
\label{sec:summary}
We have investigated the gas rotation curves, morphologies, and detailed kinematics of isolated galaxies in a suite of 24 baryonic cosmological zoom-in simulations run with the FIRE-2 model \citep{Hopkins_2017} for star formation and stellar feedback. We explored how rotation versus dispersion support in galaxies is related to the mass and angular momentum content of their CGM and quantified how the ratio of galaxies' specific angular momentum to that of their dark matter halos scales with stellar mass. Our main results are as follows: 

\begin{enumerate}[leftmargin=*]
\item \textit{The gas in galaxies becomes diskier and more rotationally-supported with increasing mass.} Morphologically, none of our lowest-mass galaxies ($\rm M_{\rm star} \lesssim 10^{8}\rm M_{\odot}$) form classical gas disks (Figure~\ref{fig:vel_maps}). In many of these galaxies, the gas is primarily supported by dispersion (Figure~\ref{fig:all_v_sig}). Some systems do show kinematic rotational support, but they rotate at well below the halo circular velocity, and none are flattened by rotation (Figure~\ref{fig:shape}). Intermediate-mass galaxies ($\rm M_{\rm star} = 10^{8 - 10} \rm M_{\odot}$) exhibit a range of kinematics and morphologies, from thin, rotationally supported disks to dispersion-supported spheroids with negligible velocity gradients. All of the highest-mass galaxies ($\rm M_{\rm star} = 10^{10-11} \rm M_{\odot}$) form rotationally supported gas disks. 

\item \textit{Gas is more rotationally supported than stars, especially in low-mass galaxies:} None of our lowest-mass galaxies ($\rm M_{\rm star} \lesssim 10^{8} \rm M_{\odot}$) -- even those with significant gas rotation -- show any rotation in their stellar component (Figure~\ref{fig:circularity_distributions}). This occurs because, at least in the FIRE simulations, stars in low-mass galaxies form almost exclusively from high-density gas in the galactic centre (see Figure~\ref{fig:sf_gas_circ} and \citealt{ElBadry_2016}), which necessarily has low angular momentum. The stellar component does rotate in higher mass galaxies, but even systems with thin gas disks have significant non-rotating bulge components consisting of primarily older stars (Figure~\ref{fig:young_old_stars}).

\item \textit{In low-mass halos, baryons have lower specific angular momentum than dark matter}: For galaxies with $\rm M_{\rm star} > 10^{10} \rm M_{\odot}$, we recover the result from previous studies that on average, $\left\langle j_{{\rm gas}}\right\rangle > \left\langle j_{{\rm DM}}\right\rangle$ of the host halo. However, the opposite is true at lower masses: most galaxies with $\rm M_{\rm star} < 10^{8} \rm M_{\odot}$ have $\left\langle j_{{\rm gas}}\right\rangle < \left\langle j_{{\rm DM}}\right\rangle $, sometimes by more than an order of magnitude (Figure~\ref{fig:jgas_over_jdark}). The reduced angular momentum of baryons in low-mass halos is a result of the fact that low-mass halos accrete gas less efficiently at late times, when the mean specific angular momentum of accreted gas is highest (Figure~\ref{fig:j_evolution}).

\item \textit{The CGM has high specific angular momentum}: Almost all the halos in our suite have higher mean specific angular momentum in their diffuse, ionized CGM than in the cool gas and stars in the central galaxy (Figure~\ref{fig:cumulative_angmom}). In many dispersion-supported galaxies, the specific angular momentum of the CGM exceeds that of the central galaxy by more than a factor of 5; if this gas could efficiently cool, these systems would likely be significantly more rotationally supported. 

\item \textit{Halos with a smaller fraction of their baryons in the CGM have more rotationally supported gas in their central galaxies}: Low-mass halos have significantly less centrally concentrated baryon distributions than high mass galaxies (Figure~\ref{fig:cum_mass_profiles}) due to their strong outflows and inefficient cooling, which prevent high-angular momentum gas from being accreted onto the galaxy at late times (Figure~\ref{fig:j_evolution}). We find that halos with more centrally concentrated baryon distributions (i.e., higher $M_{{\rm bar}}(<0.1R_{{\rm 200m}})/M_{{\rm bar}}(<R_{{\rm 200m}})$) have more rotationally supported gas, even at fixed halo mass (Figure~\ref{fig:mass_frac_vs_circ}). Additional work is required to clarify the precise causal relation between mass, baryon concentration, and rotational support. 

\item \textit{Comparison with observations:} We approximately recover the observed scaling relations between galaxy mass and angular momentum (Figure~\ref{fig:j_vs_mass_baryons}), rotation velocity (Figure~\ref{fig:SAMI}), and size (Figure~\ref{fig:D_HI}) when we measure these quantities similarly to how they are measured in observational works. At $\rm M_{\rm star} > 10^{10} M_{\odot}$, our galaxies' HI disks are systematically smaller than the observed HI mass-size relation by a factor of $\sim 1.5$. We appear to produce fewer high-angular momentum, low mass galaxies than are represented in observational studies. However, the extent of this disagreement is difficult to quantify due to our small sample size and selection effects in observational studies, which are often biased towards rotating systems. A larger suite of simulations and a morphologically blind sample of observed galaxy kinematics is required to determine conclusively whether the tentative disagreement we find here is statistically significant. 

We caution that the standard practice of using the gas rotation curve to measure stellar angular momentum may significantly overestimate angular momentum (Figure~\ref{fig:j_star_err}), particularly at low stellar masses. In our simulated galaxies, stars are without exception less rotationally supported than gas (Figure~\ref{fig:circularity_distributions}).
\end{enumerate}

In future work, we will investigate in more detail the redshift evolution of these galaxies' specific angular momentum in order to elucidate what determines whether or not galaxies with intermediate masses form disks. 

\section*{Acknowledgements}
We thank the anonymous referee for useful comments.
We thank Alyson Brooks, Michael Fall, Marla Geha, Chris Hayward, Ryan Leaman, Anne Medling, Matt Orr, Jessica Werk, and John Wise for helpful discussions, Alex Richings and Alex Gurvich for help in setting up initial conditions, Peter Nugent for assistance optimizing GIZMO on high-performance computing centres, Luca Cortese for sharing observational data, and Jing Wang for making her data publicly available. 
KE gratefully acknowledges support from a Berkeley graduate fellowship, a Hellman award for graduate study, and an NSF Graduate Research Fellowship.
EQ was supported by NASA ATP grant 12-ATP-120183, a Simons Investigator award from the Simons Foundation, and the David and Lucile Packard Foundation. 
AW was supported by a Caltech-Carnegie Fellowship, in part through the Moore Center for Theoretical Cosmology and Physics at Caltech, and by NASA through grant HST-GO-14734 from STScI.
DK was supported by NSF grant AST-1412153 and the Cottrell Scholar Award from the Research Corporation for Science Advancement.
CAFG was supported by NSF through grants AST-1412836 and AST-1517491, and by NASA through grant NNX15AB22G.
We ran numerical calculations on the Caltech compute cluster ``Zwicky'' (NSF MRI award \#PHY-0960291). This research used resources of the National Energy Research Scientific Computing Center, a DOE Office of Science User Facility supported by the Office of Science of the U.S. Department of Energy under Contract No. DE-AC02-05CH11231. This research made use of Astropy, a community-developed core Python package for Astronomy \citep{Astropy_2013} and the \texttt{sauron} colormap developed by Michele Cappellari \citep{Cappellari_2008}. 

%%%%%%%%%%%%%%%%%%%%%%%%%%%%%%%%%%%%%%%%%%%%%%%%%%

%%%%%%%%%%%%%%%%%%%% REFERENCES %%%%%%%%%%%%%%%%%%

\bibliographystyle{mnras}

\begin{thebibliography}{}
\makeatletter
\relax
\def\mn@urlcharsother{\let\do\@makeother \do\$\do\&\do\#\do\^\do\_\do\%\do\~}
\def\mn@doi{\begingroup\mn@urlcharsother \@ifnextchar [ {\mn@doi@}
  {\mn@doi@[]}}
\def\mn@doi@[#1]#2{\def\@tempa{#1}\ifx\@tempa\@empty \href
  {http://dx.doi.org/#2} {doi:#2}\else \href {http://dx.doi.org/#2} {#1}\fi
  \endgroup}
\def\mn@eprint#1#2{\mn@eprint@#1:#2::\@nil}
\def\mn@eprint@arXiv#1{\href {http://arxiv.org/abs/#1} {{\tt arXiv:#1}}}
\def\mn@eprint@dblp#1{\href {http://dblp.uni-trier.de/rec/bibtex/#1.xml}
  {dblp:#1}}
\def\mn@eprint@#1:#2:#3:#4\@nil{\def\@tempa {#1}\def\@tempb {#2}\def\@tempc
  {#3}\ifx \@tempc \@empty \let \@tempc \@tempb \let \@tempb \@tempa \fi \ifx
  \@tempb \@empty \def\@tempb {arXiv}\fi \@ifundefined
  {mn@eprint@\@tempb}{\@tempb:\@tempc}{\expandafter \expandafter \csname
  mn@eprint@\@tempb\endcsname \expandafter{\@tempc}}}

\bibitem[\protect\citeauthoryear{{Abadi}, {Navarro}, {Steinmetz}  \&
  {Eke}}{{Abadi} et~al.}{2003}]{Abadi_2003}
{Abadi} M.~G.,  {Navarro} J.~F.,  {Steinmetz} M.,   {Eke} V.~R.,  2003, \mn@doi
  [\apj] {10.1086/378316}, \href
  {http://adsabs.harvard.edu/abs/2003ApJ...597...21A} {597, 21}

\bibitem[\protect\citeauthoryear{{Adams} et~al.,}{{Adams}
  et~al.}{2014}]{Adams_2014}
{Adams} J.~J.,  et~al., 2014, \mn@doi [\apj] {10.1088/0004-637X/789/1/63},
  \href {http://adsabs.harvard.edu/abs/2014ApJ...789...63A} {789, 63}

\bibitem[\protect\citeauthoryear{{Agertz} \& {Kravtsov}}{{Agertz} \&
  {Kravtsov}}{2016}]{Agertz_2016}
{Agertz} O.,  {Kravtsov} A.~V.,  2016, \mn@doi [\apj]
  {10.3847/0004-637X/824/2/79}, \href
  {http://adsabs.harvard.edu/abs/2016ApJ...824...79A} {824, 79}

\bibitem[\protect\citeauthoryear{{Agertz} et~al.,}{{Agertz}
  et~al.}{2007}]{Agertz_2007}
{Agertz} O.,  et~al., 2007, \mn@doi [\mnras]
  {10.1111/j.1365-2966.2007.12183.x}, \href
  {http://adsabs.harvard.edu/abs/2007MNRAS.380..963A} {380, 963}

\bibitem[\protect\citeauthoryear{{Agertz}, {Teyssier}  \& {Moore}}{{Agertz}
  et~al.}{2011}]{Agertz_2011}
{Agertz} O.,  {Teyssier} R.,   {Moore} B.,  2011, \mn@doi [\mnras]
  {10.1111/j.1365-2966.2010.17530.x}, \href
  {http://adsabs.harvard.edu/abs/2011MNRAS.410.1391A} {410, 1391}

\bibitem[\protect\citeauthoryear{{Andersen}, {Bershady}, {Sparke}, {Gallagher},
  {Wilcots}, {van Driel}  \& {Monnier-Ragaigne}}{{Andersen}
  et~al.}{2006}]{Andersen_2006}
{Andersen} D.~R.,  {Bershady} M.~A.,  {Sparke} L.~S.,  {Gallagher} III J.~S.,
  {Wilcots} E.~M.,  {van Driel} W.,   {Monnier-Ragaigne} D.,  2006, \mn@doi
  [\apjs] {10.1086/506609}, \href
  {http://adsabs.harvard.edu/abs/2006ApJS..166..505A} {166, 505}

\bibitem[\protect\citeauthoryear{{Angl{\'e}s-Alc{\'a}zar},
  {Faucher-Gigu{\`e}re}, {Kere{\v s}}, {Hopkins}, {Quataert}  \&
  {Murray}}{{Angl{\'e}s-Alc{\'a}zar} et~al.}{2016}]{AnglesAlcazar_2016}
{Angl{\'e}s-Alc{\'a}zar} D.,  {Faucher-Gigu{\`e}re} C.-A.,  {Kere{\v s}} D.,
  {Hopkins} P.~F.,  {Quataert} E.,   {Murray} N.,  2016, preprint, \href
  {http://adsabs.harvard.edu/abs/2016arXiv161008523A} {} (\mn@eprint {arXiv}
  {1610.08523})

\bibitem[\protect\citeauthoryear{{Astropy Collaboration} et~al.,}{{Astropy
  Collaboration} et~al.}{2013}]{Astropy_2013}
{Astropy Collaboration} et~al., 2013, \mn@doi [\aap]
  {10.1051/0004-6361/201322068}, \href
  {http://adsabs.harvard.edu/abs/2013A%26A...558A..33A} {558, A33}

\bibitem[\protect\citeauthoryear{{Aumer}, {White}, {Naab}  \&
  {Scannapieco}}{{Aumer} et~al.}{2013}]{Aumer_2013}
{Aumer} M.,  {White} S.~D.~M.,  {Naab} T.,   {Scannapieco} C.,  2013, \mn@doi
  [\mnras] {10.1093/mnras/stt1230}, \href
  {http://adsabs.harvard.edu/abs/2013MNRAS.434.3142A} {434, 3142}

\bibitem[\protect\citeauthoryear{{Begum}, {Chengalur}, {Karachentsev},
  {Sharina}  \& {Kaisin}}{{Begum} et~al.}{2008}]{Begum_2008}
{Begum} A.,  {Chengalur} J.~N.,  {Karachentsev} I.~D.,  {Sharina} M.~E.,
  {Kaisin} S.~S.,  2008, \mn@doi [\mnras] {10.1111/j.1365-2966.2008.13150.x},
  \href {http://adsabs.harvard.edu/abs/2008MNRAS.386.1667B} {386, 1667}

\bibitem[\protect\citeauthoryear{{Binney} \& {Tremaine}}{{Binney} \&
  {Tremaine}}{2008}]{Binney_2008}
{Binney} J.,  {Tremaine} S.,  2008, {Galactic Dynamics: Second Edition}.
Princeton University Press

\bibitem[\protect\citeauthoryear{{Binney}, {Gerhard}  \& {Silk}}{{Binney}
  et~al.}{2001}]{Binney_2001}
{Binney} J.,  {Gerhard} O.,   {Silk} J.,  2001, \mn@doi [\mnras]
  {10.1046/j.1365-8711.2001.04024.x}, \href
  {http://adsabs.harvard.edu/abs/2001MNRAS.321..471B} {321, 471}

\bibitem[\protect\citeauthoryear{{Bordoloi} et~al.,}{{Bordoloi}
  et~al.}{2014}]{Bordoloi_2014}
{Bordoloi} R.,  et~al., 2014, \mn@doi [\apj] {10.1088/0004-637X/796/2/136},
  \href {http://adsabs.harvard.edu/abs/2014ApJ...796..136B} {796, 136}

\bibitem[\protect\citeauthoryear{{Bradford}, {Geha}  \& {van den
  Bosch}}{{Bradford} et~al.}{2016}]{Bradford_2016}
{Bradford} J.~D.,  {Geha} M.~C.,   {van den Bosch} F.~C.,  2016, \mn@doi [\apj]
  {10.3847/0004-637X/832/1/11}, \href
  {http://adsabs.harvard.edu/abs/2016ApJ...832...11B} {832, 11}

\bibitem[\protect\citeauthoryear{{Bressan}, {Marigo}, {Girardi}, {Salasnich},
  {Dal Cero}, {Rubele}  \& {Nanni}}{{Bressan} et~al.}{2012}]{Bressan_2012}
{Bressan} A.,  {Marigo} P.,  {Girardi} L.,  {Salasnich} B.,  {Dal Cero} C.,
  {Rubele} S.,   {Nanni} A.,  2012, \mn@doi [\mnras]
  {10.1111/j.1365-2966.2012.21948.x}, \href
  {http://adsabs.harvard.edu/abs/2012MNRAS.427..127B} {427, 127}

\bibitem[\protect\citeauthoryear{{Brook} et~al.,}{{Brook}
  et~al.}{2011}]{Brook_2011}
{Brook} C.~B.,  et~al., 2011, \mn@doi [\mnras]
  {10.1111/j.1365-2966.2011.18545.x}, \href
  {http://adsabs.harvard.edu/abs/2011MNRAS.415.1051B} {415, 1051}

\bibitem[\protect\citeauthoryear{{Brook}, {Stinson}, {Gibson}, {Ro{\v s}kar},
  {Wadsley}  \& {Quinn}}{{Brook} et~al.}{2012}]{Brook_2012}
{Brook} C.~B.,  {Stinson} G.,  {Gibson} B.~K.,  {Ro{\v s}kar} R.,  {Wadsley}
  J.,   {Quinn} T.,  2012, \mn@doi [\mnras] {10.1111/j.1365-2966.2011.19740.x},
  \href {http://adsabs.harvard.edu/abs/2012MNRAS.419..771B} {419, 771}

\bibitem[\protect\citeauthoryear{{Brooks} et~al.,}{{Brooks}
  et~al.}{2011}]{Brooks_2011}
{Brooks} A.~M.,  et~al., 2011, \mn@doi [\apj] {10.1088/0004-637X/728/1/51},
  \href {http://adsabs.harvard.edu/abs/2011ApJ...728...51B} {728, 51}

\bibitem[\protect\citeauthoryear{{Bullock}, {Kravtsov}  \&
  {Weinberg}}{{Bullock} et~al.}{2000}]{Bullock_2000}
{Bullock} J.~S.,  {Kravtsov} A.~V.,   {Weinberg} D.~H.,  2000, \mn@doi [\apj]
  {10.1086/309279}, \href {http://adsabs.harvard.edu/abs/2000ApJ...539..517B}
  {539, 517}

\bibitem[\protect\citeauthoryear{{Bundy} et~al.,}{{Bundy}
  et~al.}{2015}]{Bundy_2015}
{Bundy} K.,  et~al., 2015, \mn@doi [\apj] {10.1088/0004-637X/798/1/7}, \href
  {http://adsabs.harvard.edu/abs/2015ApJ...798....7B} {798, 7}

\bibitem[\protect\citeauthoryear{{Butler}, {Obreschkow}  \& {Oh}}{{Butler}
  et~al.}{2017}]{Butler_2016}
{Butler} K.~M.,  {Obreschkow} D.,   {Oh} S.-H.,  2017, \mn@doi [\apjl]
  {10.3847/2041-8213/834/1/L4}, \href
  {http://adsabs.harvard.edu/abs/2017ApJ...834L...4B} {834, L4}

\bibitem[\protect\citeauthoryear{{Cappellari}}{{Cappellari}}{2008}]{Cappellari_2008}
{Cappellari} M.,  2008, \mn@doi [\mnras] {10.1111/j.1365-2966.2008.13754.x},
  \href {http://adsabs.harvard.edu/abs/2008MNRAS.390...71C} {390, 71}

\bibitem[\protect\citeauthoryear{{Ceverino}, {Primack}, {Dekel}  \&
  {Kassin}}{{Ceverino} et~al.}{2017}]{Ceverino_2017}
{Ceverino} D.,  {Primack} J.,  {Dekel} A.,   {Kassin} S.~A.,  2017, \mn@doi
  [\mnras] {10.1093/mnras/stx289}, \href
  {http://adsabs.harvard.edu/abs/2017MNRAS.467.2664C} {467, 2664}

\bibitem[\protect\citeauthoryear{{Chan}, {Kere{\v s}}, {O{\~n}orbe}, {Hopkins},
  {Muratov}, {Faucher-Gigu{\`e}re}  \& {Quataert}}{{Chan}
  et~al.}{2015}]{Chan_2015}
{Chan} T.~K.,  {Kere{\v s}} D.,  {O{\~n}orbe} J.,  {Hopkins} P.~F.,  {Muratov}
  A.~L.,  {Faucher-Gigu{\`e}re} C.-A.,   {Quataert} E.,  2015, \mn@doi [\mnras]
  {10.1093/mnras/stv2165}, \href
  {http://adsabs.harvard.edu/abs/2015MNRAS.454.2981C} {454, 2981}

\bibitem[\protect\citeauthoryear{{Chowdhury} \& {Chengalur}}{{Chowdhury} \&
  {Chengalur}}{2017}]{Chowdhury_2017}
{Chowdhury} A.,  {Chengalur} J.~N.,  2017, \mn@doi [\mnras]
  {10.1093/mnras/stx355}, \href
  {http://adsabs.harvard.edu/abs/2017MNRAS.467.3856C} {467, 3856}

\bibitem[\protect\citeauthoryear{{Cortese} et~al.,}{{Cortese}
  et~al.}{2014}]{Cortese_2014}
{Cortese} L.,  et~al., 2014, \mn@doi [\apjl] {10.1088/2041-8205/795/2/L37},
  \href {http://adsabs.harvard.edu/abs/2014ApJ...795L..37C} {795, L37}

\bibitem[\protect\citeauthoryear{{Cortese} et~al.,}{{Cortese}
  et~al.}{2016}]{Cortese_2016}
{Cortese} L.,  et~al., 2016, \mn@doi [\mnras] {10.1093/mnras/stw1891}, \href
  {http://adsabs.harvard.edu/abs/2016MNRAS.463..170C} {463, 170}

\bibitem[\protect\citeauthoryear{{Dalcanton} \& {Stilp}}{{Dalcanton} \&
  {Stilp}}{2010}]{Dalcanton_2010}
{Dalcanton} J.~J.,  {Stilp} A.~M.,  2010, \mn@doi [\apj]
  {10.1088/0004-637X/721/1/547}, \href
  {http://adsabs.harvard.edu/abs/2010ApJ...721..547D} {721, 547}

\bibitem[\protect\citeauthoryear{{Dalcanton}, {Spergel}  \&
  {Summers}}{{Dalcanton} et~al.}{1997}]{Dalcanton_1997}
{Dalcanton} J.~J.,  {Spergel} D.~N.,   {Summers} F.~J.,  1997, \mn@doi [\apj]
  {10.1086/304182}, \href {http://adsabs.harvard.edu/abs/1997ApJ...482..659D}
  {482, 659}

\bibitem[\protect\citeauthoryear{{Danovich}, {Dekel}, {Hahn}, {Ceverino}  \&
  {Primack}}{{Danovich} et~al.}{2015}]{Danovich_2015}
{Danovich} M.,  {Dekel} A.,  {Hahn} O.,  {Ceverino} D.,   {Primack} J.,  2015,
  \mn@doi [\mnras] {10.1093/mnras/stv270}, \href
  {http://adsabs.harvard.edu/abs/2015MNRAS.449.2087D} {449, 2087}

\bibitem[\protect\citeauthoryear{{DeFelippis}, {Genel}, {Bryan}  \&
  {Fall}}{{DeFelippis} et~al.}{2017}]{DeFelippis_2017}
{DeFelippis} D.,  {Genel} S.,  {Bryan} G.~L.,   {Fall} S.~M.,  2017, \mn@doi
  [\apj] {10.3847/1538-4357/aa6dfc}, \href
  {http://adsabs.harvard.edu/abs/2017ApJ...841...16D} {841, 16}

\bibitem[\protect\citeauthoryear{{Di Cintio}, {Brook}, {Dutton}, {Macci{\`o}},
  {Obreja}  \& {Dekel}}{{Di Cintio} et~al.}{2017}]{DiCintio_2017}
{Di Cintio} A.,  {Brook} C.~B.,  {Dutton} A.~A.,  {Macci{\`o}} A.~V.,  {Obreja}
  A.,   {Dekel} A.,  2017, \mn@doi [\mnras] {10.1093/mnrasl/slw210}, \href
  {http://adsabs.harvard.edu/abs/2017MNRAS.466L...1D} {466, L1}

\bibitem[\protect\citeauthoryear{{Di Teodoro} \& {Fraternali}}{{Di Teodoro} \&
  {Fraternali}}{2015}]{DiTeodoro_2015}
{Di Teodoro} E.~M.,  {Fraternali} F.,  2015, \mn@doi [\mnras]
  {10.1093/mnras/stv1213}, \href
  {http://adsabs.harvard.edu/abs/2015MNRAS.451.3021D} {451, 3021}

\bibitem[\protect\citeauthoryear{{Dutton}}{{Dutton}}{2009}]{Dutton_2009}
{Dutton} A.~A.,  2009, \mn@doi [\mnras] {10.1111/j.1365-2966.2009.14741.x},
  \href {http://adsabs.harvard.edu/abs/2009MNRAS.396..121D} {396, 121}

\bibitem[\protect\citeauthoryear{{Dutton} \& {van den Bosch}}{{Dutton} \& {van
  den Bosch}}{2012}]{Dutton_2012}
{Dutton} A.~A.,  {van den Bosch} F.~C.,  2012, \mn@doi [\mnras]
  {10.1111/j.1365-2966.2011.20339.x}, \href
  {http://adsabs.harvard.edu/abs/2012MNRAS.421..608D} {421, 608}

\bibitem[\protect\citeauthoryear{{Dutton} et~al.,}{{Dutton}
  et~al.}{2016}]{Dutton_2016}
{Dutton} A.~A.,  et~al., 2016, \mn@doi [\mnras] {10.1093/mnras/stw1537}, \href
  {http://adsabs.harvard.edu/abs/2016MNRAS.461.2658D} {461, 2658}

\bibitem[\protect\citeauthoryear{{El-Badry}, {Wetzel}, {Geha}, {Hopkins},
  {Kere{\v s}}, {Chan}  \& {Faucher-Gigu{\`e}re}}{{El-Badry}
  et~al.}{2016}]{ElBadry_2016}
{El-Badry} K.,  {Wetzel} A.,  {Geha} M.,  {Hopkins} P.~F.,  {Kere{\v s}} D.,
  {Chan} T.~K.,   {Faucher-Gigu{\`e}re} C.-A.,  2016, \mn@doi [\apj]
  {10.3847/0004-637X/820/2/131}, \href
  {http://adsabs.harvard.edu/abs/2016ApJ...820..131E} {820, 131}

\bibitem[\protect\citeauthoryear{{El-Badry}, {Wetzel}, {Geha}, {Quataert},
  {Hopkins}, {Kere{\v s}}, {Chan}  \& {Faucher-Gigu{\`e}re}}{{El-Badry}
  et~al.}{2017}]{ElBadry_2017}
{El-Badry} K.,  {Wetzel} A.~R.,  {Geha} M.,  {Quataert} E.,  {Hopkins} P.~F.,
  {Kere{\v s}} D.,  {Chan} T.~K.,   {Faucher-Gigu{\`e}re} C.-A.,  2017, \mn@doi
  [\apj] {10.3847/1538-4357/835/2/193}, \href
  {http://adsabs.harvard.edu/abs/2017ApJ...835..193E} {835, 193}

\bibitem[\protect\citeauthoryear{{Emsellem} et~al.,}{{Emsellem}
  et~al.}{2007}]{Emsellem_2007}
{Emsellem} E.,  et~al., 2007, \mn@doi [\mnras]
  {10.1111/j.1365-2966.2007.11752.x}, \href
  {http://adsabs.harvard.edu/abs/2007MNRAS.379..401E} {379, 401}

\bibitem[\protect\citeauthoryear{{Fall}}{{Fall}}{1983}]{Fall_1983}
{Fall} S.~M.,  1983, in {Athanassoula} E.,  ed.,  IAU Symposium Vol. 100,
  Internal Kinematics and Dynamics of Galaxies. pp 391--398

\bibitem[\protect\citeauthoryear{{Fall} \& {Efstathiou}}{{Fall} \&
  {Efstathiou}}{1980}]{Fall_1980}
{Fall} S.~M.,  {Efstathiou} G.,  1980, \mn@doi [\mnras]
  {10.1093/mnras/193.2.189}, \href
  {http://adsabs.harvard.edu/abs/1980MNRAS.193..189F} {193, 189}

\bibitem[\protect\citeauthoryear{{Fall} \& {Romanowsky}}{{Fall} \&
  {Romanowsky}}{2013}]{Fall_2013}
{Fall} S.~M.,  {Romanowsky} A.~J.,  2013, \mn@doi [\apjl]
  {10.1088/2041-8205/769/2/L26}, \href
  {http://adsabs.harvard.edu/abs/2013ApJ...769L..26F} {769, L26}

\bibitem[\protect\citeauthoryear{{Faucher-Giguere}}{{Faucher-Giguere}}{2017}]{FaucherGiguere_2017}
{Faucher-Giguere} C.-A.,  2017, preprint, \href
  {http://adsabs.harvard.edu/abs/2017arXiv170104824F} {} (\mn@eprint {arXiv}
  {1701.04824})

\bibitem[\protect\citeauthoryear{{Faucher-Gigu{\`e}re}, {Lidz}, {Zaldarriaga}
  \& {Hernquist}}{{Faucher-Gigu{\`e}re} et~al.}{2009}]{FaucherGiguere_2009}
{Faucher-Gigu{\`e}re} C.-A.,  {Lidz} A.,  {Zaldarriaga} M.,   {Hernquist} L.,
  2009, \mn@doi [\apj] {10.1088/0004-637X/703/2/1416}, \href
  {http://adsabs.harvard.edu/abs/2009ApJ...703.1416F} {703, 1416}

\bibitem[\protect\citeauthoryear{{Faucher-Gigu{\`e}re}, {Kere{\v s}}  \&
  {Ma}}{{Faucher-Gigu{\`e}re} et~al.}{2011}]{FG_2011}
{Faucher-Gigu{\`e}re} C.-A.,  {Kere{\v s}} D.,   {Ma} C.-P.,  2011, \mn@doi
  [\mnras] {10.1111/j.1365-2966.2011.19457.x}, \href
  {http://adsabs.harvard.edu/abs/2011MNRAS.417.2982F} {417, 2982}

\bibitem[\protect\citeauthoryear{{Faucher-Gigu{\`e}re}, {Hopkins}, {Kere{\v
  s}}, {Muratov}, {Quataert}  \& {Murray}}{{Faucher-Gigu{\`e}re}
  et~al.}{2015}]{FaucherGiguere_2015}
{Faucher-Gigu{\`e}re} C.-A.,  {Hopkins} P.~F.,  {Kere{\v s}} D.,  {Muratov}
  A.~L.,  {Quataert} E.,   {Murray} N.,  2015, \mn@doi [\mnras]
  {10.1093/mnras/stv336}, \href
  {http://adsabs.harvard.edu/abs/2015MNRAS.449..987F} {449, 987}

\bibitem[\protect\citeauthoryear{{Faucher-Gigu{\`e}re}, {Feldmann}, {Quataert},
  {Kere{\v s}}, {Hopkins}  \& {Murray}}{{Faucher-Gigu{\`e}re}
  et~al.}{2016}]{FaucherGiguere_2016}
{Faucher-Gigu{\`e}re} C.-A.,  {Feldmann} R.,  {Quataert} E.,  {Kere{\v s}} D.,
  {Hopkins} P.~F.,   {Murray} N.,  2016, \mn@doi [\mnras]
  {10.1093/mnrasl/slw091}, \href
  {http://adsabs.harvard.edu/abs/2016MNRAS.461L..32F} {461, L32}

\bibitem[\protect\citeauthoryear{{Feldmann}, {Quataert}, {Hopkins},
  {Faucher-Gigu{\`e}re}  \& {Kere{\v s}}}{{Feldmann}
  et~al.}{2017}]{Feldmann_2016}
{Feldmann} R.,  {Quataert} E.,  {Hopkins} P.~F.,  {Faucher-Gigu{\`e}re} C.-A.,
   {Kere{\v s}} D.,  2017, \mn@doi [\mnras] {10.1093/mnras/stx1120}, \href
  {http://adsabs.harvard.edu/abs/2017MNRAS.470.1050F} {470, 1050}

\bibitem[\protect\citeauthoryear{{Ferland} et~al.,}{{Ferland}
  et~al.}{2013}]{Ferland_2013}
{Ferland} G.~J.,  et~al., 2013, \rmxaa, \href
  {http://adsabs.harvard.edu/abs/2013RMxAA..49..137F} {49, 137}

\bibitem[\protect\citeauthoryear{{Few}, {Dobbs}, {Pettitt}  \&
  {Konstandin}}{{Few} et~al.}{2016}]{Few_2016}
{Few} C.~G.,  {Dobbs} C.,  {Pettitt} A.,   {Konstandin} L.,  2016, \mn@doi
  [\mnras] {10.1093/mnras/stw1226}, \href
  {http://adsabs.harvard.edu/abs/2016MNRAS.460.4382F} {460, 4382}

\bibitem[\protect\citeauthoryear{{Fielding}, {Quataert}, {McCourt}  \&
  {Thompson}}{{Fielding} et~al.}{2017}]{Fielding_2017}
{Fielding} D.,  {Quataert} E.,  {McCourt} M.,   {Thompson} T.~A.,  2017,
  \mn@doi [\mnras] {10.1093/mnras/stw3326}, \href
  {http://cdsads.u-strasbg.fr/abs/2017MNRAS.466.3810F} {466, 3810}

\bibitem[\protect\citeauthoryear{{Fitts} et~al.,}{{Fitts}
  et~al.}{2016}]{Fitts_2016}
{Fitts} A.,  et~al., 2016, preprint, \href
  {http://adsabs.harvard.edu/abs/2016arXiv161102281F} {} (\mn@eprint {arXiv}
  {1611.02281})

\bibitem[\protect\citeauthoryear{{Flores} \& {Primack}}{{Flores} \&
  {Primack}}{1994}]{Flores_1994}
{Flores} R.~A.,  {Primack} J.~R.,  1994, \mn@doi [\apjl] {10.1086/187350},
  \href {http://adsabs.harvard.edu/abs/1994ApJ...427L...1F} {427, L1}

\bibitem[\protect\citeauthoryear{{Genel}, {Fall}, {Hernquist}, {Vogelsberger},
  {Snyder}, {Rodriguez-Gomez}, {Sijacki}  \& {Springel}}{{Genel}
  et~al.}{2015}]{Genel_2015}
{Genel} S.,  {Fall} S.~M.,  {Hernquist} L.,  {Vogelsberger} M.,  {Snyder}
  G.~F.,  {Rodriguez-Gomez} V.,  {Sijacki} D.,   {Springel} V.,  2015, \mn@doi
  [\apjl] {10.1088/2041-8205/804/2/L40}, \href
  {http://adsabs.harvard.edu/abs/2015ApJ...804L..40G} {804, L40}

\bibitem[\protect\citeauthoryear{{Genzel} et~al.,}{{Genzel}
  et~al.}{2017}]{Genzel_2017}
{Genzel} R.,  et~al., 2017, \mn@doi [\nat] {10.1038/nature21685}, \href
  {http://adsabs.harvard.edu/abs/2017Natur.543..397G} {543, 397}

\bibitem[\protect\citeauthoryear{{Glazebrook}}{{Glazebrook}}{2013}]{Glazebrook_2013}
{Glazebrook} K.,  2013, \mn@doi [\pasa] {10.1017/pasa.2013.34}, \href
  {http://adsabs.harvard.edu/abs/2013PASA...30...56G} {30, e056}

\bibitem[\protect\citeauthoryear{{Gonz{\'a}lez-Garc{\'{\i}}a} \& {van
  Albada}}{{Gonz{\'a}lez-Garc{\'{\i}}a} \& {van
  Albada}}{2005}]{GonzalezGarcia_2005}
{Gonz{\'a}lez-Garc{\'{\i}}a} A.~C.,  {van Albada} T.~S.,  2005, \mn@doi
  [\mnras] {10.1111/j.1365-2966.2005.09242.x}, \href
  {http://adsabs.harvard.edu/abs/2005MNRAS.361.1030G} {361, 1030}

\bibitem[\protect\citeauthoryear{{Gonz{\'a}lez-Samaniego}, {Col{\'{\i}}n},
  {Avila-Reese}, {Rodr{\'{\i}}guez-Puebla}  \&
  {Valenzuela}}{{Gonz{\'a}lez-Samaniego} et~al.}{2014}]{Gonzalez_2014}
{Gonz{\'a}lez-Samaniego} A.,  {Col{\'{\i}}n} P.,  {Avila-Reese} V.,
  {Rodr{\'{\i}}guez-Puebla} A.,   {Valenzuela} O.,  2014, \mn@doi [\apj]
  {10.1088/0004-637X/785/1/58}, \href
  {http://adsabs.harvard.edu/abs/2014ApJ...785...58G} {785, 58}

\bibitem[\protect\citeauthoryear{{Gonz{\'a}lez-Samaniego}, {Avila-Reese}  \&
  {Col{\'{\i}}n}}{{Gonz{\'a}lez-Samaniego} et~al.}{2016}]{Gozalez_2016}
{Gonz{\'a}lez-Samaniego} A.,  {Avila-Reese} V.,   {Col{\'{\i}}n} P.,  2016,
  \mn@doi [\apj] {10.3847/0004-637X/819/2/101}, \href
  {http://adsabs.harvard.edu/abs/2016ApJ...819..101G} {819, 101}

\bibitem[\protect\citeauthoryear{{Governato} et~al.,}{{Governato}
  et~al.}{2004}]{Governato_2004}
{Governato} F.,  et~al., 2004, \mn@doi [\apj] {10.1086/383516}, \href
  {http://adsabs.harvard.edu/abs/2004ApJ...607..688G} {607, 688}

\bibitem[\protect\citeauthoryear{{Governato}, {Willman}, {Mayer}, {Brooks},
  {Stinson}, {Valenzuela}, {Wadsley}  \& {Quinn}}{{Governato}
  et~al.}{2007}]{Governato_2007}
{Governato} F.,  {Willman} B.,  {Mayer} L.,  {Brooks} A.,  {Stinson} G.,
  {Valenzuela} O.,  {Wadsley} J.,   {Quinn} T.,  2007, \mn@doi [\mnras]
  {10.1111/j.1365-2966.2006.11266.x}, \href
  {http://adsabs.harvard.edu/abs/2007MNRAS.374.1479G} {374, 1479}

\bibitem[\protect\citeauthoryear{{Governato} et~al.,}{{Governato}
  et~al.}{2010}]{Governato_2010}
{Governato} F.,  et~al., 2010, \mn@doi [\nat] {10.1038/nature08640}, \href
  {http://adsabs.harvard.edu/abs/2010Natur.463..203G} {463, 203}

\bibitem[\protect\citeauthoryear{{Grand} et~al.,}{{Grand}
  et~al.}{2017}]{Grand_2017}
{Grand} R.~J.~J.,  et~al., 2017, \mn@doi [\mnras] {10.1093/mnras/stx071}, \href
  {http://adsabs.harvard.edu/abs/2017MNRAS.467..179G} {467, 179}

\bibitem[\protect\citeauthoryear{{Guedes}, {Callegari}, {Madau}  \&
  {Mayer}}{{Guedes} et~al.}{2011}]{Guedes_2011}
{Guedes} J.,  {Callegari} S.,  {Madau} P.,   {Mayer} L.,  2011, \mn@doi [\apj]
  {10.1088/0004-637X/742/2/76}, \href
  {http://adsabs.harvard.edu/abs/2011ApJ...742...76G} {742, 76}

\bibitem[\protect\citeauthoryear{{Hahn} \& {Abel}}{{Hahn} \&
  {Abel}}{2011}]{Hahn_2011}
{Hahn} O.,  {Abel} T.,  2011, \mn@doi [\mnras]
  {10.1111/j.1365-2966.2011.18820.x}, \href
  {http://adsabs.harvard.edu/abs/2011MNRAS.415.2101H} {415, 2101}

\bibitem[\protect\citeauthoryear{{Hayward} \& {Hopkins}}{{Hayward} \&
  {Hopkins}}{2017}]{Hayward_2017}
{Hayward} C.~C.,  {Hopkins} P.~F.,  2017, \mn@doi [\mnras]
  {10.1093/mnras/stw2888}, \href
  {http://adsabs.harvard.edu/abs/2017MNRAS.465.1682H} {465, 1682}

\bibitem[\protect\citeauthoryear{{Heavens} \& {Peacock}}{{Heavens} \&
  {Peacock}}{1988}]{Heavens_1988}
{Heavens} A.,  {Peacock} J.,  1988, \mn@doi [\mnras] {10.1093/mnras/232.2.339},
  \href {http://adsabs.harvard.edu/abs/1988MNRAS.232..339H} {232, 339}

\bibitem[\protect\citeauthoryear{{Hernandez} \& {Cervantes-Sodi}}{{Hernandez}
  \& {Cervantes-Sodi}}{2006}]{Hernandez_2006}
{Hernandez} X.,  {Cervantes-Sodi} B.,  2006, \mn@doi [\mnras]
  {10.1111/j.1365-2966.2006.10115.x}, \href
  {http://adsabs.harvard.edu/abs/2006MNRAS.368..351H} {368, 351}

\bibitem[\protect\citeauthoryear{{Ho} et~al.,}{{Ho} et~al.}{2014}]{Ho_2014}
{Ho} I.-T.,  et~al., 2014, \mn@doi [\mnras] {10.1093/mnras/stu1653}, \href
  {http://adsabs.harvard.edu/abs/2014MNRAS.444.3894H} {444, 3894}

\bibitem[\protect\citeauthoryear{{Hoeft}, {Yepes}, {Gottl{\"o}ber}  \&
  {Springel}}{{Hoeft} et~al.}{2006}]{Hoeft_2006}
{Hoeft} M.,  {Yepes} G.,  {Gottl{\"o}ber} S.,   {Springel} V.,  2006, \mn@doi
  [\mnras] {10.1111/j.1365-2966.2006.10678.x}, \href
  {http://adsabs.harvard.edu/abs/2006MNRAS.371..401H} {371, 401}

\bibitem[\protect\citeauthoryear{{Hopkins}}{{Hopkins}}{2015}]{Hopkins_2015}
{Hopkins} P.~F.,  2015, \mn@doi [\mnras] {10.1093/mnras/stv195}, \href
  {http://adsabs.harvard.edu/abs/2015MNRAS.450...53H} {450, 53}

\bibitem[\protect\citeauthoryear{{Hopkins}, {Quataert}  \& {Murray}}{{Hopkins}
  et~al.}{2012}]{Hopkins_2012}
{Hopkins} P.~F.,  {Quataert} E.,   {Murray} N.,  2012, \mn@doi [\mnras]
  {10.1111/j.1365-2966.2012.20578.x}, \href
  {http://adsabs.harvard.edu/abs/2012MNRAS.421.3488H} {421, 3488}

\bibitem[\protect\citeauthoryear{{Hopkins}, {Narayanan}  \& {Murray}}{{Hopkins}
  et~al.}{2013}]{Hopkins_2013}
{Hopkins} P.~F.,  {Narayanan} D.,   {Murray} N.,  2013, \mn@doi [\mnras]
  {10.1093/mnras/stt723}, \href
  {http://adsabs.harvard.edu/abs/2013MNRAS.432.2647H} {432, 2647}

\bibitem[\protect\citeauthoryear{{Hopkins}, {Kere{\v s}}, {O{\~n}orbe},
  {Faucher-Gigu{\`e}re}, {Quataert}, {Murray}  \& {Bullock}}{{Hopkins}
  et~al.}{2014}]{Hopkins_2014}
{Hopkins} P.~F.,  {Kere{\v s}} D.,  {O{\~n}orbe} J.,  {Faucher-Gigu{\`e}re}
  C.-A.,  {Quataert} E.,  {Murray} N.,   {Bullock} J.~S.,  2014, \mn@doi
  [\mnras] {10.1093/mnras/stu1738}, \href
  {http://adsabs.harvard.edu/abs/2014MNRAS.445..581H} {445, 581}

\bibitem[\protect\citeauthoryear{{Hopkins} et~al.,}{{Hopkins}
  et~al.}{2017}]{Hopkins_2017}
{Hopkins} P.~F.,  et~al., 2017, preprint, \href
  {http://adsabs.harvard.edu/abs/2017arXiv170206148H} {} (\mn@eprint {arXiv}
  {1702.06148})

\bibitem[\protect\citeauthoryear{{Hunter} et~al.,}{{Hunter}
  et~al.}{2012}]{Hunter_2012}
{Hunter} D.~A.,  et~al., 2012, \mn@doi [\aj] {10.1088/0004-6256/144/5/134},
  \href {http://adsabs.harvard.edu/abs/2012AJ....144..134H} {144, 134}

\bibitem[\protect\citeauthoryear{{Iorio}, {Fraternali}, {Nipoti}, {Di Teodoro},
  {Read}  \& {Battaglia}}{{Iorio} et~al.}{2017}]{Iorio_2017}
{Iorio} G.,  {Fraternali} F.,  {Nipoti} C.,  {Di Teodoro} E.,  {Read} J.~I.,
  {Battaglia} G.,  2017, \mn@doi [\mnras] {10.1093/mnras/stw3285}, \href
  {http://adsabs.harvard.edu/abs/2017MNRAS.466.4159I} {466, 4159}

\bibitem[\protect\citeauthoryear{{Johnson}, {Hunter}, {Kamphuis}  \&
  {Wang}}{{Johnson} et~al.}{2017}]{Johnson_2017}
{Johnson} M.~C.,  {Hunter} D.~A.,  {Kamphuis} P.,   {Wang} J.,  2017, \mn@doi
  [\mnras] {10.1093/mnrasl/slw203}, \href
  {http://adsabs.harvard.edu/abs/2017MNRAS.465L..49J} {465, L49}

\bibitem[\protect\citeauthoryear{{Kannan}, {Macci{\`o}}, {Fontanot}, {Moster},
  {Karman}  \& {Somerville}}{{Kannan} et~al.}{2015}]{Kannan_2015}
{Kannan} R.,  {Macci{\`o}} A.~V.,  {Fontanot} F.,  {Moster} B.~P.,  {Karman}
  W.,   {Somerville} R.~S.,  2015, \mn@doi [\mnras] {10.1093/mnras/stv1633},
  \href {http://adsabs.harvard.edu/abs/2015MNRAS.452.4347K} {452, 4347}

\bibitem[\protect\citeauthoryear{{Kassin} et~al.,}{{Kassin}
  et~al.}{2012}]{Kassin_2012}
{Kassin} S.~A.,  et~al., 2012, \mn@doi [\apj] {10.1088/0004-637X/758/2/106},
  \href {http://adsabs.harvard.edu/abs/2012ApJ...758..106K} {758, 106}

\bibitem[\protect\citeauthoryear{{Kassin}, {Brooks}, {Governato}, {Weiner}  \&
  {Gardner}}{{Kassin} et~al.}{2014}]{Kassin_2014}
{Kassin} S.~A.,  {Brooks} A.,  {Governato} F.,  {Weiner} B.~J.,   {Gardner}
  J.~P.,  2014, \mn@doi [\apj] {10.1088/0004-637X/790/2/89}, \href
  {http://adsabs.harvard.edu/abs/2014ApJ...790...89K} {790, 89}

\bibitem[\protect\citeauthoryear{{Katz} \& {Gunn}}{{Katz} \&
  {Gunn}}{1991}]{Katz_1991}
{Katz} N.,  {Gunn} J.~E.,  1991, \mn@doi [\apj] {10.1086/170367}, \href
  {http://adsabs.harvard.edu/abs/1991ApJ...377..365K} {377, 365}

\bibitem[\protect\citeauthoryear{{Kaufmann}, {Mayer}, {Wadsley}, {Stadel}  \&
  {Moore}}{{Kaufmann} et~al.}{2007a}]{Kaufmann_2007}
{Kaufmann} T.,  {Mayer} L.,  {Wadsley} J.,  {Stadel} J.,   {Moore} B.,  2007a,
  \mn@doi [\mnras] {10.1111/j.1365-2966.2006.11314.x}, \href
  {http://adsabs.harvard.edu/abs/2007MNRAS.375...53K} {375, 53}

\bibitem[\protect\citeauthoryear{{Kaufmann}, {Wheeler}  \&
  {Bullock}}{{Kaufmann} et~al.}{2007b}]{Kaufmann_2007_disk}
{Kaufmann} T.,  {Wheeler} C.,   {Bullock} J.~S.,  2007b, \mn@doi [\mnras]
  {10.1111/j.1365-2966.2007.12436.x}, \href
  {http://adsabs.harvard.edu/abs/2007MNRAS.382.1187K} {382, 1187}

\bibitem[\protect\citeauthoryear{{Kere{\v s}}, {Katz}, {Weinberg}  \&
  {Dav{\'e}}}{{Kere{\v s}} et~al.}{2005}]{Keres_2005}
{Kere{\v s}} D.,  {Katz} N.,  {Weinberg} D.~H.,   {Dav{\'e}} R.,  2005, \mn@doi
  [\mnras] {10.1111/j.1365-2966.2005.09451.x}, \href
  {http://adsabs.harvard.edu/abs/2005MNRAS.363....2K} {363, 2}

\bibitem[\protect\citeauthoryear{{Kere{\v s}}, {Vogelsberger}, {Sijacki},
  {Springel}  \& {Hernquist}}{{Kere{\v s}} et~al.}{2012}]{Keres_2012}
{Kere{\v s}} D.,  {Vogelsberger} M.,  {Sijacki} D.,  {Springel} V.,
  {Hernquist} L.,  2012, \mn@doi [\mnras] {10.1111/j.1365-2966.2012.21548.x},
  \href {http://adsabs.harvard.edu/abs/2012MNRAS.425.2027K} {425, 2027}

\bibitem[\protect\citeauthoryear{{Knebe} et~al.,}{{Knebe}
  et~al.}{2013}]{Knebe_2013}
{Knebe} A.,  et~al., 2013, \mn@doi [\mnras] {10.1093/mnras/sts173}, \href
  {http://adsabs.harvard.edu/abs/2013MNRAS.428.2039K} {428, 2039}

\bibitem[\protect\citeauthoryear{{Kormendy}}{{Kormendy}}{2016}]{Kormendy_2016}
{Kormendy} J.,  2016, \mn@doi [Galactic Bulges] {10.1007/978-3-319-19378-6_16},
  \href {http://adsabs.harvard.edu/abs/2016ASSL..418..431K} {418, 431}

\bibitem[\protect\citeauthoryear{{Kormendy}, {Drory}, {Bender}  \&
  {Cornell}}{{Kormendy} et~al.}{2010}]{Kormendy_2010}
{Kormendy} J.,  {Drory} N.,  {Bender} R.,   {Cornell} M.~E.,  2010, \mn@doi
  [\apj] {10.1088/0004-637X/723/1/54}, \href
  {http://adsabs.harvard.edu/abs/2010ApJ...723...54K} {723, 54}

\bibitem[\protect\citeauthoryear{{Kroupa}}{{Kroupa}}{2001}]{Kroupa_2001}
{Kroupa} P.,  2001, \mn@doi [\mnras] {10.1046/j.1365-8711.2001.04022.x}, \href
  {http://adsabs.harvard.edu/abs/2001MNRAS.322..231K} {322, 231}

\bibitem[\protect\citeauthoryear{{Krumholz} \& {Gnedin}}{{Krumholz} \&
  {Gnedin}}{2011}]{Krumholz_2011}
{Krumholz} M.~R.,  {Gnedin} N.~Y.,  2011, \mn@doi [\apj]
  {10.1088/0004-637X/729/1/36}, \href
  {http://adsabs.harvard.edu/abs/2011ApJ...729...36K} {729, 36}

\bibitem[\protect\citeauthoryear{{Kuzio de Naray} \& {Kaufmann}}{{Kuzio de
  Naray} \& {Kaufmann}}{2011}]{KuzioDeNaray_2011}
{Kuzio de Naray} R.,  {Kaufmann} T.,  2011, \mn@doi [\mnras]
  {10.1111/j.1365-2966.2011.18656.x}, \href
  {http://adsabs.harvard.edu/abs/2011MNRAS.414.3617K} {414, 3617}

\bibitem[\protect\citeauthoryear{{Lagos} et~al.,}{{Lagos}
  et~al.}{2017a}]{Lagos_2017b}
{Lagos} C.~d.~P.,  et~al., 2017a, preprint, \href
  {http://adsabs.harvard.edu/abs/2017arXiv170104407L} {} (\mn@eprint {arXiv}
  {1701.04407})

\bibitem[\protect\citeauthoryear{{Lagos}, {Theuns}, {Stevens}, {Cortese},
  {Padilla}, {Davis}, {Contreras}  \& {Croton}}{{Lagos}
  et~al.}{2017b}]{Lagos_2017}
{Lagos} C.~d.~P.,  {Theuns} T.,  {Stevens} A.~R.~H.,  {Cortese} L.,  {Padilla}
  N.~D.,  {Davis} T.~A.,  {Contreras} S.,   {Croton} D.,  2017b, \mn@doi
  [\mnras] {10.1093/mnras/stw2610}, \href
  {http://adsabs.harvard.edu/abs/2017MNRAS.464.3850L} {464, 3850}

\bibitem[\protect\citeauthoryear{{Leitherer} et~al.,}{{Leitherer}
  et~al.}{1999}]{Leitherer_1999}
{Leitherer} C.,  et~al., 1999, \mn@doi [\apjs] {10.1086/313233}, \href
  {http://adsabs.harvard.edu/abs/1999ApJS..123....3L} {123, 3}

\bibitem[\protect\citeauthoryear{{Leitherer}, {Ortiz Ot{\'a}lvaro}, {Bresolin},
  {Kudritzki}, {Lo Faro}, {Pauldrach}, {Pettini}  \& {Rix}}{{Leitherer}
  et~al.}{2010}]{Leitherer_2010}
{Leitherer} C.,  {Ortiz Ot{\'a}lvaro} P.~A.,  {Bresolin} F.,  {Kudritzki}
  R.-P.,  {Lo Faro} B.,  {Pauldrach} A.~W.~A.,  {Pettini} M.,   {Rix} S.~A.,
  2010, \mn@doi [\apjs] {10.1088/0067-0049/189/2/309}, \href
  {http://adsabs.harvard.edu/abs/2010ApJS..189..309L} {189, 309}

\bibitem[\protect\citeauthoryear{{Leitherer}, {Ekstr{\"o}m}, {Meynet},
  {Schaerer}, {Agienko}  \& {Levesque}}{{Leitherer}
  et~al.}{2014}]{Leitherer_2014}
{Leitherer} C.,  {Ekstr{\"o}m} S.,  {Meynet} G.,  {Schaerer} D.,  {Agienko}
  K.~B.,   {Levesque} E.~M.,  2014, \mn@doi [\apjs]
  {10.1088/0067-0049/212/1/14}, \href
  {http://adsabs.harvard.edu/abs/2014ApJS..212...14L} {212, 14}

\bibitem[\protect\citeauthoryear{{Leroy}, {Walter}, {Brinks}, {Bigiel}, {de
  Blok}, {Madore}  \& {Thornley}}{{Leroy} et~al.}{2008}]{Leroy_2008}
{Leroy} A.~K.,  {Walter} F.,  {Brinks} E.,  {Bigiel} F.,  {de Blok} W.~J.~G.,
  {Madore} B.,   {Thornley} M.~D.,  2008, \mn@doi [\aj]
  {10.1088/0004-6256/136/6/2782}, \href
  {http://adsabs.harvard.edu/abs/2008AJ....136.2782L} {136, 2782}

\bibitem[\protect\citeauthoryear{{Ma}, {Hopkins}, {Faucher-Gigu{\`e}re},
  {Zolman}, {Muratov}, {Kere{\v s}}  \& {Quataert}}{{Ma}
  et~al.}{2016}]{Ma_2016}
{Ma} X.,  {Hopkins} P.~F.,  {Faucher-Gigu{\`e}re} C.-A.,  {Zolman} N.,
  {Muratov} A.~L.,  {Kere{\v s}} D.,   {Quataert} E.,  2016, \mn@doi [\mnras]
  {10.1093/mnras/stv2659}, \href
  {http://adsabs.harvard.edu/abs/2016MNRAS.456.2140M} {456, 2140}

\bibitem[\protect\citeauthoryear{{Ma}, {Hopkins}, {Wetzel}, {Kirby},
  {Angl{\'e}s-Alc{\'a}zar}, {Faucher-Gigu{\`e}re}, {Kere{\v s}}  \&
  {Quataert}}{{Ma} et~al.}{2017}]{Ma_2017}
{Ma} X.,  {Hopkins} P.~F.,  {Wetzel} A.~R.,  {Kirby} E.~N.,
  {Angl{\'e}s-Alc{\'a}zar} D.,  {Faucher-Gigu{\`e}re} C.-A.,  {Kere{\v s}} D.,
   {Quataert} E.,  2017, \mn@doi [\mnras] {10.1093/mnras/stx273}, \href
  {http://adsabs.harvard.edu/abs/2017MNRAS.467.2430M} {467, 2430}

\bibitem[\protect\citeauthoryear{{Martig}, {Bournaud}, {Croton}, {Dekel}  \&
  {Teyssier}}{{Martig} et~al.}{2012}]{Martig_2012}
{Martig} M.,  {Bournaud} F.,  {Croton} D.~J.,  {Dekel} A.,   {Teyssier} R.,
  2012, \mn@doi [\apj] {10.1088/0004-637X/756/1/26}, \href
  {http://adsabs.harvard.edu/abs/2012ApJ...756...26M} {756, 26}

\bibitem[\protect\citeauthoryear{{Martinez-Medina}, {Pichardo},
  {P{\'e}rez-Villegas}  \& {Moreno}}{{Martinez-Medina}
  et~al.}{2015}]{MartinezMedina_2015}
{Martinez-Medina} L.~A.,  {Pichardo} B.,  {P{\'e}rez-Villegas} A.,   {Moreno}
  E.,  2015, \mn@doi [\apj] {10.1088/0004-637X/802/2/109}, \href
  {http://adsabs.harvard.edu/abs/2015ApJ...802..109M} {802, 109}

\bibitem[\protect\citeauthoryear{{McGaugh}, {Schombert}, {Bothun}  \& {de
  Blok}}{{McGaugh} et~al.}{2000}]{McGaugh_2000}
{McGaugh} S.~S.,  {Schombert} J.~M.,  {Bothun} G.~D.,   {de Blok} W.~J.~G.,
  2000, \mn@doi [\apjl] {10.1086/312628}, \href
  {http://adsabs.harvard.edu/abs/2000ApJ...533L..99M} {533, L99}

\bibitem[\protect\citeauthoryear{{Meurer}, {Carignan}, {Beaulieu}  \&
  {Freeman}}{{Meurer} et~al.}{1996}]{Meurer_1996}
{Meurer} G.~R.,  {Carignan} C.,  {Beaulieu} S.~F.,   {Freeman} K.~C.,  1996,
  \mn@doi [\aj] {10.1086/117895}, \href
  {http://adsabs.harvard.edu/abs/1996AJ....111.1551M} {111, 1551}

\bibitem[\protect\citeauthoryear{{Mo}, {Mao}  \& {White}}{{Mo}
  et~al.}{1998}]{Mo_1998}
{Mo} H.~J.,  {Mao} S.,   {White} S.~D.~M.,  1998, \mn@doi [\mnras]
  {10.1046/j.1365-8711.1998.01227.x}, \href
  {http://adsabs.harvard.edu/abs/1998MNRAS.295..319M} {295, 319}

\bibitem[\protect\citeauthoryear{{Moiseev}}{{Moiseev}}{2014}]{Moiseev_2014}
{Moiseev} A.~V.,  2014, \mn@doi [Astrophysical Bulletin]
  {10.1134/S1990341314010015}, \href
  {http://adsabs.harvard.edu/abs/2014AstBu..69....1M} {69, 1}

\bibitem[\protect\citeauthoryear{{Moore}}{{Moore}}{1994}]{Moore_1994}
{Moore} B.,  1994, \mn@doi [\nat] {10.1038/370629a0}, \href
  {http://adsabs.harvard.edu/abs/1994Natur.370..629M} {370, 629}

\bibitem[\protect\citeauthoryear{{Muratov}, {Kere{\v s}},
  {Faucher-Gigu{\`e}re}, {Hopkins}, {Quataert}  \& {Murray}}{{Muratov}
  et~al.}{2015}]{Muratov_2015}
{Muratov} A.~L.,  {Kere{\v s}} D.,  {Faucher-Gigu{\`e}re} C.-A.,  {Hopkins}
  P.~F.,  {Quataert} E.,   {Murray} N.,  2015, \mn@doi [\mnras]
  {10.1093/mnras/stv2126}, \href
  {http://adsabs.harvard.edu/abs/2015MNRAS.454.2691M} {454, 2691}

\bibitem[\protect\citeauthoryear{{Muratov} et~al.,}{{Muratov}
  et~al.}{2017}]{Muratov_2016}
{Muratov} A.~L.,  et~al., 2017, \mn@doi [\mnras] {10.1093/mnras/stx667}, \href
  {http://adsabs.harvard.edu/abs/2017MNRAS.468.4170M} {468, 4170}

\bibitem[\protect\citeauthoryear{{Naab} et~al.,}{{Naab}
  et~al.}{2014}]{Naab_2014}
{Naab} T.,  et~al., 2014, \mn@doi [\mnras] {10.1093/mnras/stt1919}, \href
  {http://adsabs.harvard.edu/abs/2014MNRAS.444.3357N} {444, 3357}

\bibitem[\protect\citeauthoryear{{Navarro} \& {Steinmetz}}{{Navarro} \&
  {Steinmetz}}{1997}]{Navarro_1997}
{Navarro} J.~F.,  {Steinmetz} M.,  1997, \mn@doi [\apj] {10.1086/303763}, \href
  {http://adsabs.harvard.edu/abs/1997ApJ...478...13N} {478, 13}

\bibitem[\protect\citeauthoryear{{Noh} \& {McQuinn}}{{Noh} \&
  {McQuinn}}{2014}]{Noh_2014}
{Noh} Y.,  {McQuinn} M.,  2014, \mn@doi [\mnras] {10.1093/mnras/stu1412}, \href
  {http://adsabs.harvard.edu/abs/2014MNRAS.444..503N} {444, 503}

\bibitem[\protect\citeauthoryear{{Nordstr{\"o}m} et~al.,}{{Nordstr{\"o}m}
  et~al.}{2004}]{Nordstrom_2004}
{Nordstr{\"o}m} B.,  et~al., 2004, \mn@doi [\aap] {10.1051/0004-6361:20035959},
  \href {http://adsabs.harvard.edu/abs/2004A%26A...418..989N} {418, 989}

\bibitem[\protect\citeauthoryear{{O{\~n}orbe}, {Garrison-Kimmel}, {Maller},
  {Bullock}, {Rocha}  \& {Hahn}}{{O{\~n}orbe} et~al.}{2014}]{Onorbe_2014}
{O{\~n}orbe} J.,  {Garrison-Kimmel} S.,  {Maller} A.~H.,  {Bullock} J.~S.,
  {Rocha} M.,   {Hahn} O.,  2014, \mn@doi [\mnras] {10.1093/mnras/stt2020},
  \href {http://adsabs.harvard.edu/abs/2014MNRAS.437.1894O} {437, 1894}

\bibitem[\protect\citeauthoryear{{O{\~n}orbe}, {Boylan-Kolchin}, {Bullock},
  {Hopkins}, {Kere{\v s}}, {Faucher-Gigu{\`e}re}, {Quataert}  \&
  {Murray}}{{O{\~n}orbe} et~al.}{2015}]{Onorbe_2015}
{O{\~n}orbe} J.,  {Boylan-Kolchin} M.,  {Bullock} J.~S.,  {Hopkins} P.~F.,
  {Kere{\v s}} D.,  {Faucher-Gigu{\`e}re} C.-A.,  {Quataert} E.,   {Murray} N.,
   2015, \mn@doi [\mnras] {10.1093/mnras/stv2072}, \href
  {http://adsabs.harvard.edu/abs/2015MNRAS.454.2092O} {454, 2092}

\bibitem[\protect\citeauthoryear{{Obreja}, {Stinson}, {Dutton}, {Macci{\`o}},
  {Wang}  \& {Kang}}{{Obreja} et~al.}{2016}]{Obreja_2016}
{Obreja} A.,  {Stinson} G.~S.,  {Dutton} A.~A.,  {Macci{\`o}} A.~V.,  {Wang}
  L.,   {Kang} X.,  2016, \mn@doi [\mnras] {10.1093/mnras/stw690}, \href
  {http://adsabs.harvard.edu/abs/2016MNRAS.459..467O} {459, 467}

\bibitem[\protect\citeauthoryear{{Obreschkow} \& {Glazebrook}}{{Obreschkow} \&
  {Glazebrook}}{2014}]{Obreschkow_2014}
{Obreschkow} D.,  {Glazebrook} K.,  2014, \mn@doi [\apj]
  {10.1088/0004-637X/784/1/26}, \href
  {http://adsabs.harvard.edu/abs/2014ApJ...784...26O} {784, 26}

\bibitem[\protect\citeauthoryear{{Oh}, {Brook}, {Governato}, {Brinks}, {Mayer},
  {de Blok}, {Brooks}  \& {Walter}}{{Oh} et~al.}{2011}]{Oh_2011}
{Oh} S.-H.,  {Brook} C.,  {Governato} F.,  {Brinks} E.,  {Mayer} L.,  {de Blok}
  W.~J.~G.,  {Brooks} A.,   {Walter} F.,  2011, \mn@doi [\aj]
  {10.1088/0004-6256/142/1/24}, \href
  {http://adsabs.harvard.edu/abs/2011AJ....142...24O} {142, 24}

\bibitem[\protect\citeauthoryear{{Oh} et~al.,}{{Oh} et~al.}{2015}]{Oh_2015}
{Oh} S.-H.,  et~al., 2015, \mn@doi [\aj] {10.1088/0004-6256/149/6/180}, \href
  {http://adsabs.harvard.edu/abs/2015AJ....149..180O} {149, 180}

\bibitem[\protect\citeauthoryear{{Okamoto}, {Jenkins}, {Eke}, {Quilis}  \&
  {Frenk}}{{Okamoto} et~al.}{2003}]{Okamoto_2003}
{Okamoto} T.,  {Jenkins} A.,  {Eke} V.~R.,  {Quilis} V.,   {Frenk} C.~S.,
  2003, \mn@doi [\mnras] {10.1046/j.1365-8711.2003.06948.x}, \href
  {http://adsabs.harvard.edu/abs/2003MNRAS.345..429O} {345, 429}

\bibitem[\protect\citeauthoryear{{Okamoto}, {Frenk}, {Jenkins}  \&
  {Theuns}}{{Okamoto} et~al.}{2010}]{Okamoto_2010}
{Okamoto} T.,  {Frenk} C.~S.,  {Jenkins} A.,   {Theuns} T.,  2010, \mn@doi
  [\mnras] {10.1111/j.1365-2966.2010.16690.x}, \href
  {http://adsabs.harvard.edu/abs/2010MNRAS.406..208O} {406, 208}

\bibitem[\protect\citeauthoryear{{Oman} et~al.,}{{Oman}
  et~al.}{2015}]{Oman_2015}
{Oman} K.~A.,  et~al., 2015, \mn@doi [\mnras] {10.1093/mnras/stv1504}, \href
  {http://adsabs.harvard.edu/abs/2015MNRAS.452.3650O} {452, 3650}

\bibitem[\protect\citeauthoryear{{Orr} et~al.,}{{Orr} et~al.}{2017}]{Orr_2017}
{Orr} M.,  et~al., 2017, preprint, \href
  {http://adsabs.harvard.edu/abs/2017arXiv170101788O} {} (\mn@eprint {arXiv}
  {1701.01788})

\bibitem[\protect\citeauthoryear{{Ott} et~al.,}{{Ott} et~al.}{2012}]{Ott_2012}
{Ott} J.,  et~al., 2012, \mn@doi [\aj] {10.1088/0004-6256/144/4/123}, \href
  {http://adsabs.harvard.edu/abs/2012AJ....144..123O} {144, 123}

\bibitem[\protect\citeauthoryear{{Pawlik} \& {Schaye}}{{Pawlik} \&
  {Schaye}}{2009}]{Pawlik_2009}
{Pawlik} A.~H.,  {Schaye} J.,  2009, \mn@doi [\mnras]
  {10.1111/j.1745-3933.2009.00659.x}, \href
  {http://adsabs.harvard.edu/abs/2009MNRAS.396L..46P} {396, L46}

\bibitem[\protect\citeauthoryear{{Pedrosa} \& {Tissera}}{{Pedrosa} \&
  {Tissera}}{2015}]{Pedrosa_2015}
{Pedrosa} S.~E.,  {Tissera} P.~B.,  2015, \mn@doi [\aap]
  {10.1051/0004-6361/201526440}, \href
  {http://adsabs.harvard.edu/abs/2015A%26A...584A..43P} {584, A43}

\bibitem[\protect\citeauthoryear{{Peebles}}{{Peebles}}{1969}]{Peebles_1969}
{Peebles} P.~J.~E.,  1969, \mn@doi [\apj] {10.1086/149876}, \href
  {http://adsabs.harvard.edu/abs/1969ApJ...155..393P} {155, 393}

\bibitem[\protect\citeauthoryear{{Peeples}, {Werk}, {Tumlinson}, {Oppenheimer},
  {Prochaska}, {Katz}  \& {Weinberg}}{{Peeples} et~al.}{2014}]{Peeples_2014}
{Peeples} M.~S.,  {Werk} J.~K.,  {Tumlinson} J.,  {Oppenheimer} B.~D.,
  {Prochaska} J.~X.,  {Katz} N.,   {Weinberg} D.~H.,  2014, \mn@doi [\apj]
  {10.1088/0004-637X/786/1/54}, \href
  {http://adsabs.harvard.edu/abs/2014ApJ...786...54P} {786, 54}

\bibitem[\protect\citeauthoryear{{Penoyre}, {Moster}, {Sijacki}  \&
  {Genel}}{{Penoyre} et~al.}{2017}]{Penoyre_2017}
{Penoyre} Z.,  {Moster} B.~P.,  {Sijacki} D.,   {Genel} S.,  2017, \mn@doi
  [\mnras] {10.1093/mnras/stx762}, \href
  {http://adsabs.harvard.edu/abs/2017MNRAS.468.3883P} {468, 3883}

\bibitem[\protect\citeauthoryear{{Peschken}, {Athanassoula}  \&
  {Rodionov}}{{Peschken} et~al.}{2017}]{Peschken_2017}
{Peschken} N.,  {Athanassoula} E.,   {Rodionov} S.~A.,  2017, \mn@doi [\mnras]
  {10.1093/mnras/stx481}, \href
  {http://adsabs.harvard.edu/abs/2017MNRAS.468..994P} {468, 994}

\bibitem[\protect\citeauthoryear{{Pichon}, {Pogosyan}, {Kimm}, {Slyz},
  {Devriendt}  \& {Dubois}}{{Pichon} et~al.}{2011}]{Pichon_2011}
{Pichon} C.,  {Pogosyan} D.,  {Kimm} T.,  {Slyz} A.,  {Devriendt} J.,
  {Dubois} Y.,  2011, \mn@doi [\mnras] {10.1111/j.1365-2966.2011.19640.x},
  \href {http://adsabs.harvard.edu/abs/2011MNRAS.418.2493P} {418, 2493}

\bibitem[\protect\citeauthoryear{{Pineda}, {Hayward}, {Springel}  \& {Mendes de
  Oliveira}}{{Pineda} et~al.}{2017}]{Pineda_2017}
{Pineda} J.~C.~B.,  {Hayward} C.~C.,  {Springel} V.,   {Mendes de Oliveira} C.,
   2017, \mn@doi [\mnras] {10.1093/mnras/stw3004}, \href
  {http://adsabs.harvard.edu/abs/2017MNRAS.466...63P} {466, 63}

\bibitem[\protect\citeauthoryear{{Pontzen} \& {Governato}}{{Pontzen} \&
  {Governato}}{2012}]{Pontzen_2012}
{Pontzen} A.,  {Governato} F.,  2012, \mn@doi [\mnras]
  {10.1111/j.1365-2966.2012.20571.x}, \href
  {http://adsabs.harvard.edu/abs/2012MNRAS.421.3464P} {421, 3464}

\bibitem[\protect\citeauthoryear{{Pontzen} \& {Governato}}{{Pontzen} \&
  {Governato}}{2014}]{Pontzen_2014}
{Pontzen} A.,  {Governato} F.,  2014, \mn@doi [\nat] {10.1038/nature12953},
  \href {http://adsabs.harvard.edu/abs/2014Natur.506..171P} {506, 171}

\bibitem[\protect\citeauthoryear{{Porter}}{{Porter}}{1985}]{Porter_1985}
{Porter} D.~H.,  1985, PhD thesis, California Univ., Berkeley.

\bibitem[\protect\citeauthoryear{{Power}, {Navarro}, {Jenkins}, {Frenk},
  {White}, {Springel}, {Stadel}  \& {Quinn}}{{Power} et~al.}{2003}]{Power_2003}
{Power} C.,  {Navarro} J.~F.,  {Jenkins} A.,  {Frenk} C.~S.,  {White} S.~D.~M.,
   {Springel} V.,  {Stadel} J.,   {Quinn} T.,  2003, \mn@doi [\mnras]
  {10.1046/j.1365-8711.2003.05925.x}, \href
  {http://adsabs.harvard.edu/abs/2003MNRAS.338...14P} {338, 14}

\bibitem[\protect\citeauthoryear{{Price} \& {Monaghan}}{{Price} \&
  {Monaghan}}{2007}]{Price_2007}
{Price} D.~J.,  {Monaghan} J.~J.,  2007, \mn@doi [\mnras]
  {10.1111/j.1365-2966.2006.11241.x}, \href
  {http://adsabs.harvard.edu/abs/2007MNRAS.374.1347P} {374, 1347}

\bibitem[\protect\citeauthoryear{{Quinn} \& {Binney}}{{Quinn} \&
  {Binney}}{1992}]{Quinn_1992}
{Quinn} T.,  {Binney} J.,  1992, \mn@doi [\mnras] {10.1093/mnras/255.4.729},
  \href {http://adsabs.harvard.edu/abs/1992MNRAS.255..729Q} {255, 729}

\bibitem[\protect\citeauthoryear{{Read}, {Pontzen}  \& {Viel}}{{Read}
  et~al.}{2006}]{Read_2006}
{Read} J.~I.,  {Pontzen} A.~P.,   {Viel} M.,  2006, \mn@doi [\mnras]
  {10.1111/j.1365-2966.2006.10720.x}, \href
  {http://adsabs.harvard.edu/abs/2006MNRAS.371..885R} {371, 885}

\bibitem[\protect\citeauthoryear{{Read}, {Iorio}, {Agertz}  \&
  {Fraternali}}{{Read} et~al.}{2016}]{Read_2016}
{Read} J.~I.,  {Iorio} G.,  {Agertz} O.,   {Fraternali} F.,  2016, \mn@doi
  [\mnras] {10.1093/mnras/stw1876}, \href
  {http://adsabs.harvard.edu/abs/2016MNRAS.462.3628R} {462, 3628}

\bibitem[\protect\citeauthoryear{{Rodriguez-Gomez} et~al.,}{{Rodriguez-Gomez}
  et~al.}{2017}]{Rodriguez_2016}
{Rodriguez-Gomez} V.,  et~al., 2017, \mn@doi [\mnras] {10.1093/mnras/stx305},
  \href {http://adsabs.harvard.edu/abs/2017MNRAS.467.3083R} {467, 3083}

\bibitem[\protect\citeauthoryear{{Romanowsky} \& {Fall}}{{Romanowsky} \&
  {Fall}}{2012}]{Romanowsky_2012}
{Romanowsky} A.~J.,  {Fall} S.~M.,  2012, \mn@doi [\apjs]
  {10.1088/0067-0049/203/2/17}, \href
  {http://adsabs.harvard.edu/abs/2012ApJS..203...17R} {203, 17}

\bibitem[\protect\citeauthoryear{{Ro{\v s}kar}, {Teyssier}, {Agertz},
  {Wetzstein}  \& {Moore}}{{Ro{\v s}kar} et~al.}{2014}]{Roskar_2014}
{Ro{\v s}kar} R.,  {Teyssier} R.,  {Agertz} O.,  {Wetzstein} M.,   {Moore} B.,
  2014, \mn@doi [\mnras] {10.1093/mnras/stu1548}, \href
  {http://adsabs.harvard.edu/abs/2014MNRAS.444.2837R} {444, 2837}

\bibitem[\protect\citeauthoryear{{Roychowdhury}, {Chengalur}, {Begum}  \&
  {Karachentsev}}{{Roychowdhury} et~al.}{2010}]{Roychowdhury_2010}
{Roychowdhury} S.,  {Chengalur} J.~N.,  {Begum} A.,   {Karachentsev} I.~D.,
  2010, \mn@doi [\mnras] {10.1111/j.1745-3933.2010.00835.x}, \href
  {http://adsabs.harvard.edu/abs/2010MNRAS.404L..60R} {404, L60}

\bibitem[\protect\citeauthoryear{{Roychowdhury}, {Chengalur}, {Karachentsev}
  \& {Kaisina}}{{Roychowdhury} et~al.}{2013}]{Roychowdhury_2013}
{Roychowdhury} S.,  {Chengalur} J.~N.,  {Karachentsev} I.~D.,   {Kaisina}
  E.~I.,  2013, \mn@doi [\mnras] {10.1093/mnrasl/slt123}, \href
  {http://adsabs.harvard.edu/abs/2013MNRAS.436L.104R} {436, L104}

\bibitem[\protect\citeauthoryear{{Ryden}}{{Ryden}}{1988}]{Ryden_1988}
{Ryden} B.~S.,  1988, \mn@doi [\apj] {10.1086/166406}, \href
  {http://adsabs.harvard.edu/abs/1988ApJ...329..589R} {329, 589}

\bibitem[\protect\citeauthoryear{{Sales}, {Navarro}, {Theuns}, {Schaye},
  {White}, {Frenk}, {Crain}  \& {Dalla Vecchia}}{{Sales}
  et~al.}{2012}]{Sales_2012}
{Sales} L.~V.,  {Navarro} J.~F.,  {Theuns} T.,  {Schaye} J.,  {White} S.~D.~M.,
   {Frenk} C.~S.,  {Crain} R.~A.,   {Dalla Vecchia} C.,  2012, \mn@doi [\mnras]
  {10.1111/j.1365-2966.2012.20975.x}, \href
  {http://adsabs.harvard.edu/abs/2012MNRAS.423.1544S} {423, 1544}

\bibitem[\protect\citeauthoryear{{S{\'a}nchez-Janssen}, {M{\'e}ndez-Abreu}  \&
  {Aguerri}}{{S{\'a}nchez-Janssen} et~al.}{2010}]{SanchezJanssen_2010}
{S{\'a}nchez-Janssen} R.,  {M{\'e}ndez-Abreu} J.,   {Aguerri} J.~A.~L.,  2010,
  \mn@doi [\mnras] {10.1111/j.1745-3933.2010.00883.x}, \href
  {http://adsabs.harvard.edu/abs/2010MNRAS.406L..65S} {406, L65}

\bibitem[\protect\citeauthoryear{{Sharma} \& {Steinmetz}}{{Sharma} \&
  {Steinmetz}}{2005}]{Sharma_2005}
{Sharma} S.,  {Steinmetz} M.,  2005, \mn@doi [\apj] {10.1086/430660}, \href
  {http://adsabs.harvard.edu/abs/2005ApJ...628...21S} {628, 21}

\bibitem[\protect\citeauthoryear{{Simons}, {Kassin}, {Weiner}, {Heckman},
  {Lee}, {Lotz}, {Peth}  \& {Tchernyshyov}}{{Simons}
  et~al.}{2015}]{Simons_2015}
{Simons} R.~C.,  {Kassin} S.~A.,  {Weiner} B.~J.,  {Heckman} T.~M.,  {Lee}
  J.~C.,  {Lotz} J.~M.,  {Peth} M.,   {Tchernyshyov} K.,  2015, \mn@doi
  [\mnras] {10.1093/mnras/stv1298}, \href
  {http://adsabs.harvard.edu/abs/2015MNRAS.452..986S} {452, 986}

\bibitem[\protect\citeauthoryear{{Simons} et~al.,}{{Simons}
  et~al.}{2016}]{Simons_2016}
{Simons} R.~C.,  et~al., 2016, \mn@doi [\apj] {10.3847/0004-637X/830/1/14},
  \href {http://adsabs.harvard.edu/abs/2016ApJ...830...14S} {830, 14}

\bibitem[\protect\citeauthoryear{{Sokolowska}, {Capelo}, {Fall}, {Mayer},
  {Shen}  \& {Bonoli}}{{Sokolowska} et~al.}{2016}]{Sokolowska_2016}
{Sokolowska} A.,  {Capelo} P.~R.,  {Fall} S.~M.,  {Mayer} L.,  {Shen} S.,
  {Bonoli} S.,  2016, preprint, \href
  {http://adsabs.harvard.edu/abs/2016arXiv161207362S} {} (\mn@eprint {arXiv}
  {1612.07362})

\bibitem[\protect\citeauthoryear{{Soko{\l}owska}, {Capelo}, {Fall}, {Mayer},
  {Shen}  \& {Bonoli}}{{Soko{\l}owska} et~al.}{2017}]{Sokolowska_2017}
{Soko{\l}owska} A.,  {Capelo} P.~R.,  {Fall} S.~M.,  {Mayer} L.,  {Shen} S.,
  {Bonoli} S.,  2017, \mn@doi [\apj] {10.3847/1538-4357/835/2/289}, \href
  {http://adsabs.harvard.edu/abs/2017ApJ...835..289S} {835, 289}

\bibitem[\protect\citeauthoryear{{Sommer-Larsen}, {Gelato}  \&
  {Vedel}}{{Sommer-Larsen} et~al.}{1999}]{SommerLarsen_1999}
{Sommer-Larsen} J.,  {Gelato} S.,   {Vedel} H.,  1999, \mn@doi [\apj]
  {10.1086/307374}, \href {http://adsabs.harvard.edu/abs/1999ApJ...519..501S}
  {519, 501}

\bibitem[\protect\citeauthoryear{{Springel}}{{Springel}}{2005}]{Springel_2005}
{Springel} V.,  2005, \mn@doi [\mnras] {10.1111/j.1365-2966.2005.09655.x},
  \href {http://adsabs.harvard.edu/abs/2005MNRAS.364.1105S} {364, 1105}

\bibitem[\protect\citeauthoryear{{Stanimirovi{\'c}}, {Staveley-Smith}  \&
  {Jones}}{{Stanimirovi{\'c}} et~al.}{2004}]{Stanimirovic_2004}
{Stanimirovi{\'c}} S.,  {Staveley-Smith} L.,   {Jones} P.~A.,  2004, \mn@doi
  [\apj] {10.1086/381869}, \href
  {http://adsabs.harvard.edu/abs/2004ApJ...604..176S} {604, 176}

\bibitem[\protect\citeauthoryear{{Steinmetz} \& {Navarro}}{{Steinmetz} \&
  {Navarro}}{1999}]{Steinmetz_1999}
{Steinmetz} M.,  {Navarro} J.~F.,  1999, \mn@doi [\apj] {10.1086/306904}, \href
  {http://adsabs.harvard.edu/abs/1999ApJ...513..555S} {513, 555}

\bibitem[\protect\citeauthoryear{{Stern}, {Hennawi}, {Prochaska}  \&
  {Werk}}{{Stern} et~al.}{2016}]{Stern_2016}
{Stern} J.,  {Hennawi} J.~F.,  {Prochaska} J.~X.,   {Werk} J.~K.,  2016,
  \mn@doi [\apj] {10.3847/0004-637X/830/2/87}, \href
  {http://adsabs.harvard.edu/abs/2016ApJ...830...87S} {830, 87}

\bibitem[\protect\citeauthoryear{{Stewart}, {Brooks}, {Bullock}, {Maller},
  {Diemand}, {Wadsley}  \& {Moustakas}}{{Stewart} et~al.}{2013}]{Stewart_2013}
{Stewart} K.~R.,  {Brooks} A.~M.,  {Bullock} J.~S.,  {Maller} A.~H.,  {Diemand}
  J.,  {Wadsley} J.,   {Moustakas} L.~A.,  2013, \mn@doi [\apj]
  {10.1088/0004-637X/769/1/74}, \href
  {http://adsabs.harvard.edu/abs/2013ApJ...769...74S} {769, 74}

\bibitem[\protect\citeauthoryear{{Stewart} et~al.,}{{Stewart}
  et~al.}{2017}]{Stewart_2016}
{Stewart} K.~R.,  et~al., 2017, \mn@doi [\apj] {10.3847/1538-4357/aa6dff},
  \href {http://adsabs.harvard.edu/abs/2017ApJ...843...47S} {843, 47}

\bibitem[\protect\citeauthoryear{{Stinson}, {Seth}, {Katz}, {Wadsley},
  {Governato}  \& {Quinn}}{{Stinson} et~al.}{2006}]{Stinson_2006}
{Stinson} G.,  {Seth} A.,  {Katz} N.,  {Wadsley} J.,  {Governato} F.,   {Quinn}
  T.,  2006, \mn@doi [\mnras] {10.1111/j.1365-2966.2006.11097.x}, \href
  {http://adsabs.harvard.edu/abs/2006MNRAS.373.1074S} {373, 1074}

\bibitem[\protect\citeauthoryear{{Stinson}, {Bailin}, {Couchman}, {Wadsley},
  {Shen}, {Nickerson}, {Brook}  \& {Quinn}}{{Stinson}
  et~al.}{2010}]{Stinson_2010}
{Stinson} G.~S.,  {Bailin} J.,  {Couchman} H.,  {Wadsley} J.,  {Shen} S.,
  {Nickerson} S.,  {Brook} C.,   {Quinn} T.,  2010, \mn@doi [\mnras]
  {10.1111/j.1365-2966.2010.17187.x}, \href
  {http://adsabs.harvard.edu/abs/2010MNRAS.408..812S} {408, 812}

\bibitem[\protect\citeauthoryear{{Swaters}}{{Swaters}}{1999}]{Swaters_1999}
{Swaters} R.~A.,  1999, PhD thesis, , Rijksuniversiteit Groningen, (1999)

\bibitem[\protect\citeauthoryear{{Teklu}, {Remus}, {Dolag}, {Beck}, {Burkert},
  {Schmidt}, {Schulze}  \& {Steinborn}}{{Teklu} et~al.}{2015}]{Teklu_2015}
{Teklu} A.~F.,  {Remus} R.-S.,  {Dolag} K.,  {Beck} A.~M.,  {Burkert} A.,
  {Schmidt} A.~S.,  {Schulze} F.,   {Steinborn} L.~K.,  2015, \mn@doi [\apj]
  {10.1088/0004-637X/812/1/29}, \href
  {http://adsabs.harvard.edu/abs/2015ApJ...812...29T} {812, 29}

\bibitem[\protect\citeauthoryear{{Teyssier}, {Pontzen}, {Dubois}  \&
  {Read}}{{Teyssier} et~al.}{2013}]{Teyssier_2013}
{Teyssier} R.,  {Pontzen} A.,  {Dubois} Y.,   {Read} J.~I.,  2013, \mn@doi
  [\mnras] {10.1093/mnras/sts563}, \href
  {http://adsabs.harvard.edu/abs/2013MNRAS.429.3068T} {429, 3068}

\bibitem[\protect\citeauthoryear{{Thacker} \& {Couchman}}{{Thacker} \&
  {Couchman}}{2001}]{Thacker_2001}
{Thacker} R.~J.,  {Couchman} H.~M.~P.,  2001, \mn@doi [\apjl] {10.1086/321739},
  \href {http://adsabs.harvard.edu/abs/2001ApJ...555L..17T} {555, L17}

\bibitem[\protect\citeauthoryear{{Torrey}, {Vogelsberger}, {Sijacki},
  {Springel}  \& {Hernquist}}{{Torrey} et~al.}{2012}]{Torrey_2012}
{Torrey} P.,  {Vogelsberger} M.,  {Sijacki} D.,  {Springel} V.,   {Hernquist}
  L.,  2012, \mn@doi [\mnras] {10.1111/j.1365-2966.2012.22082.x}, \href
  {http://adsabs.harvard.edu/abs/2012MNRAS.427.2224T} {427, 2224}

\bibitem[\protect\citeauthoryear{{Torrey}, {Hopkins}, {Faucher-Gigu{\`e}re},
  {Vogelsberger}, {Quataert}, {Kere{\v s}}  \& {Murray}}{{Torrey}
  et~al.}{2017}]{Torrey_2017}
{Torrey} P.,  {Hopkins} P.~F.,  {Faucher-Gigu{\`e}re} C.-A.,  {Vogelsberger}
  M.,  {Quataert} E.,  {Kere{\v s}} D.,   {Murray} N.,  2017, \mn@doi [\mnras]
  {10.1093/mnras/stx254}, \href
  {http://adsabs.harvard.edu/abs/2017MNRAS.467.2301T} {467, 2301}

\bibitem[\protect\citeauthoryear{{Tully}, {Bottinelli}, {Fisher}, {Gougenheim},
  {Sancisi}  \& {van Woerden}}{{Tully} et~al.}{1978}]{Tully_1978}
{Tully} R.~B.,  {Bottinelli} L.,  {Fisher} J.~R.,  {Gougenheim} L.,  {Sancisi}
  R.,   {van Woerden} H.,  1978, \aap, \href
  {http://adsabs.harvard.edu/abs/1978A%26A....63...37T} {63, 37}

\bibitem[\protect\citeauthoryear{{{\"U}bler}, {Naab}, {Oser}, {Aumer}, {Sales}
  \& {White}}{{{\"U}bler} et~al.}{2014}]{Ubler_2014}
{{\"U}bler} H.,  {Naab} T.,  {Oser} L.,  {Aumer} M.,  {Sales} L.~V.,   {White}
  S.~D.~M.,  2014, \mn@doi [\mnras] {10.1093/mnras/stu1275}, \href
  {http://adsabs.harvard.edu/abs/2014MNRAS.443.2092U} {443, 2092}

\bibitem[\protect\citeauthoryear{{Valenzuela}, {Rhee}, {Klypin}, {Governato},
  {Stinson}, {Quinn}  \& {Wadsley}}{{Valenzuela}
  et~al.}{2007}]{Valenzuela_2007}
{Valenzuela} O.,  {Rhee} G.,  {Klypin} A.,  {Governato} F.,  {Stinson} G.,
  {Quinn} T.,   {Wadsley} J.,  2007, \mn@doi [\apj] {10.1086/508674}, \href
  {http://adsabs.harvard.edu/abs/2007ApJ...657..773V} {657, 773}

\bibitem[\protect\citeauthoryear{{Verbeke}, {Papastergis}, {Ponomareva},
  {Rathi}  \& {De Rijcke}}{{Verbeke} et~al.}{2017}]{Verbeke_2017}
{Verbeke} R.,  {Papastergis} E.,  {Ponomareva} A.~A.,  {Rathi} S.,   {De
  Rijcke} S.,  2017, preprint, \href
  {http://adsabs.harvard.edu/abs/2017arXiv170303810V} {} (\mn@eprint {arXiv}
  {1703.03810})

\bibitem[\protect\citeauthoryear{{Walter}, {Brinks}, {de Blok}, {Bigiel},
  {Kennicutt}, {Thornley}  \& {Leroy}}{{Walter} et~al.}{2008}]{Walter_2008}
{Walter} F.,  {Brinks} E.,  {de Blok} W.~J.~G.,  {Bigiel} F.,  {Kennicutt} Jr.
  R.~C.,  {Thornley} M.~D.,   {Leroy} A.,  2008, \mn@doi [\aj]
  {10.1088/0004-6256/136/6/2563}, \href
  {http://adsabs.harvard.edu/abs/2008AJ....136.2563W} {136, 2563}

\bibitem[\protect\citeauthoryear{{Wang}, {Dutton}, {Stinson}, {Macci{\`o}},
  {Penzo}, {Kang}, {Keller}  \& {Wadsley}}{{Wang} et~al.}{2015}]{Wang_2015}
{Wang} L.,  {Dutton} A.~A.,  {Stinson} G.~S.,  {Macci{\`o}} A.~V.,  {Penzo} C.,
   {Kang} X.,  {Keller} B.~W.,   {Wadsley} J.,  2015, \mn@doi [\mnras]
  {10.1093/mnras/stv1937}, \href
  {http://adsabs.harvard.edu/abs/2015MNRAS.454...83W} {454, 83}

\bibitem[\protect\citeauthoryear{{Wang}, {Koribalski}, {Serra}, {van der
  Hulst}, {Roychowdhury}, {Kamphuis}  \& {Chengalur}}{{Wang}
  et~al.}{2016}]{Wang_2016}
{Wang} J.,  {Koribalski} B.~S.,  {Serra} P.,  {van der Hulst} T.,
  {Roychowdhury} S.,  {Kamphuis} P.,   {Chengalur} J.~N.,  2016, \mn@doi
  [\mnras] {10.1093/mnras/stw1099}, \href
  {http://adsabs.harvard.edu/abs/2016MNRAS.460.2143W} {460, 2143}

\bibitem[\protect\citeauthoryear{{Werk} et~al.,}{{Werk}
  et~al.}{2016}]{Werk_2016}
{Werk} J.~K.,  et~al., 2016, \mn@doi [\apj] {10.3847/1538-4357/833/1/54}, \href
  {http://adsabs.harvard.edu/abs/2016ApJ...833...54W} {833, 54}

\bibitem[\protect\citeauthoryear{{Wetzel} \& {Nagai}}{{Wetzel} \&
  {Nagai}}{2015}]{Wetzel_2015}
{Wetzel} A.~R.,  {Nagai} D.,  2015, \mn@doi [\apj]
  {10.1088/0004-637X/808/1/40}, \href
  {http://adsabs.harvard.edu/abs/2015ApJ...808...40W} {808, 40}

\bibitem[\protect\citeauthoryear{{Wetzel}, {Hopkins}, {Kim},
  {Faucher-Gigu{\`e}re}, {Kere{\v s}}  \& {Quataert}}{{Wetzel}
  et~al.}{2016}]{Wetzel_2016}
{Wetzel} A.~R.,  {Hopkins} P.~F.,  {Kim} J.-h.,  {Faucher-Gigu{\`e}re} C.-A.,
  {Kere{\v s}} D.,   {Quataert} E.,  2016, \mn@doi [\apjl]
  {10.3847/2041-8205/827/2/L23}, \href
  {http://adsabs.harvard.edu/abs/2016ApJ...827L..23W} {827, L23}

\bibitem[\protect\citeauthoryear{{Wheeler} et~al.,}{{Wheeler}
  et~al.}{2017}]{Wheeler_2017}
{Wheeler} C.,  et~al., 2017, \mn@doi [\mnras] {10.1093/mnras/stw2583}, \href
  {http://adsabs.harvard.edu/abs/2017MNRAS.465.2420W} {465, 2420}

\bibitem[\protect\citeauthoryear{{White}}{{White}}{1984}]{White_1984}
{White} S.~D.~M.,  1984, \mn@doi [\apj] {10.1086/162573}, \href
  {http://adsabs.harvard.edu/abs/1984ApJ...286...38W} {286, 38}

\bibitem[\protect\citeauthoryear{{Yoachim} \& {Dalcanton}}{{Yoachim} \&
  {Dalcanton}}{2006}]{Yoachim_2006}
{Yoachim} P.,  {Dalcanton} J.~J.,  2006, \mn@doi [\aj] {10.1086/497970}, \href
  {http://adsabs.harvard.edu/abs/2006AJ....131..226Y} {131, 226}

\bibitem[\protect\citeauthoryear{{Zasov} \& {Zaitseva}}{{Zasov} \&
  {Zaitseva}}{2017}]{Zasov_2017}
{Zasov} A.~V.,  {Zaitseva} N.~A.,  2017, preprint, \href
  {http://adsabs.harvard.edu/abs/2017arXiv170507659Z} {} (\mn@eprint {arXiv}
  {1705.07659})

\bibitem[\protect\citeauthoryear{{Zavala} et~al.,}{{Zavala}
  et~al.}{2016}]{Zavala_2016}
{Zavala} J.,  et~al., 2016, \mn@doi [\mnras] {10.1093/mnras/stw1286}, \href
  {http://adsabs.harvard.edu/abs/2016MNRAS.460.4466Z} {460, 4466}

\bibitem[\protect\citeauthoryear{{Zhu} \& {Li}}{{Zhu} \& {Li}}{2016}]{Zhu_2016}
{Zhu} Q.,  {Li} Y.,  2016, \mn@doi [\apj] {10.3847/0004-637X/831/1/52}, \href
  {http://adsabs.harvard.edu/abs/2016ApJ...831...52Z} {831, 52}

\bibitem[\protect\citeauthoryear{{Zjupa} \& {Springel}}{{Zjupa} \&
  {Springel}}{2017}]{Zjupa_2017}
{Zjupa} J.,  {Springel} V.,  2017, \mn@doi [\mnras] {10.1093/mnras/stw2945},
  \href {http://adsabs.harvard.edu/abs/2017MNRAS.466.1625Z} {466, 1625}

\bibitem[\protect\citeauthoryear{{van den Bosch} \& {Swaters}}{{van den Bosch}
  \& {Swaters}}{2001}]{vdBosch2001b}
{van den Bosch} F.~C.,  {Swaters} R.~A.,  2001, \mn@doi [\mnras]
  {10.1046/j.1365-8711.2001.04456.x}, \href
  {http://adsabs.harvard.edu/abs/2001MNRAS.325.1017V} {325, 1017}

\bibitem[\protect\citeauthoryear{{van den Bosch}, {Burkert}  \& {Swaters}}{{van
  den Bosch} et~al.}{2001}]{vandenBosch_2001}
{van den Bosch} F.~C.,  {Burkert} A.,   {Swaters} R.~A.,  2001, \mn@doi
  [\mnras] {10.1046/j.1365-8711.2001.04656.x}, \href
  {http://adsabs.harvard.edu/abs/2001MNRAS.326.1205V} {326, 1205}

\bibitem[\protect\citeauthoryear{{van den Bosch}, {Abel}, {Croft}, {Hernquist}
  \& {White}}{{van den Bosch} et~al.}{2002}]{vandenBosch_2002}
{van den Bosch} F.~C.,  {Abel} T.,  {Croft} R.~A.~C.,  {Hernquist} L.,
  {White} S.~D.~M.,  2002, \mn@doi [\apj] {10.1086/341619}, \href
  {http://adsabs.harvard.edu/abs/2002ApJ...576...21V} {576, 21}

\bibitem[\protect\citeauthoryear{{van der Marel} \& {Kallivayalil}}{{van der
  Marel} \& {Kallivayalil}}{2014}]{vanderMarel_2014}
{van der Marel} R.~P.,  {Kallivayalil} N.,  2014, \mn@doi [\apj]
  {10.1088/0004-637X/781/2/121}, \href
  {http://adsabs.harvard.edu/abs/2014ApJ...781..121V} {781, 121}

\makeatother
\end{thebibliography}

%%%%%%%%%%%%%%%%%%%%%%%%%%%%%%%%%%%%%%%%%%%%%%%%%%

%%%%%%%%%%%%%%%%% APPENDICES %%%%%%%%%%%%%%%%%%%%%

\appendix

\section{Data for individual galaxies}
\label{sec:idvid_gals}
In this Appendix, we present mock HI moment maps, rotation curves, and orbital circularity histograms for all the galaxies in our sample. For each galaxy, we show the following: 

\noindent\textbf{Top}: Mock HI velocity moment maps. These are produced at a fixed spatial resolution of 0.1 kpc for an ``edge-on'' viewing angle; i.e, the galaxy's  HI net angular momentum vector is aligned with the vertical axis. Only pixels in which the total HI column density exceeds $N_{\rm HI} = 5\times10^{19}\,\rm cm^{-2}$ are shown, comparable to the sensitivity of VLA surveys of nearby galaxies.

\textit{Left}: Projected neutral hydrogen column density.

\textit{Middle}: Mean line-of-sight velocity of HI gas in each pixel.

\textit{Right}: Line-of-sight velocity dispersion of HI gas in each pixel. Note that this does not include thermal broadening.

\noindent\textbf{Bottom}: Kinematic diagnostics of rotation vs dispersion. 

\textit{Left}: Orbital circularity distributions for stars and HI gas. Circularity is a measure of rotational support: a coherently rotating disk will have a highly skewed $P(\epsilon)$ distribution with $\overline{\epsilon} \to 1$, while a disordered system without net rotation will have a symmetric $P(\epsilon)$ distribution with $\overline{\epsilon} \sim 0$ (see Section~\ref{sec:circ}). 

\textit{Middle}: Rotation curves. Black line shows the halo circular velocity, $v_c(r) = \sqrt{GM(<r)/r}$. Red line shows the mean azimuthal velocity $v_{\phi}$ of HI gas in cylindrical bins. Blue line shows the mean line-of-sight velocity in a slit aligned with the major axis when the galaxy is viewed edge-on. We plot out to the major radius where the mean HI surface density falls below $\Sigma_{\rm HI} = \rm  0.1 \rm M_{\odot}\,pc^{-2}\approx 10^{19}\,cm^{-2}$, comparable to observational studies.

\textit{Right}: Velocity dispersion curves. Black line shows the halo circular velocity, $v_c(r) = \sqrt{GM(<r)/r}$, for comparison. Red line shows the azimuthal velocity dispersion $\sigma_{\phi}$ of HI gas in cylindrical bins. Blue line shows the line-of-sight velocity dispersion in a slit aligned with the major axis when the galaxy is viewed edge-on. Note that this does not include thermal broadening. 

\begin{figure*}
\centering
\begin{minipage}[b]{.4\textwidth}
\includegraphics[width=\columnwidth]{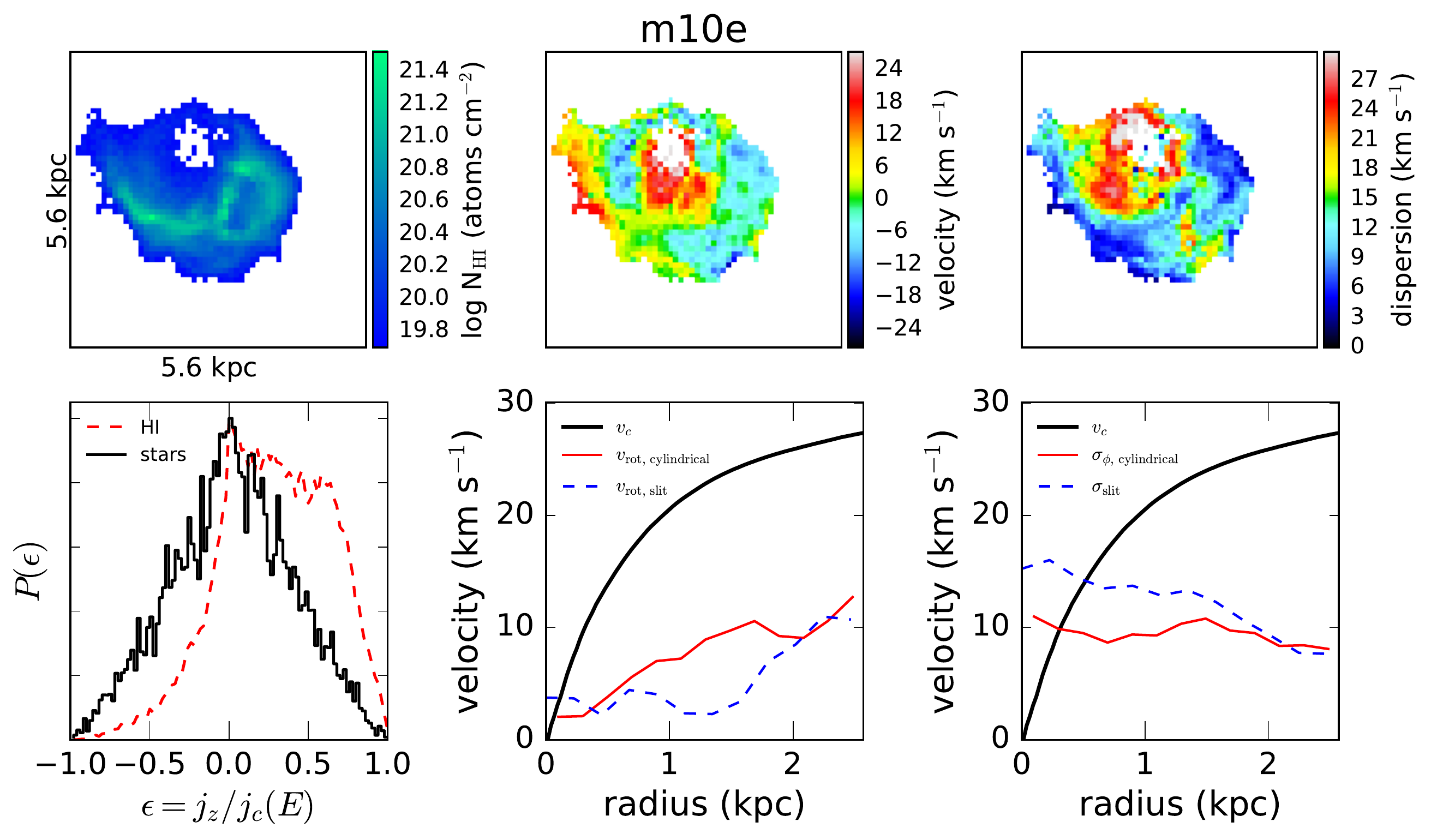}
\caption{\textbf{Top}: HI column density (left), line-of-sight velocity (middle) and velocity dispersion (right) for \texttt{m10e} ($\rm M_{\rm star}=10^{6.3} \rm M_{\odot}$). \textbf{Bottom}: Left: orbital circularity distribution (Equation~\ref{eqn:circularity}). Middle: HI rotation curves, compared to the halo circular velocity, $v_c = \sqrt{GM(<r)/r}$. Right: HI velocity dispersion. See Appendix~\ref{sec:idvid_gals} for details.}
\label{fig:m10e}
\end{minipage}\qquad
\begin{minipage}[b]{.4\textwidth}
\includegraphics[width=\columnwidth]{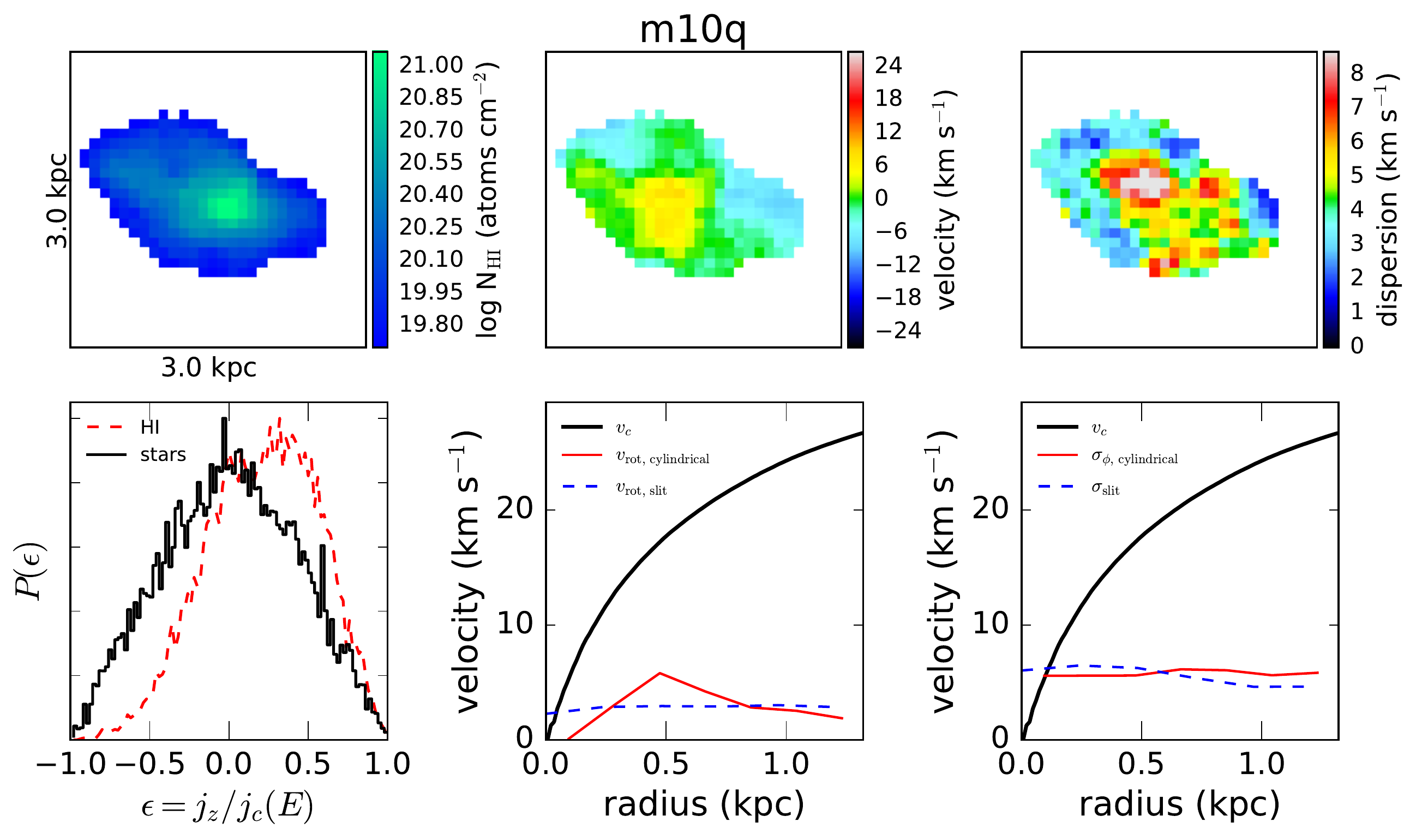}
\caption{\textbf{Top}: HI column density (left), line-of-sight velocity (middle) and velocity dispersion (right) for \texttt{m10q} ($\rm M_{\rm star}=10^{6.3} \rm M_{\odot}$). \textbf{Bottom}: Left: orbital circularity distribution (Equation~\ref{eqn:circularity}). Middle: HI rotation curves, compared to the halo circular velocity, $v_c = \sqrt{GM(<r)/r}$. Right: HI velocity dispersion. See Appendix~\ref{sec:idvid_gals} for details.}\label{fig:m10q}
\end{minipage}
\end{figure*}

\begin{figure*}
\centering
\begin{minipage}[b]{.4\textwidth}
\includegraphics[width=\columnwidth]{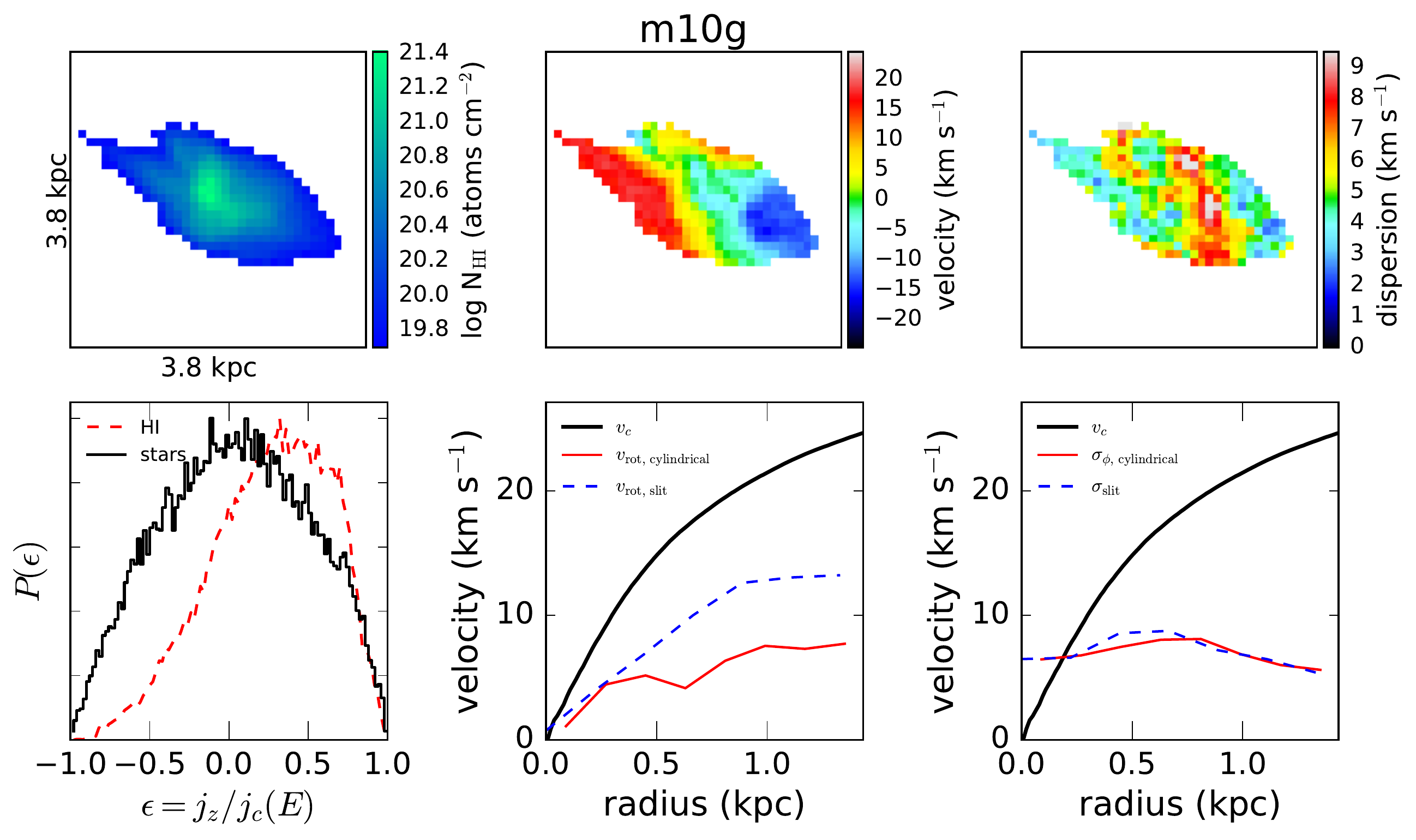}
\caption{\textbf{Top}: HI column density (left), line-of-sight velocity (middle) and velocity dispersion (right) for \texttt{m10g} ($\rm M_{\rm star}=10^{6.7} \rm M_{\odot}$). \textbf{Bottom}: Left: orbital circularity distribution (Equation~\ref{eqn:circularity}). Middle: HI rotation curves, compared to the halo circular velocity, $v_c = \sqrt{GM(<r)/r}$. Right: HI velocity dispersion. See Appendix~\ref{sec:idvid_gals} for details.}
\label{fig:m10g}
\end{minipage}\qquad
\begin{minipage}[b]{.4\textwidth}
\includegraphics[width=\columnwidth]{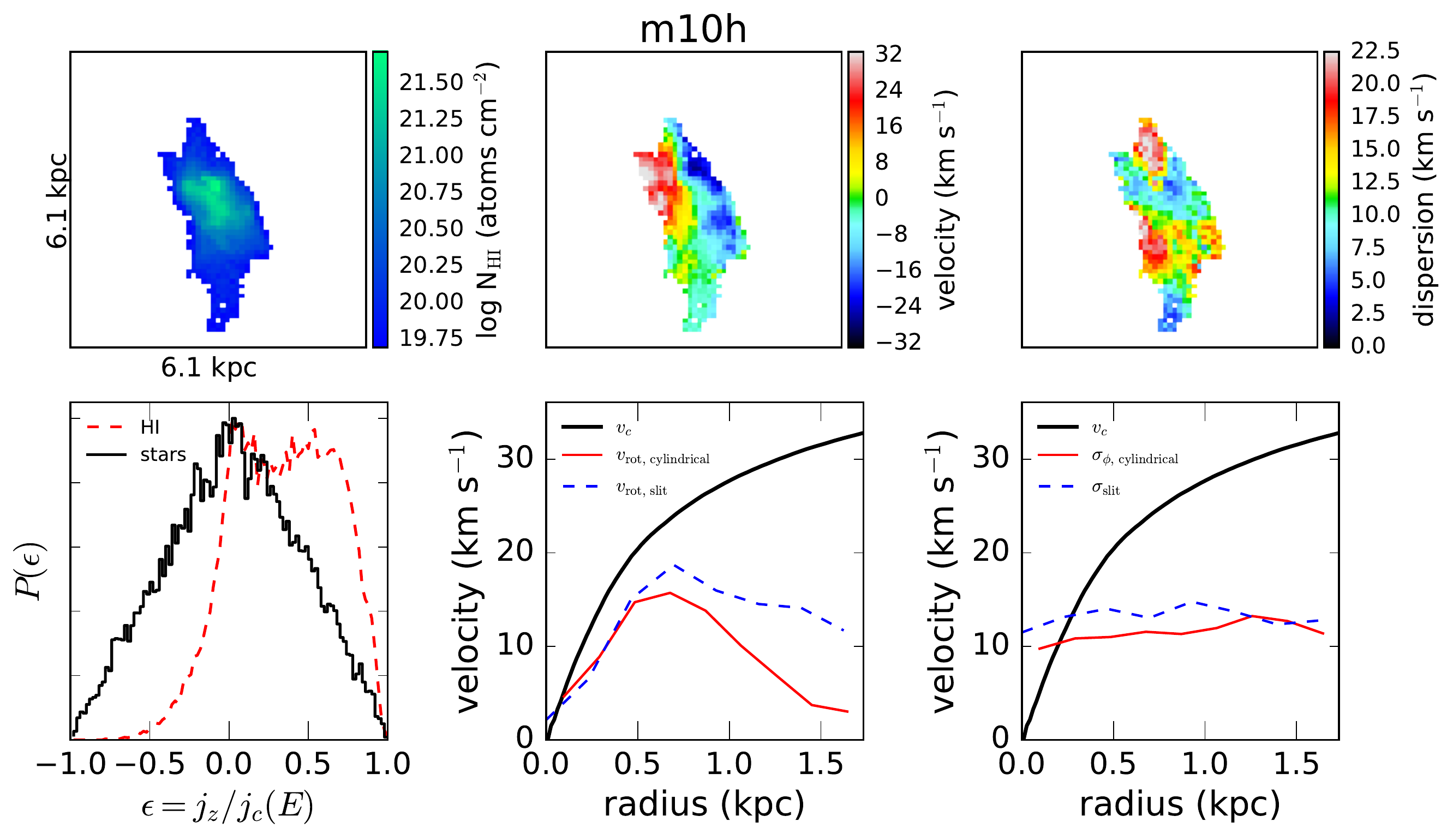}
\caption{\textbf{Top}: HI column density (left), line-of-sight velocity (middle) and velocity dispersion (right) for \texttt{m10h} ($\rm M_{\rm star}=10^{6.8} \rm M_{\odot}$). \textbf{Bottom}: Left: orbital circularity distribution (Equation~\ref{eqn:circularity}). Middle: HI rotation curves, compared to the halo circular velocity, $v_c = \sqrt{GM(<r)/r}$. Right: HI velocity dispersion. See Appendix~\ref{sec:idvid_gals} for details.}\label{fig:m10h}
\end{minipage}
\end{figure*}

\begin{figure*}
\centering
\begin{minipage}[b]{.4\textwidth}
\includegraphics[width=\columnwidth]{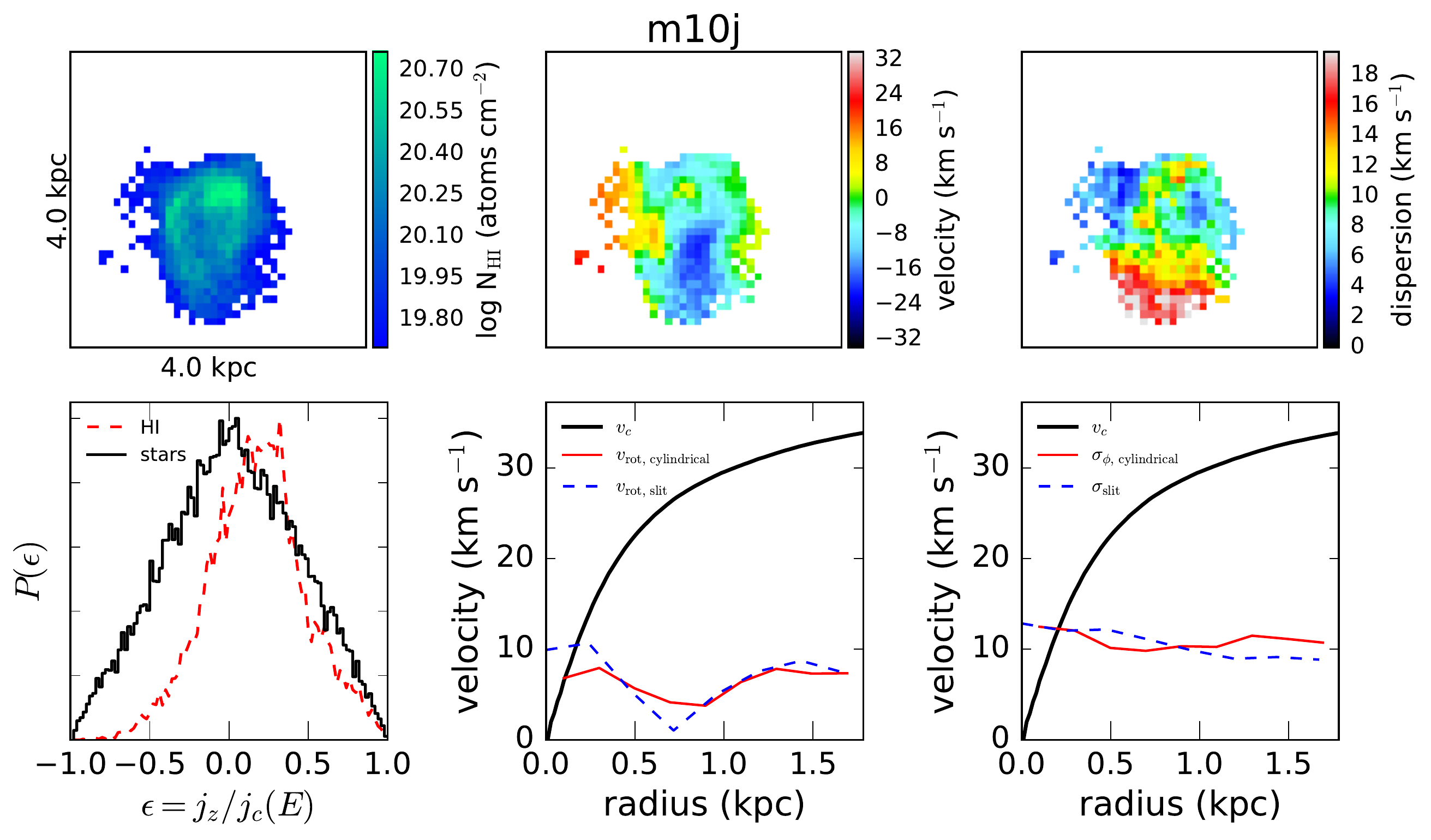}
\caption{\textbf{Top}: HI column density (left), line-of-sight velocity (middle) and velocity dispersion (right) for \texttt{m10j} ($\rm M_{\rm star}=10^{6.9} \rm M_{\odot}$). \textbf{Bottom}: Left: orbital circularity distribution (Equation~\ref{eqn:circularity}). Middle: HI rotation curves, compared to the halo circular velocity, $v_c = \sqrt{GM(<r)/r}$. Right: HI velocity dispersion. See Appendix~\ref{sec:idvid_gals} for details.}
\label{fig:m10j}
\end{minipage}\qquad
\begin{minipage}[b]{.4\textwidth}
\includegraphics[width=\columnwidth]{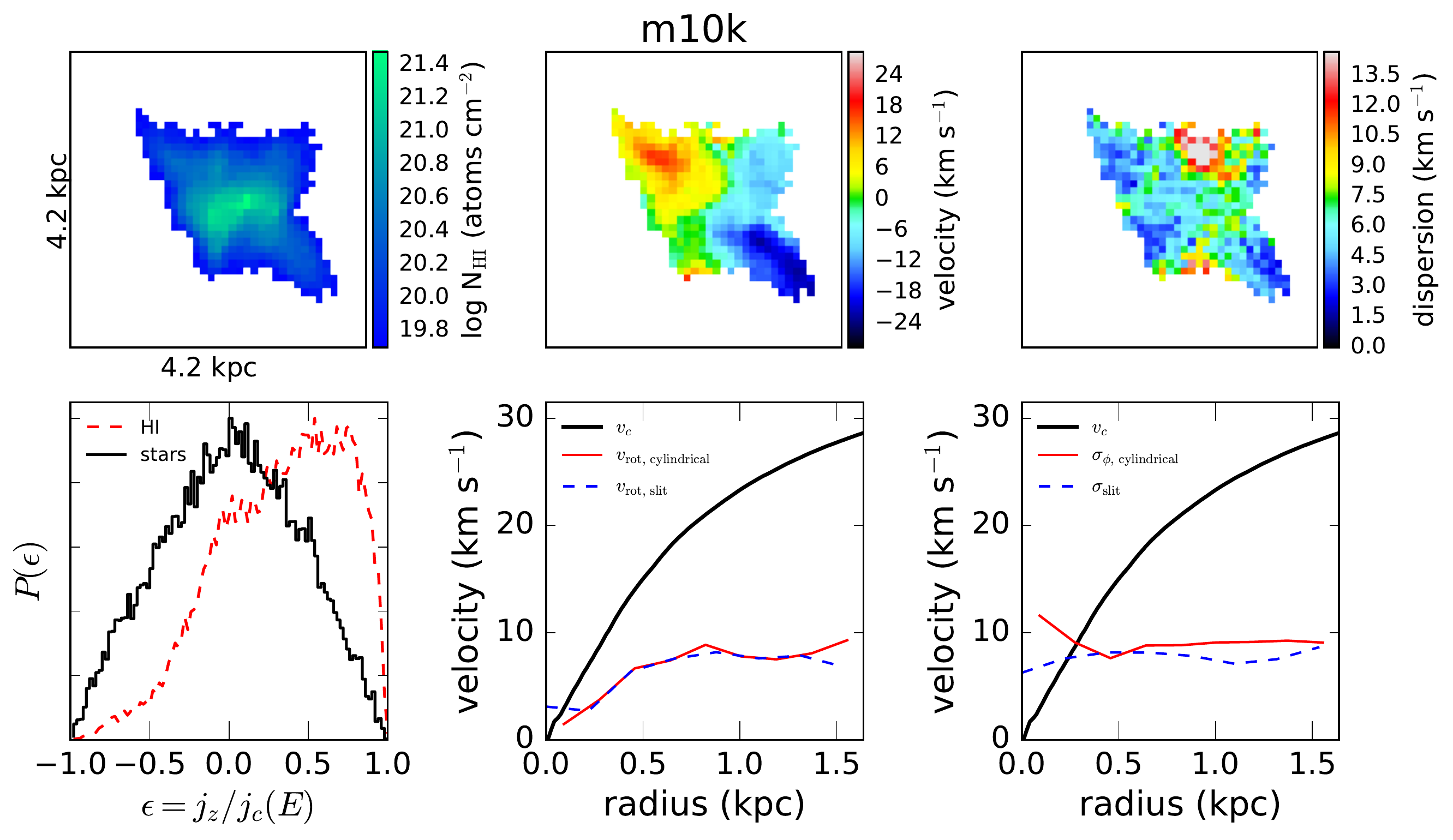}
\caption{\textbf{Top}: HI column density (left), line-of-sight velocity (middle) and velocity dispersion (right) for \texttt{m10k} ($\rm M_{\rm star}=10^{7.0} \rm M_{\odot}$). \textbf{Bottom}: Left: orbital circularity distribution (Equation~\ref{eqn:circularity}). Middle: HI rotation curves, compared to the halo circular velocity, $v_c = \sqrt{GM(<r)/r}$. Right: HI velocity dispersion. See Appendix~\ref{sec:idvid_gals} for details.}\label{fig:m10k}
\end{minipage}
\end{figure*}

\begin{figure*}
\centering
\begin{minipage}[b]{.4\textwidth}
\includegraphics[width=\columnwidth]{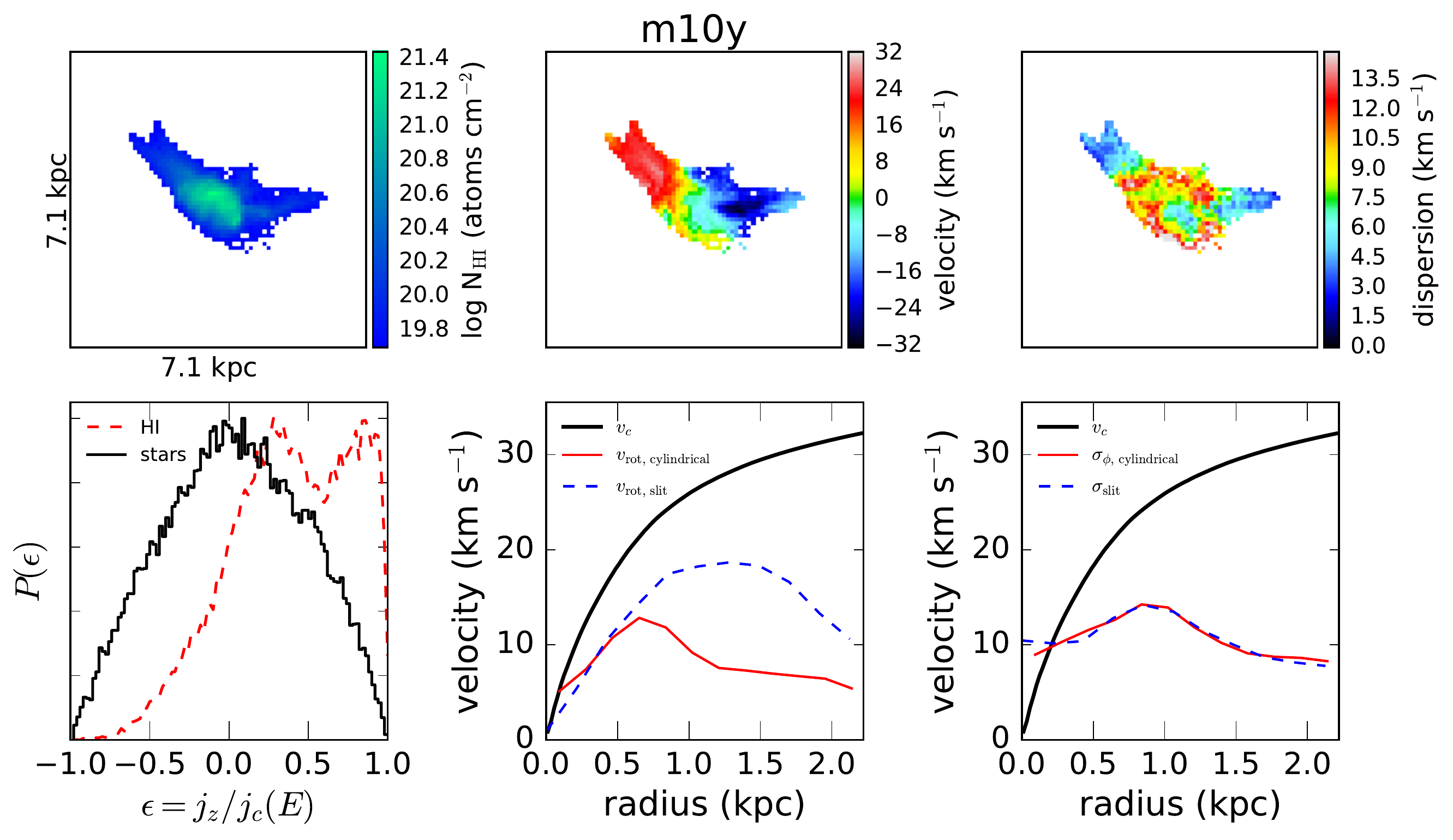}
\caption{\textbf{Top}: HI column density (left), line-of-sight velocity (middle) and velocity dispersion (right) for \texttt{m10y} ($\rm M_{\rm star}=10^{7.0} \rm M_{\odot}$). \textbf{Bottom}: Left: orbital circularity distribution (Equation~\ref{eqn:circularity}). Middle: HI rotation curves, compared to the halo circular velocity, $v_c = \sqrt{GM(<r)/r}$. Right: HI velocity dispersion. See Appendix~\ref{sec:idvid_gals} for details.}
\label{fig:m10y}
\end{minipage}\qquad
\begin{minipage}[b]{.4\textwidth}
\includegraphics[width=\columnwidth]{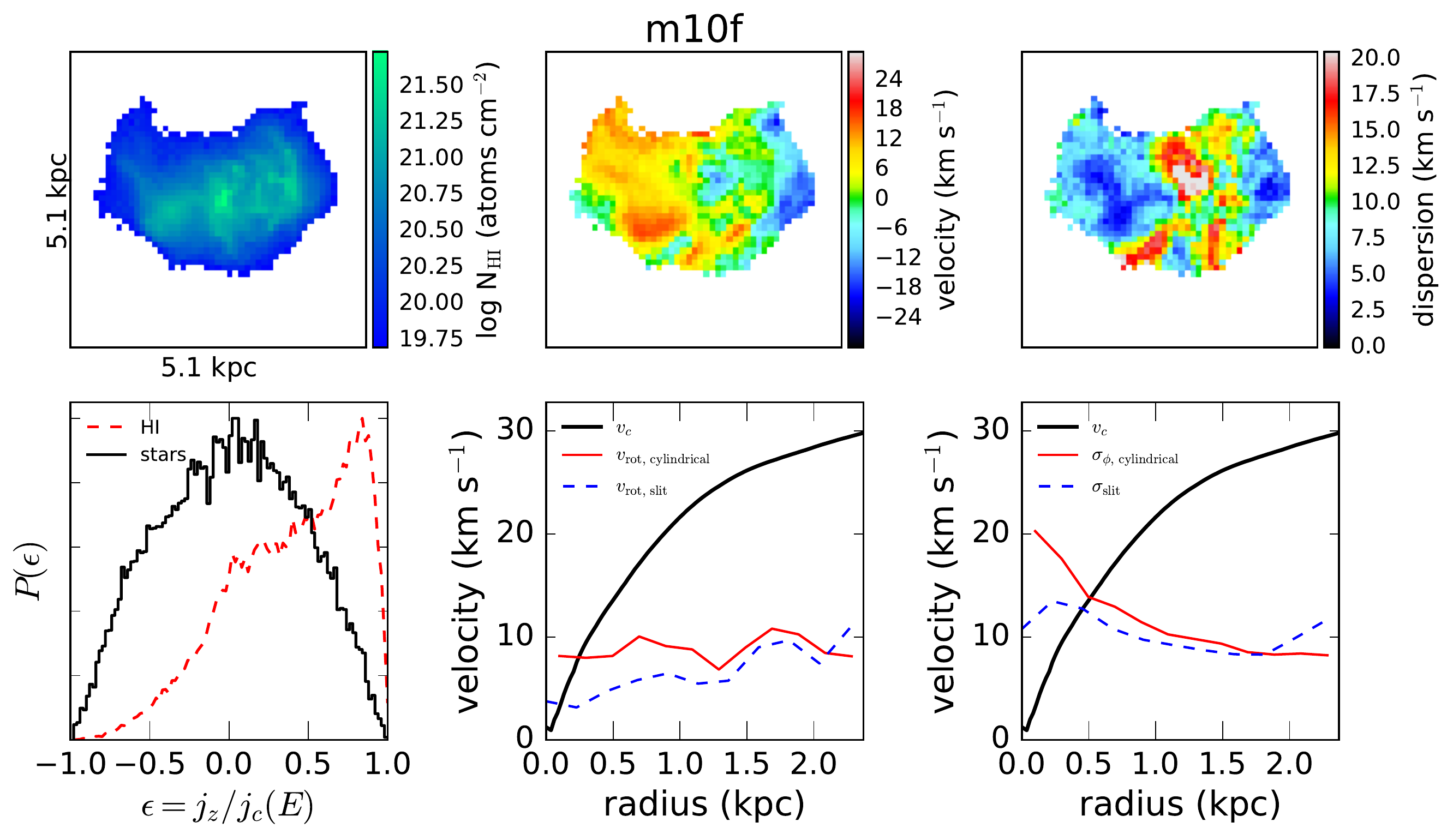}
\caption{\textbf{Top}: HI column density (left), line-of-sight velocity (middle) and velocity dispersion (right) for \texttt{m10f} ($\rm M_{\rm star}=10^{7.0} \rm M_{\odot}$). \textbf{Bottom}: Left: orbital circularity distribution (Equation~\ref{eqn:circularity}). Middle: HI rotation curves, compared to the halo circular velocity, $v_c = \sqrt{GM(<r)/r}$. Right: HI velocity dispersion. See Appendix~\ref{sec:idvid_gals} for details.}\label{fig:m10f}
\end{minipage}
\end{figure*}

\begin{figure*}
\centering
\begin{minipage}[b]{.4\textwidth}
\includegraphics[width=\columnwidth]{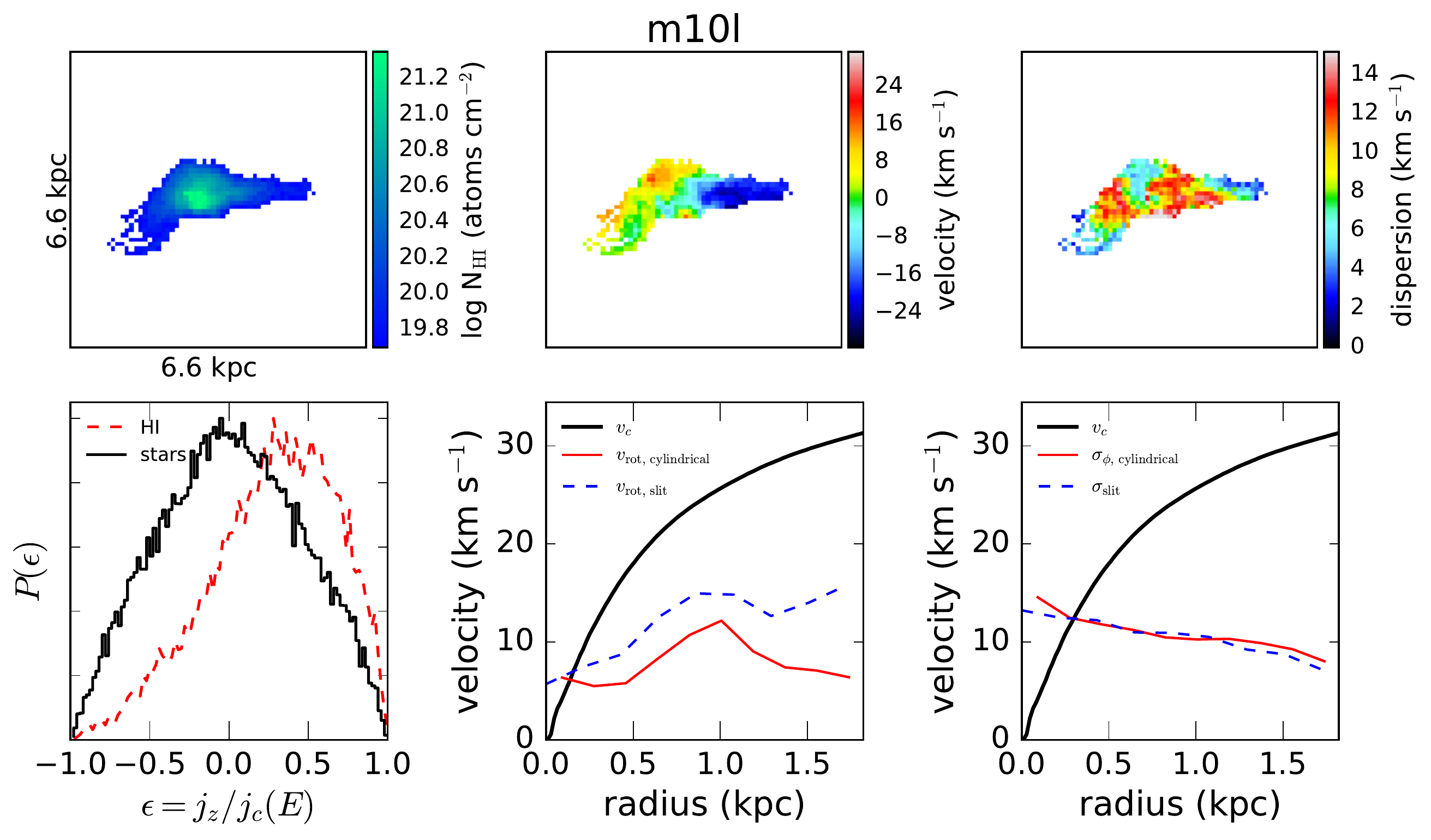}
\caption{\textbf{Top}: HI column density (left), line-of-sight velocity (middle) and velocity dispersion (right) for \texttt{m10l} ($\rm M_{\rm star}=10^{7.1} \rm M_{\odot}$). \textbf{Bottom}: Left: orbital circularity distribution (Equation~\ref{eqn:circularity}). Middle: HI rotation curves, compared to the halo circular velocity, $v_c = \sqrt{GM(<r)/r}$. Right: HI velocity dispersion. See Appendix~\ref{sec:idvid_gals} for details.}
\label{fig:m10l}
\end{minipage}\qquad
\begin{minipage}[b]{.4\textwidth}
\includegraphics[width=\columnwidth]{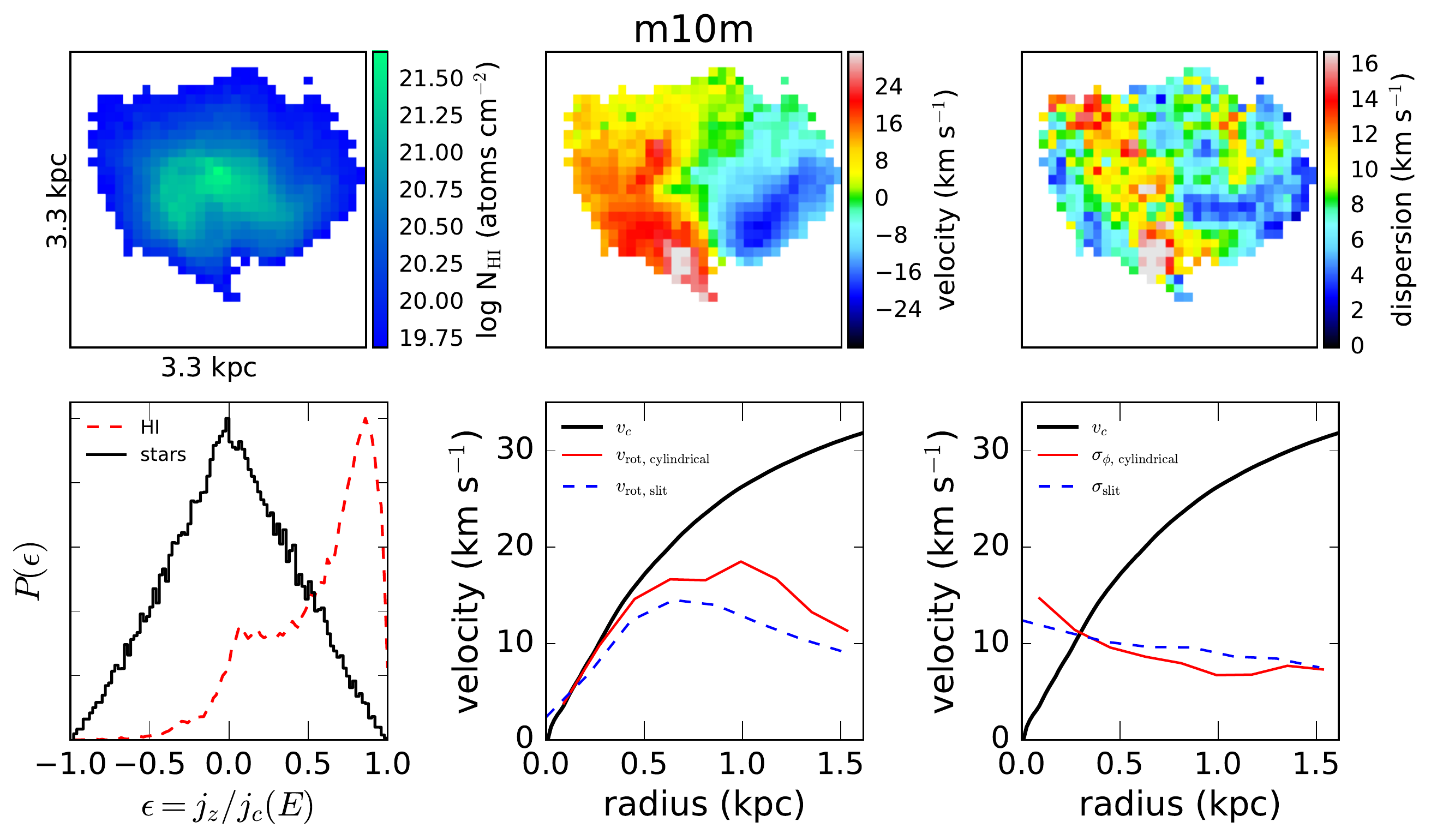}
\caption{\textbf{Top}: HI column density (left), line-of-sight velocity (middle) and velocity dispersion (right) for \texttt{m10m} ($\rm M_{\rm star}=10^{7.1} \rm M_{\odot}$). \textbf{Bottom}: Left: orbital circularity distribution (Equation~\ref{eqn:circularity}). Middle: HI rotation curves, compared to the halo circular velocity, $v_c = \sqrt{GM(<r)/r}$. Right: HI velocity dispersion. See Appendix~\ref{sec:idvid_gals} for details.}\label{fig:m10m}
\end{minipage}
\end{figure*}

\begin{figure*}
\centering
\begin{minipage}[b]{.4\textwidth}
\includegraphics[width=\columnwidth]{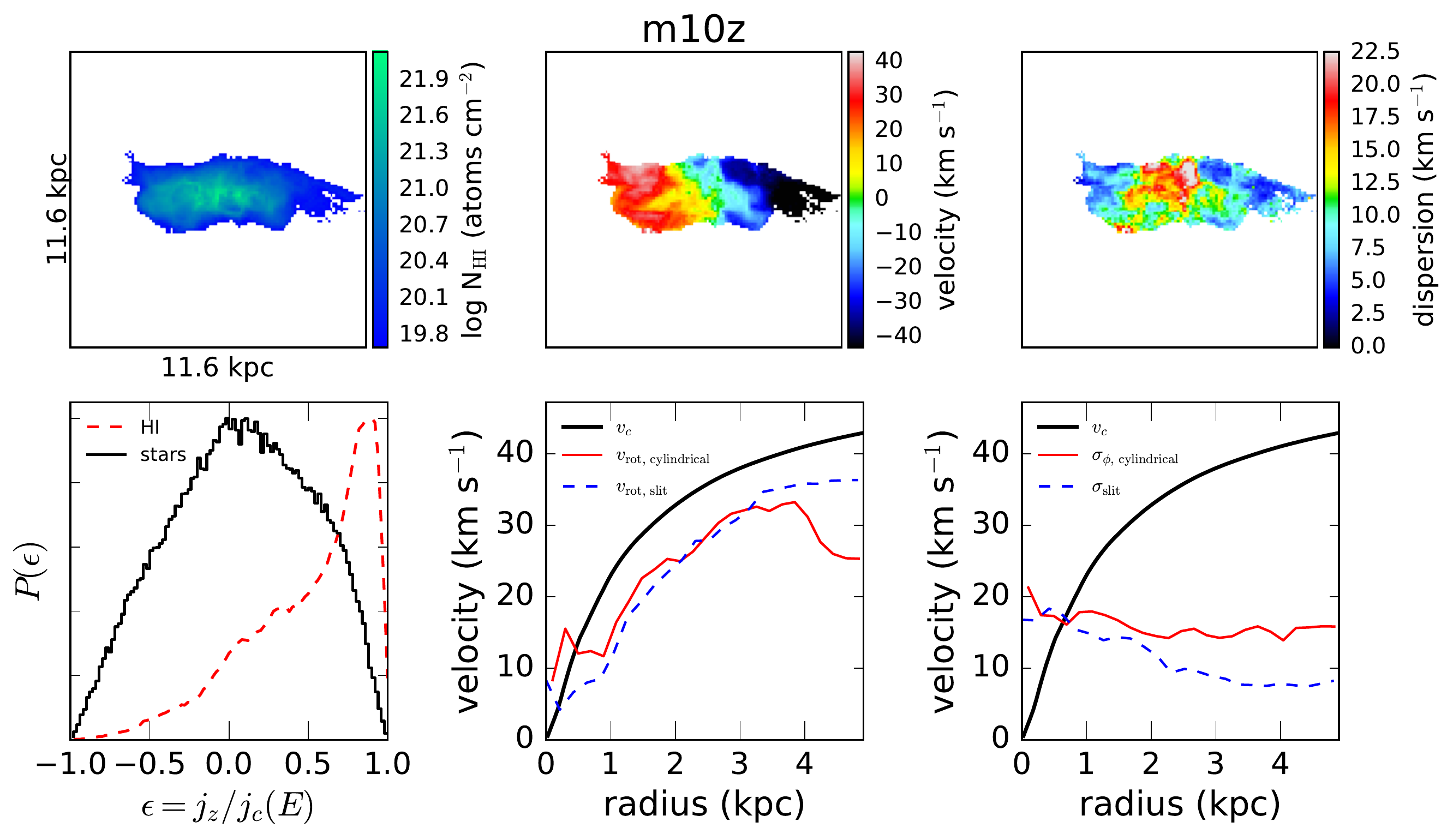}
\caption{\textbf{Top}: HI column density (left), line-of-sight velocity (middle) and velocity dispersion (right) for \texttt{m10z} ($\rm M_{\rm star}=10^{7.6} \rm M_{\odot}$). \textbf{Bottom}: Left: orbital circularity distribution (Equation~\ref{eqn:circularity}). Middle: HI rotation curves, compared to the halo circular velocity, $v_c = \sqrt{GM(<r)/r}$. Right: HI velocity dispersion. See Appendix~\ref{sec:idvid_gals} for details.}
\label{fig:m10z}
\end{minipage}\qquad
\begin{minipage}[b]{.4\textwidth}
\includegraphics[width=\columnwidth]{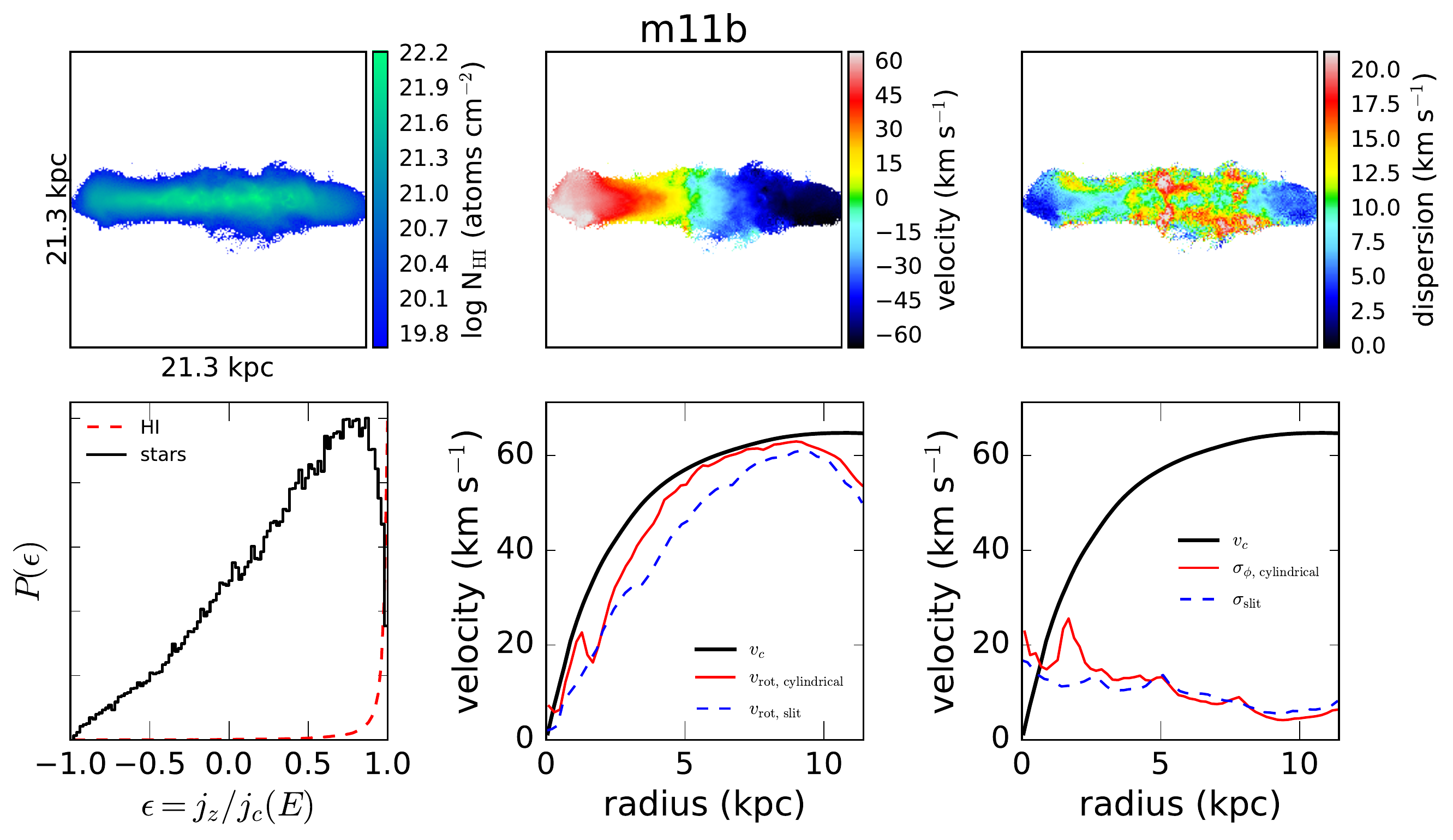}
\caption{\textbf{Top}: HI column density (left), line-of-sight velocity (middle) and velocity dispersion (right) for \texttt{m11b} ($\rm M_{\rm star}=10^{8.0} \rm M_{\odot}$). \textbf{Bottom}: Left: orbital circularity distribution (Equation~\ref{eqn:circularity}). Middle: HI rotation curves, compared to the halo circular velocity, $v_c = \sqrt{GM(<r)/r}$. Right: HI velocity dispersion. See Appendix~\ref{sec:idvid_gals} for details.}\label{fig:m11b}
\end{minipage}
\end{figure*}

\begin{figure*}
\centering
\begin{minipage}[b]{.4\textwidth}
\includegraphics[width=\columnwidth]{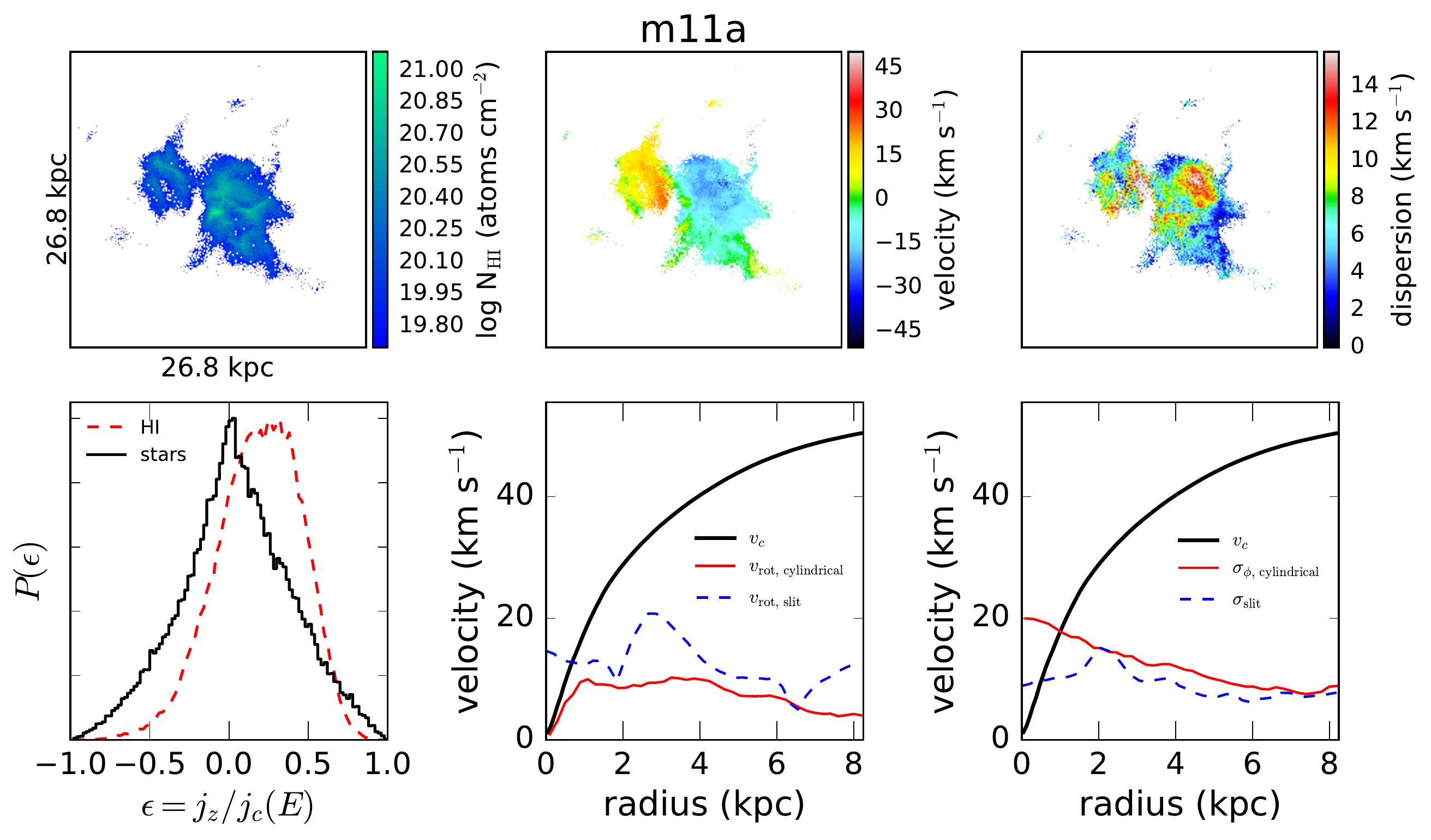}
\caption{\textbf{Top}: HI column density (left), line-of-sight velocity (middle) and velocity dispersion (right) for \texttt{m11a} ($\rm M_{\rm star}=10^{8.1} \rm M_{\odot}$). \textbf{Bottom}: Left: orbital circularity distribution (Equation~\ref{eqn:circularity}). Middle: HI rotation curves, compared to the halo circular velocity, $v_c = \sqrt{GM(<r)/r}$. Right: HI velocity dispersion. See Appendix~\ref{sec:idvid_gals} for details.}
\label{fig:m11a}
\end{minipage}\qquad
\begin{minipage}[b]{.4\textwidth}
\includegraphics[width=\columnwidth]{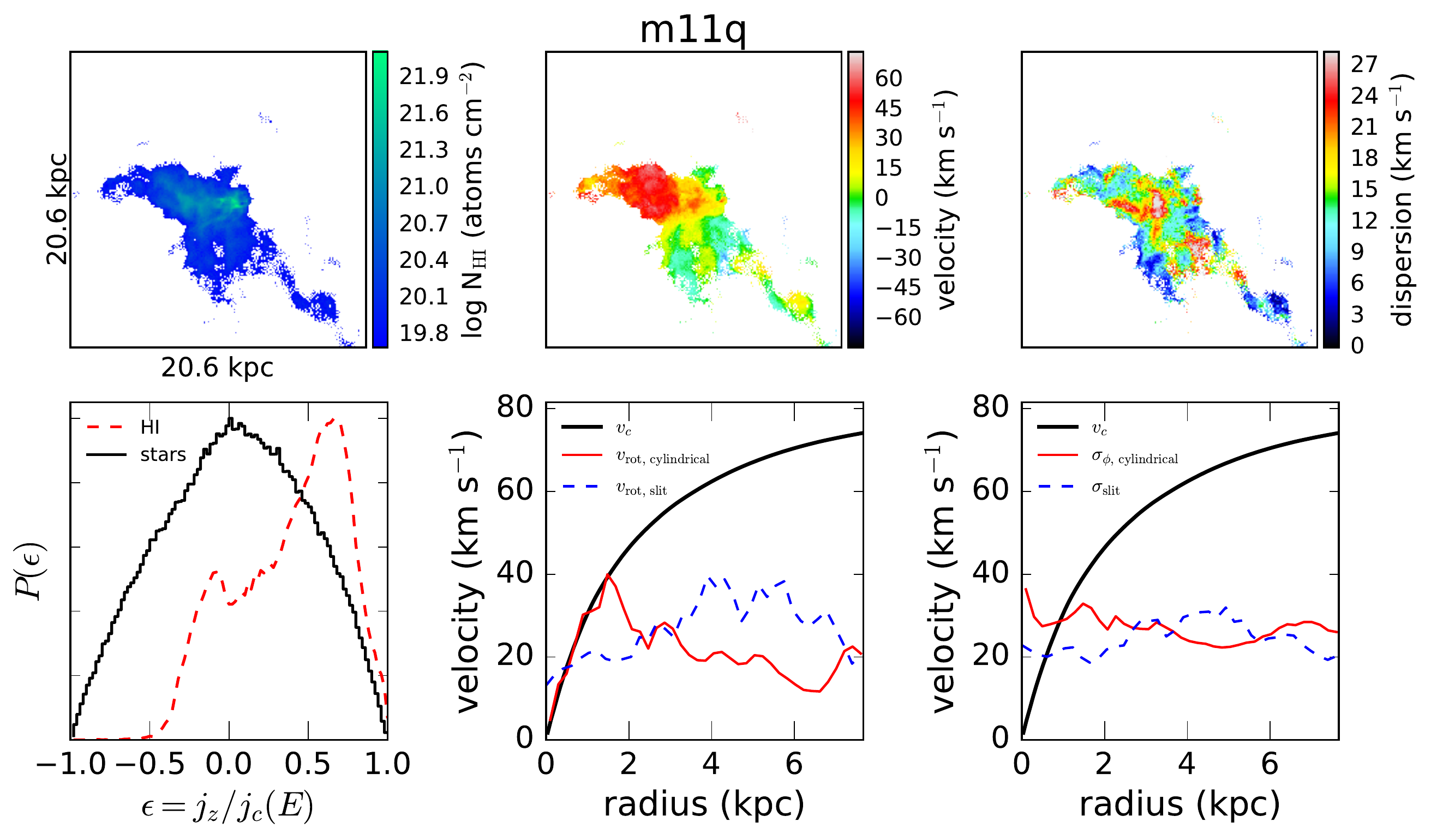}
\caption{\textbf{Top}: HI column density (left), line-of-sight velocity (middle) and velocity dispersion (right) for \texttt{m11q} ($\rm M_{\rm star}=10^{8.6} \rm M_{\odot}$). \textbf{Bottom}: Left: orbital circularity distribution (Equation~\ref{eqn:circularity}). Middle: HI rotation curves, compared to the halo circular velocity, $v_c = \sqrt{GM(<r)/r}$. Right: HI velocity dispersion. See Appendix~\ref{sec:idvid_gals} for details.}\label{fig:m11q}
\end{minipage}
\end{figure*}

\begin{figure*}
\centering
\begin{minipage}[b]{.4\textwidth}
\includegraphics[width=\columnwidth]{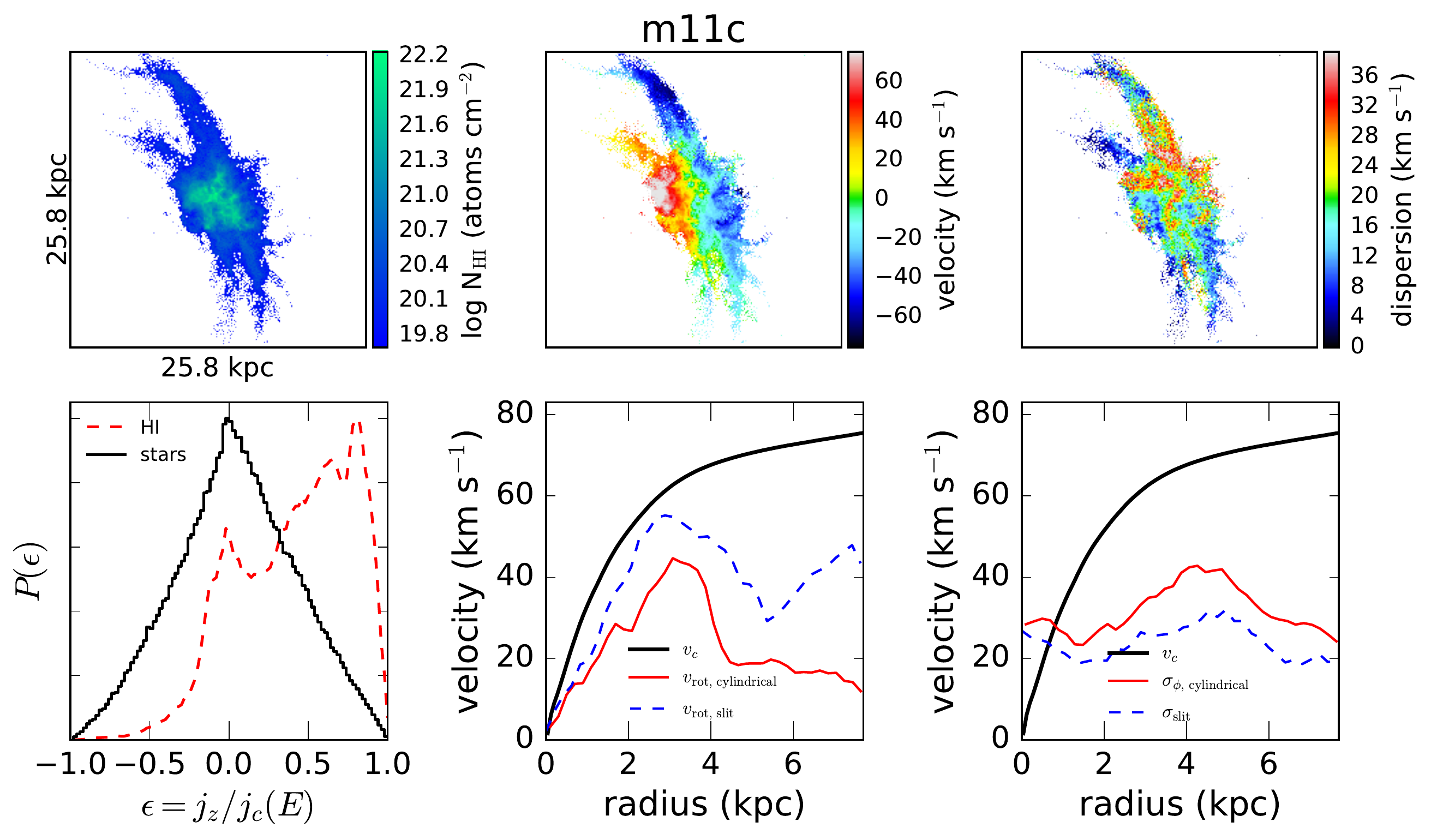}
\caption{\textbf{Top}: HI column density (left), line-of-sight velocity (middle) and velocity dispersion (right) for \texttt{m11c} ($\rm M_{\rm star}=10^{9.0} \rm M_{\odot}$). \textbf{Bottom}: Left: orbital circularity distribution (Equation~\ref{eqn:circularity}). Middle: HI rotation curves, compared to the halo circular velocity, $v_c = \sqrt{GM(<r)/r}$. Right: HI velocity dispersion. See Appendix~\ref{sec:idvid_gals} for details.}
\label{fig:m11c}
\end{minipage}\qquad
\begin{minipage}[b]{.4\textwidth}
\includegraphics[width=\columnwidth]{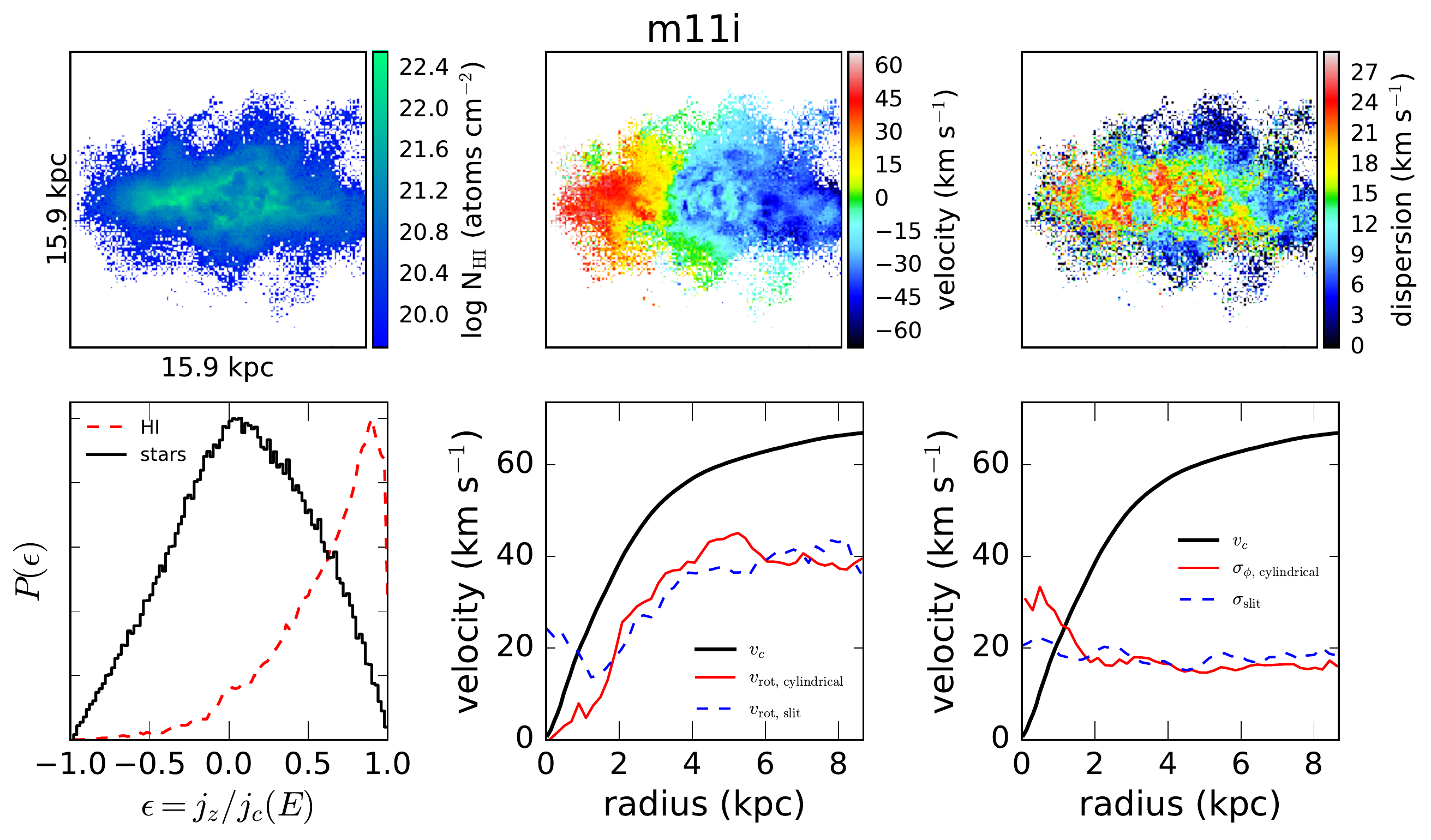}
\caption{\textbf{Top}: HI column density (left), line-of-sight velocity (middle) and velocity dispersion (right) for \texttt{m11i} ($\rm M_{\rm star}=10^{9.0} \rm M_{\odot}$). \textbf{Bottom}: Left: orbital circularity distribution (Equation~\ref{eqn:circularity}). Middle: HI rotation curves, compared to the halo circular velocity, $v_c = \sqrt{GM(<r)/r}$. Right: HI velocity dispersion. See Appendix~\ref{sec:idvid_gals} for details.}\label{fig:m11i}
\end{minipage}
\end{figure*}

\begin{figure*}
\centering
\begin{minipage}[b]{.4\textwidth}
\includegraphics[width=\columnwidth]{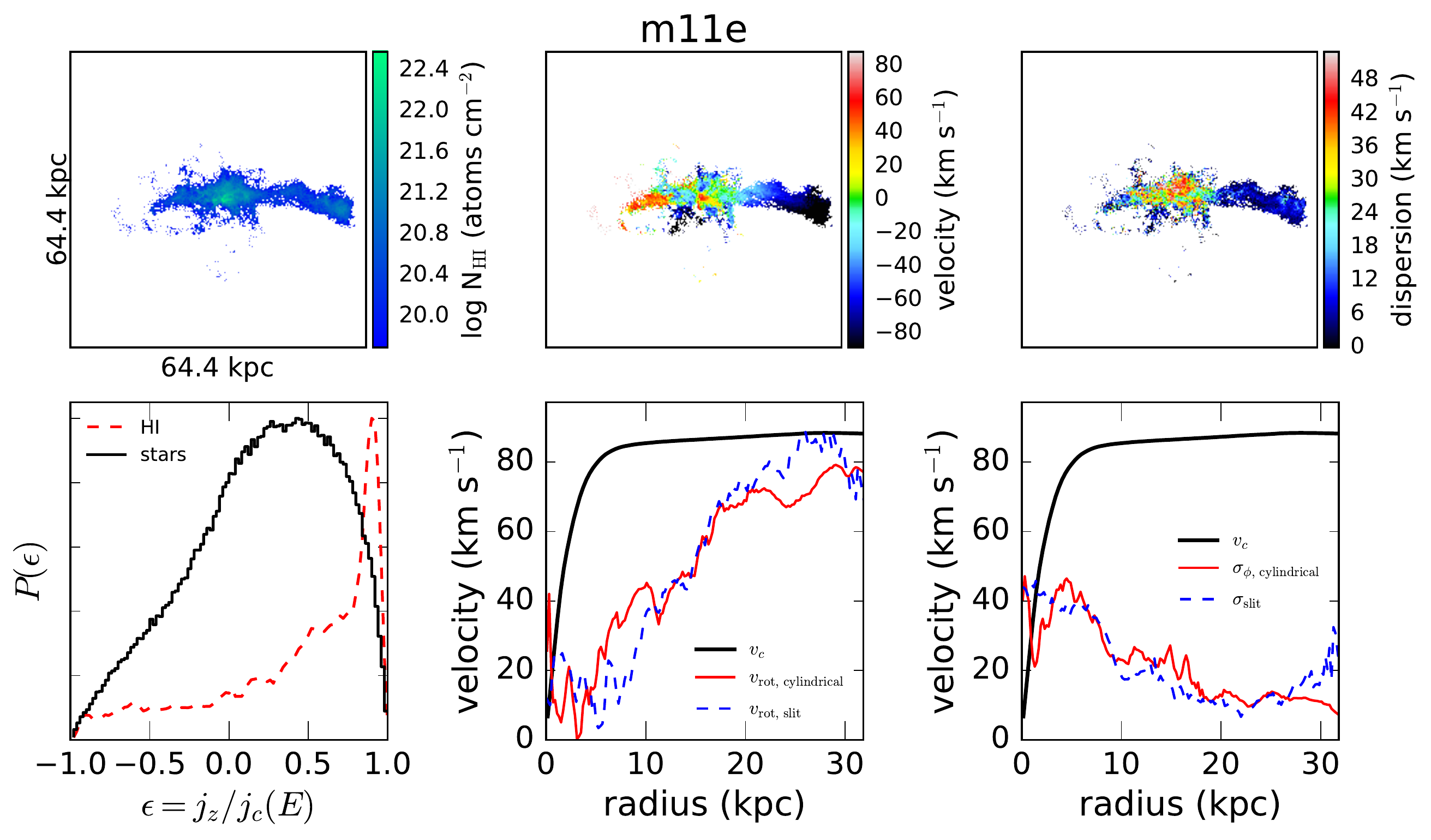}
\caption{\textbf{Top}: HI column density (left), line-of-sight velocity (middle) and velocity dispersion (right) for \texttt{m11e} ($\rm M_{\rm star}=10^{9.1} \rm M_{\odot}$). \textbf{Bottom}: Left: orbital circularity distribution (Equation~\ref{eqn:circularity}). Middle: HI rotation curves, compared to the halo circular velocity, $v_c = \sqrt{GM(<r)/r}$. Right: HI velocity dispersion. See Appendix~\ref{sec:idvid_gals} for details.}
\label{fig:m11e}
\end{minipage}\qquad
\begin{minipage}[b]{.4\textwidth}
\includegraphics[width=\columnwidth]{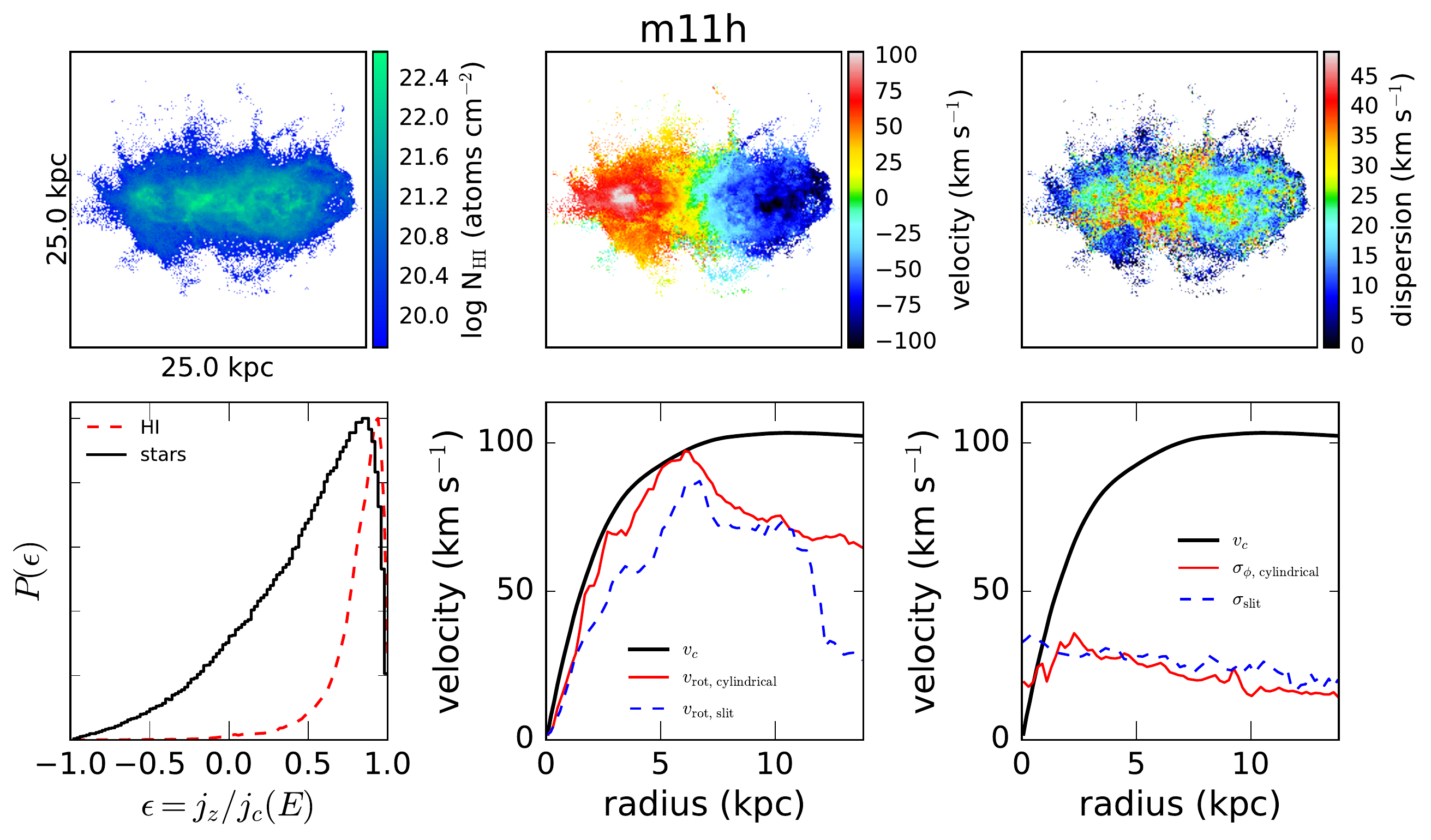}
\caption{\textbf{Top}: HI column density (left), line-of-sight velocity (middle) and velocity dispersion (right) for \texttt{m11h} ($\rm M_{\rm star}=10^{9.6} \rm M_{\odot}$). \textbf{Bottom}: Left: orbital circularity distribution (Equation~\ref{eqn:circularity}). Middle: HI rotation curves, compared to the halo circular velocity, $v_c = \sqrt{GM(<r)/r}$. Right: HI velocity dispersion. See Appendix~\ref{sec:idvid_gals} for details.}\label{fig:m11h}
\end{minipage}
\end{figure*}

\begin{figure*}
\centering
\begin{minipage}[b]{.4\textwidth}
\includegraphics[width=\columnwidth]{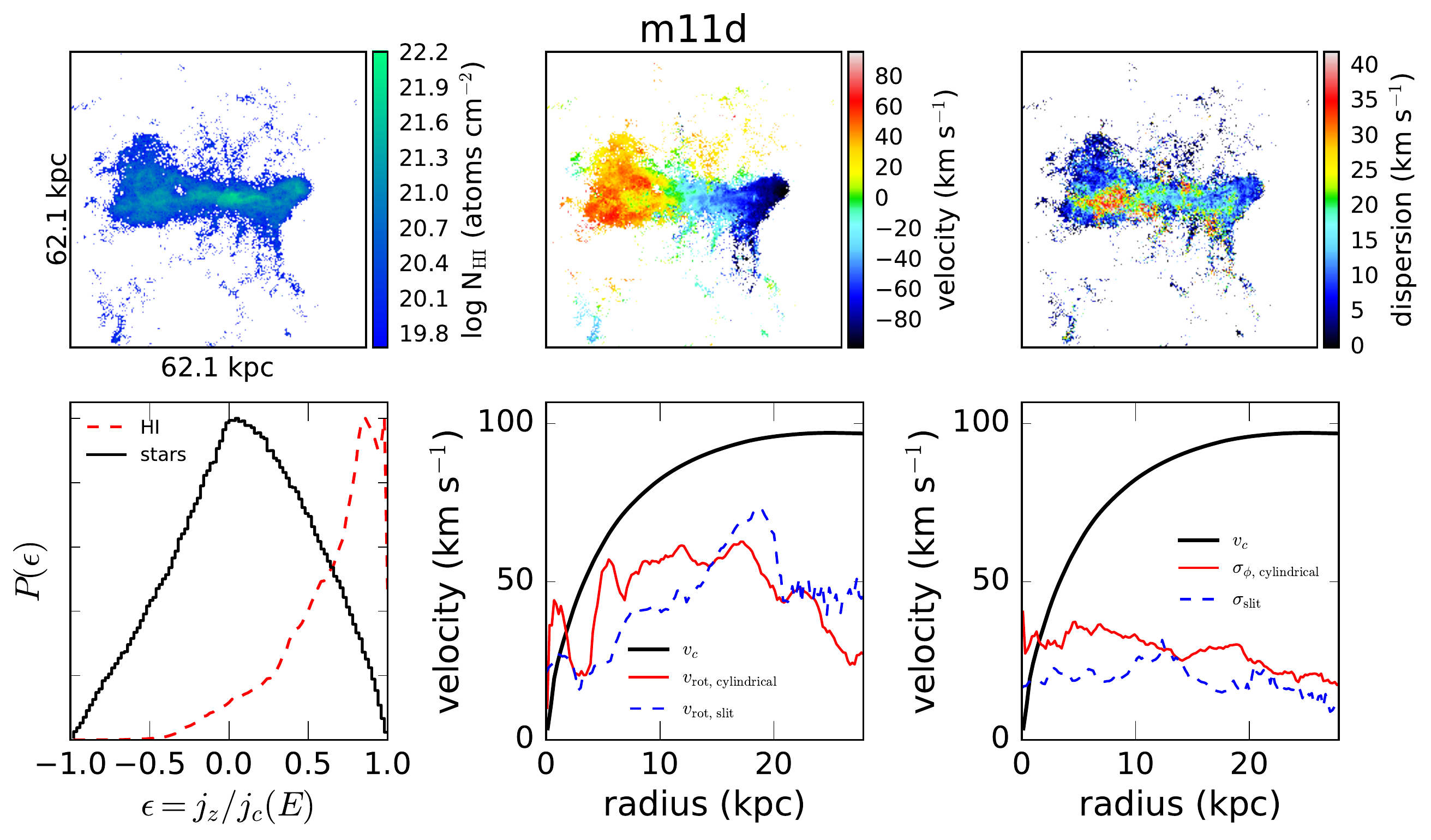}
\caption{\textbf{Top}: HI column density (left), line-of-sight velocity (middle) and velocity dispersion (right) for \texttt{m11d} ($\rm M_{\rm star}=10^{9.6} \rm M_{\odot}$). \textbf{Bottom}: Left: orbital circularity distribution (Equation~\ref{eqn:circularity}). Middle: HI rotation curves, compared to the halo circular velocity, $v_c = \sqrt{GM(<r)/r}$. Right: HI velocity dispersion. See Appendix~\ref{sec:idvid_gals} for details.}
\label{fig:m11d}
\end{minipage}\qquad
\begin{minipage}[b]{.4\textwidth}
\includegraphics[width=\columnwidth]{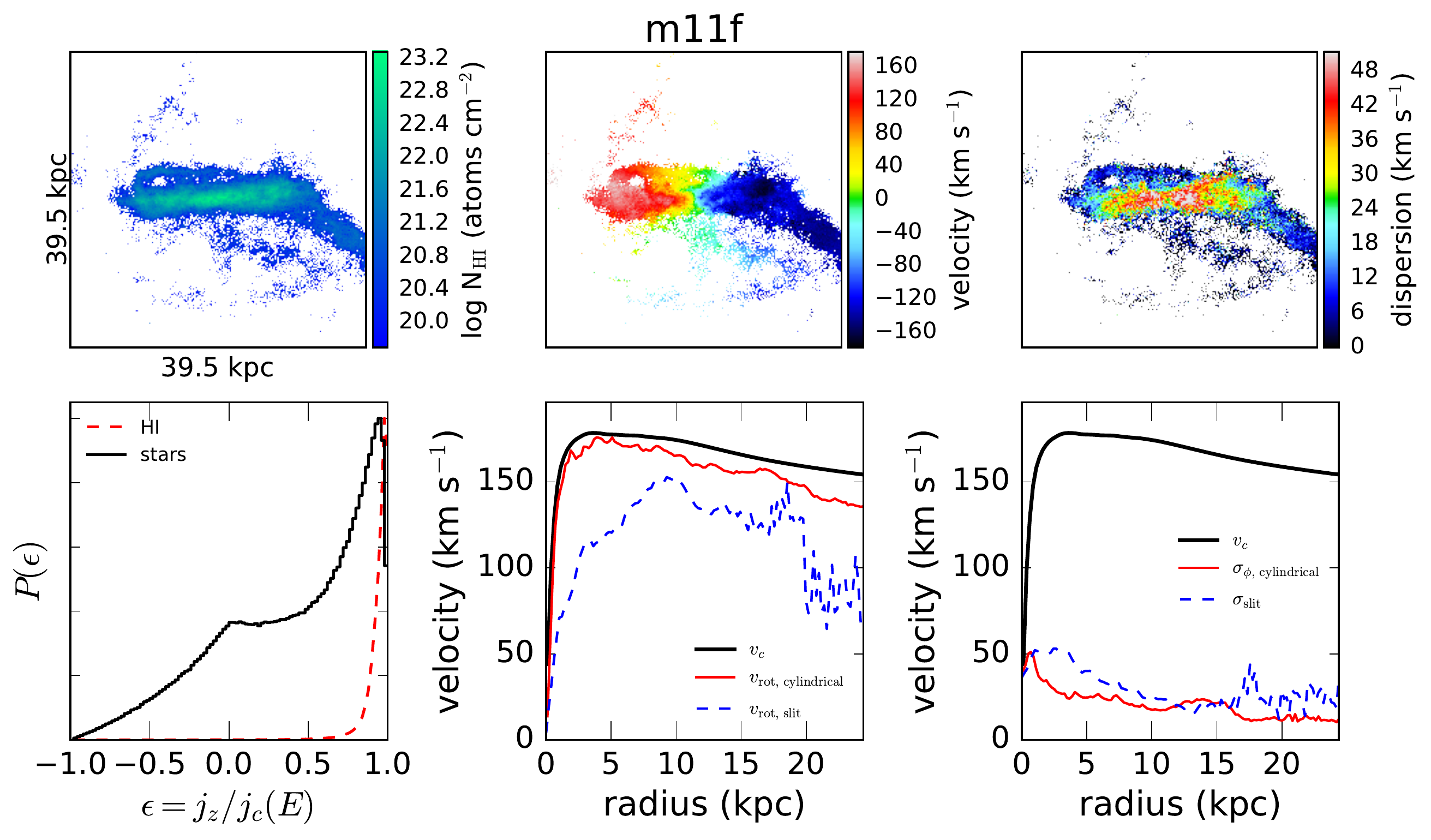}
\caption{\textbf{Top}: HI column density (left), line-of-sight velocity (middle) and velocity dispersion (right) for \texttt{m11f} ($\rm M_{\rm star}=10^{10.4} \rm M_{\odot}$). \textbf{Bottom}: Left: orbital circularity distribution (Equation~\ref{eqn:circularity}). Middle: HI rotation curves, compared to the halo circular velocity, $v_c = \sqrt{GM(<r)/r}$. Right: HI velocity dispersion. See Appendix~\ref{sec:idvid_gals} for details.}\label{fig:m11f}
\end{minipage}
\end{figure*}

\begin{figure*}
\centering
\begin{minipage}[b]{.4\textwidth}
\includegraphics[width=\columnwidth]{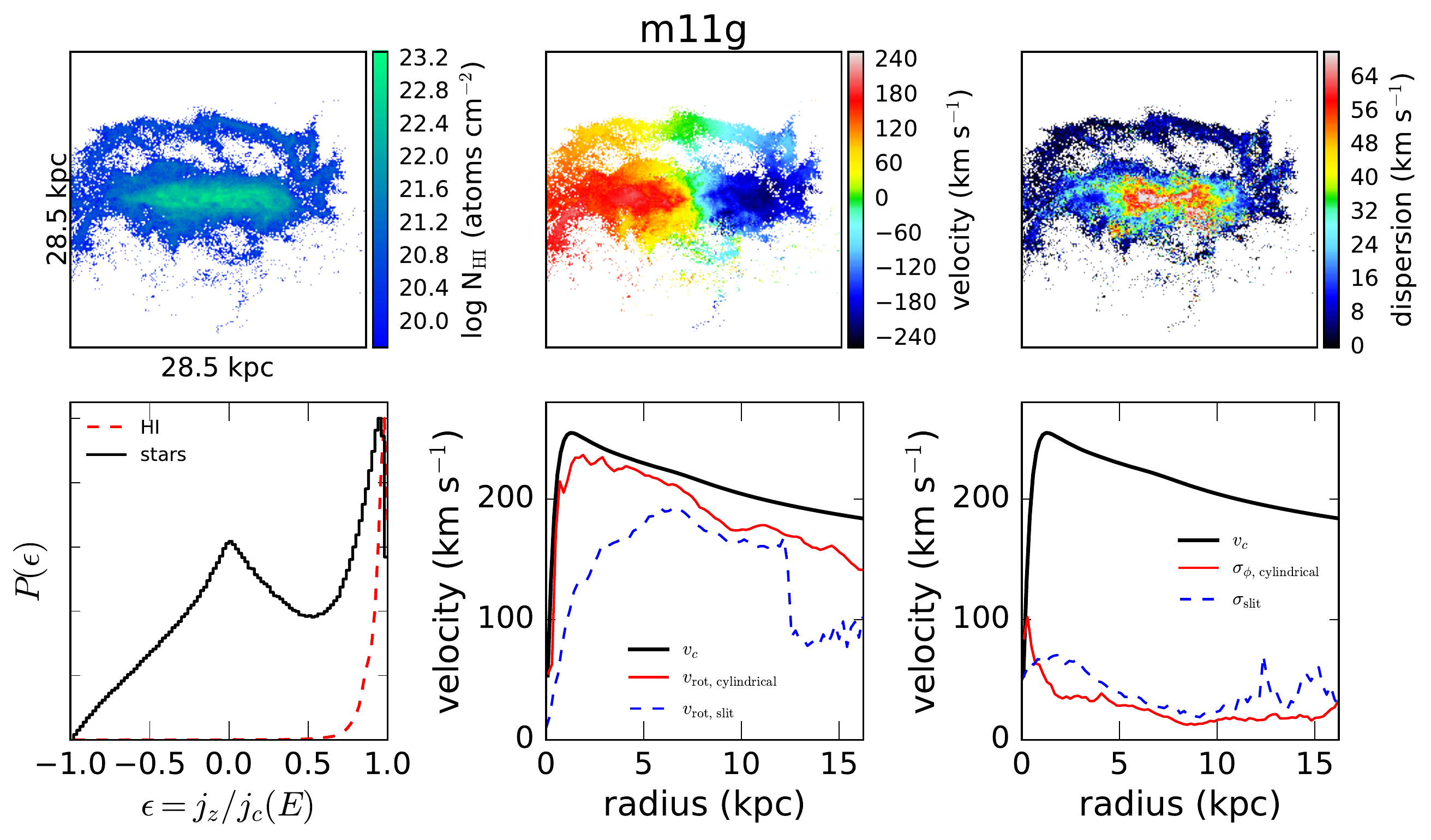}
\caption{\textbf{Top}: HI column density (left), line-of-sight velocity (middle) and velocity dispersion (right) for \texttt{m11g} ($\rm M_{\rm star}=10^{10.7} \rm M_{\odot}$). \textbf{Bottom}: Left: orbital circularity distribution (Equation~\ref{eqn:circularity}). Middle: HI rotation curves, compared to the halo circular velocity, $v_c = \sqrt{GM(<r)/r}$. Right: HI velocity dispersion. See Appendix~\ref{sec:idvid_gals} for details.}
\label{fig:m11g}
\end{minipage}\qquad
\begin{minipage}[b]{.4\textwidth}
\includegraphics[width=\columnwidth]{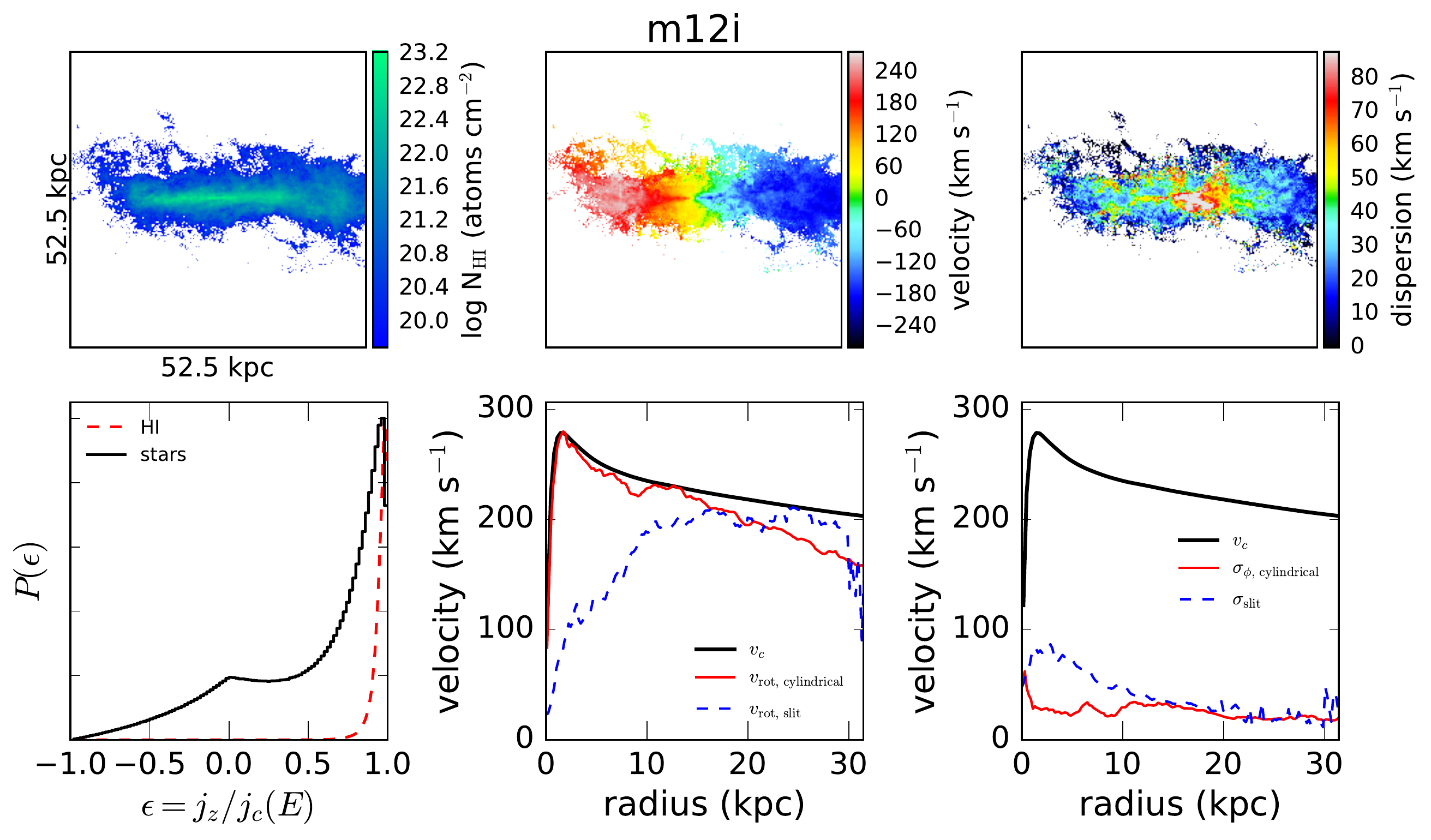}
\caption{\textbf{Top}: HI column density (left), line-of-sight velocity (middle) and velocity dispersion (right) for \texttt{m12i} ($\rm M_{\rm star}=10^{10.8} \rm M_{\odot}$). \textbf{Bottom}: Left: orbital circularity distribution (Equation~\ref{eqn:circularity}). Middle: HI rotation curves, compared to the halo circular velocity, $v_c = \sqrt{GM(<r)/r}$. Right: HI velocity dispersion. See Appendix~\ref{sec:idvid_gals} for details.}\label{fig:m12i}
\end{minipage}
\end{figure*}

\begin{figure*}
\centering
\begin{minipage}[b]{.4\textwidth}
\includegraphics[width=\columnwidth]{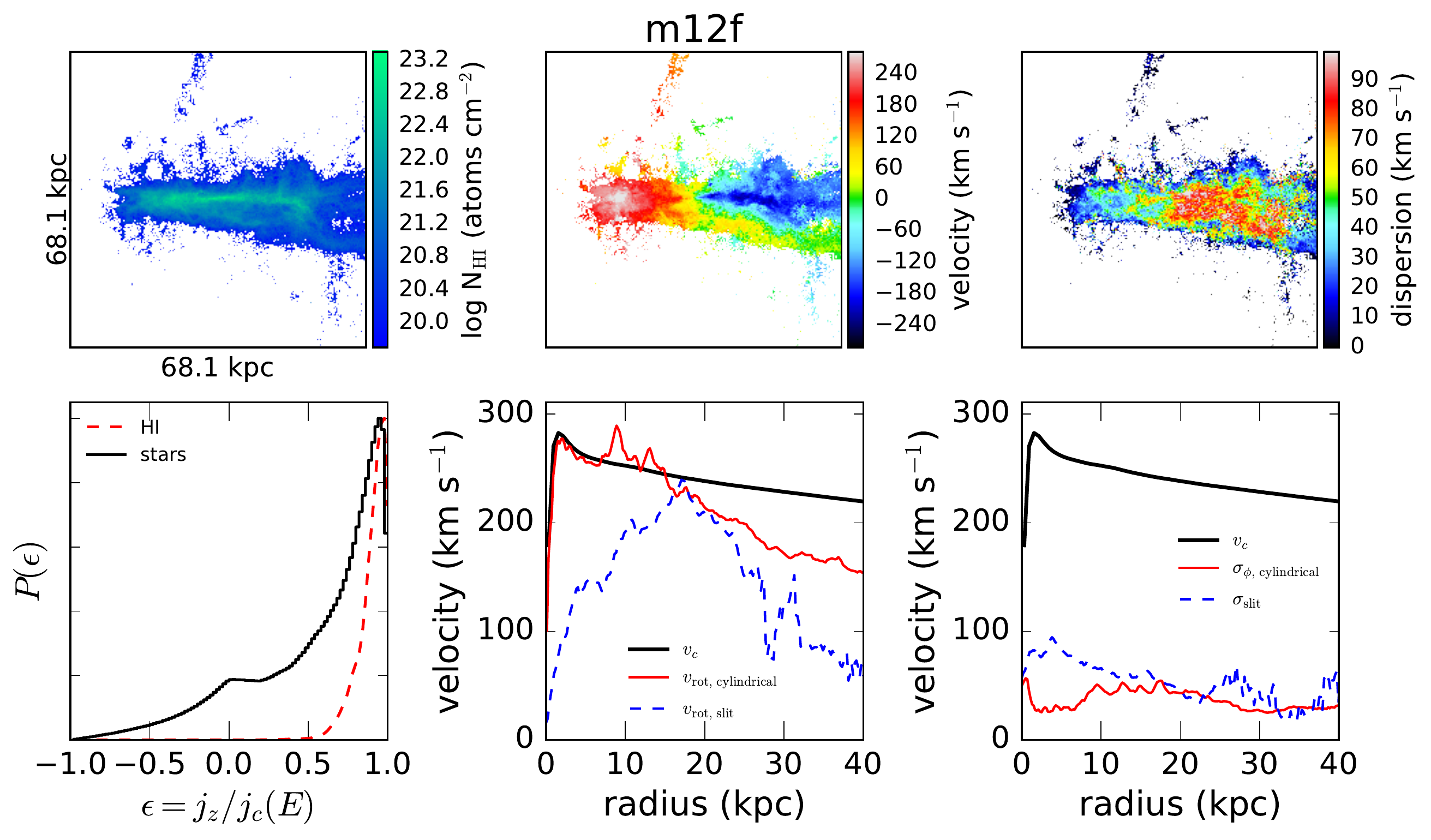}
\caption{\textbf{Top}: HI column density (left), line-of-sight velocity (middle) and velocity dispersion (right) for \texttt{m12f} ($\rm M_{\rm star}=10^{10.9} \rm M_{\odot}$). \textbf{Bottom}: Left: orbital circularity distribution (Equation~\ref{eqn:circularity}). Middle: HI rotation curves, compared to the halo circular velocity, $v_c = \sqrt{GM(<r)/r}$. Right: HI velocity dispersion. See Appendix~\ref{sec:idvid_gals} for details.}
\label{fig:m12f}
\end{minipage}\qquad
\begin{minipage}[b]{.4\textwidth}
\includegraphics[width=\columnwidth]{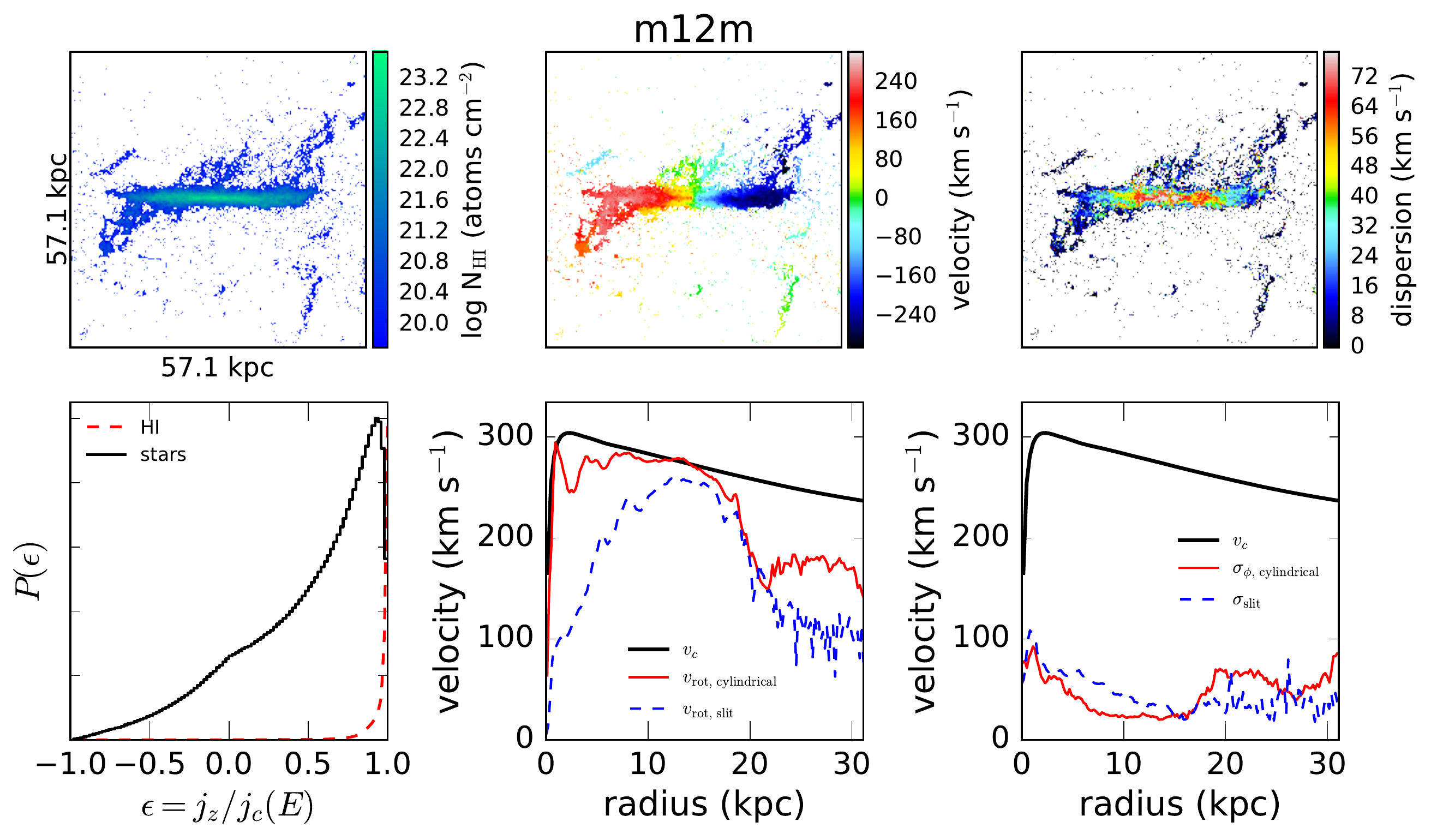}
\caption{\textbf{Top}: HI column density (left), line-of-sight velocity (middle) and velocity dispersion (right) for \texttt{m12m} ($\rm M_{\rm star}=10^{11.1} \rm M_{\odot}$). \textbf{Bottom}: Left: orbital circularity distribution (Equation~\ref{eqn:circularity}). Middle: HI rotation curves, compared to the halo circular velocity, $v_c = \sqrt{GM(<r)/r}$. Right: HI velocity dispersion. See Appendix~\ref{sec:idvid_gals} for details.}\label{fig:m12m}
\end{minipage}
\end{figure*}

Index of figures: 
\texttt{m10e}:~\ref{fig:m10e}, 
\texttt{m10q}:~\ref{fig:m10q}, 
\texttt{m10g}:~\ref{fig:m10g}, 
\texttt{m10h}:~\ref{fig:m10h}, 
\texttt{m10j}:~\ref{fig:m10j}, 
\texttt{m10k}:~\ref{fig:m10k},
\texttt{m10y}:~\ref{fig:m10y}, 
\texttt{m10f}:~\ref{fig:m10f}, 
\texttt{m10l}:~\ref{fig:m10l}, 
\texttt{m10m}:~\ref{fig:m10m}, 
\texttt{m10z}:~\ref{fig:m10z}, 
\texttt{m11b}:~\ref{fig:m11b}, 
\texttt{m11a}:~\ref{fig:m11a}, 
\texttt{m11q}:~\ref{fig:m11q}, 
\texttt{m11c}:~\ref{fig:m11c}, 
\texttt{m11i}:~\ref{fig:m11i}, 
\texttt{m11e}:~\ref{fig:m11e}, 
\texttt{m11h}:~\ref{fig:m11h}, 
\texttt{m11d}:~\ref{fig:m11d}, 
\texttt{m11f}:~\ref{fig:m11f}, 
\texttt{m11g}:~\ref{fig:m11g}, 
\texttt{m12i}:~\ref{fig:m12i}, 
\texttt{m12f}:~\ref{fig:m12f}, 
\texttt{m12m}:~\ref{fig:m12m}.

%%%%%%%%%%%%%%%%%%%%%%%%%%%%%%%%%%%%%%%%%%%%%%%%%%

% Don't change these lines
\bsp	% typesetting comment
\label{lastpage}
\end{document}